\documentclass{article}
\usepackage[english]{babel}
\usepackage{todonotes}
\usepackage{listings}
 \usepackage{color} %red, green, blue, yellow, cyan, magenta, black, white
 \definecolor{mygreen}{RGB}{28,172,0} % color values Red, Green, Blue
 \definecolor{mylilas}{RGB}{170,55,241}
 \usepackage{graphics} %new
 \usepackage{graphicx}
\usepackage{caption}
\usepackage{subcaption}
\usepackage{amsmath}
\usepackage{amssymb}
\usepackage{mathtools}
\usepackage{overpic}
\usepackage{amsthm}
\usepackage{changepage}
\usepackage{hyperref}

\providecommand{\keywords}[1]{\textbf{Key words: } #1}

\usepackage{booktabs,array}

\usepackage{pgfplots}
\pgfplotsset{grid style={dotted,gray}}
\pgfplotsset{compat=newest}
\pgfplotsset{plot coordinates/math parser=false}
\newlength\figureheight
\newlength\figurewidth

\usepackage{stfloats} % for positioning of figure* on the same page
\usepackage{array, multirow}

\usepackage[pass]{geometry}
\usepackage[outdir=./]{epstopdf}
\usepackage{url}
\usepackage{tabstackengine}
\setstacktabbedgap{1.5ex}%               sets gap between columns
\setstackgap{L}{1.2\normalbaselineskip}% sets baselineskip of rows
\theoremstyle{definition}

\newtheorem{remark}{Remark}

\definecolor{KTHblue}{RGB}{25,105,188}
\definecolor{KTHlblue}{RGB}{22,159,219}
\definecolor{KTHyellow}{RGB}{251,186,0}
\definecolor{KTHred}{RGB}{176,9,48}
\definecolor{KTHlred}{RGB}{231,51,57}
\definecolor{KTHgreen}{RGB}{98,146,46}
\definecolor{KTHlgreen}{RGB}{175,202,11}
\definecolor{KTHpink}{RGB}{219,81,151}

\colorlet{fred}{KTHlred}
\colorlet{fblue}{KTHlblue}

\addto{\captionsenglish}{}
\renewcommand{\vec}[1]{\boldsymbol{#1}}

\addto{\captionsenglish}{}

\newgeometry{left=2.5cm,right =2.5cm,top= 2.5cm,bottom = 2.5cm}

\usepackage[english]{nomencl}
\makenomenclature
\newcommand{\thickhline}{%
	\noalign {\ifnum 0=`}\fi \hrule height 1pt
	\futurelet \reserved@a \@xhline
}
\title{A Locally Corrected Multiblob Method with Hydrodynamically Matched Grids for the Stokes Mobility Problem}

\author{\underline{Anna Broms}$^{a}$, e-mail: \url{annabrom@kth.se}\\
	Mattias Sandberg$^{a}$, e-mail: \url{msandb@kth.se}\\
	Anna-Karin Tornberg$^{a}$, e-mail: \url{akto@kth.se}\\\\
	$^{a}$ Department of Mathematics\\
	KTH Royal Institute of Technology\\
	Lindstedtsv{\"a}gen 25, 114 28 Stockholm, Sweden\\
}
\date{\today}

\begin{document}
        \lstset{language=Matlab,%
                %basicstyle=\color{red},
                breaklines=true,%
                morekeywords={matlab2tikz},
                keywordstyle=\color{black},%
                morekeywords=[2]{1}, keywordstyle=[2]{\color{black}},
                identifierstyle=\color{black},%
                stringstyle=\color{mylilas},
                commentstyle=\color{mygreen},%
                showstringspaces=false,%without this there will be a symbol in the places where there is a space
                numbers=left,%
                numberstyle={\tiny \color{black}},% size of the numbers
                numbersep=9pt, % this defines how far the numbers are from the text
                emph=[1]{for,end,break},emphstyle=[1]\color{blue}, %some words to emphasise
                %emph=[2]{word1,word2}, 6 emphstyle=[2]{style},    
        }
\maketitle

\begin{abstract}
	Inexpensive numerical methods are key to enabling simulations of systems of a large number of particles of different shapes in Stokes flow. Several approximate methods have been introduced for this purpose. We study the accuracy of the multiblob method for solving the Stokes mobility problem in free space, where the 3D geometry of a particle surface is discretized with spherical blobs and the pair-wise interaction between blobs is described by the RPY-tensor. The paper aims to investigate and improve on the magnitude of the error in the solution velocities of the Stokes mobility problem using a combination of two different techniques: an optimally chosen grid of blobs and a pair-correction inspired by Stokesian dynamics. 
Different optimisation strategies to determine a grid with a certain number of blobs are presented with the aim of matching the hydrodynamic response of a single accurately described ideal particle, alone in the fluid. It is essential to obtain small errors in this self-interaction as they determine the basic error level in a system of well-separated particles. With a good match, reasonable accuracy can be obtained even with coarse blob-resolutions of the particle surfaces. The error in the self-interaction is however sensitive to the exact choice of grid parameters and simply hand-picking a suitable geometry of blobs can lead to errors several orders of magnitude larger in size.  
%Different optimisation strategies are presented to match 
%We can choose to match rotational or translational properties of the ideal particle, or both, with the centers of the blobs placed on a geometric surface interior to the surface of the model particle. 
The pair-correction is local and cheap to apply, and reduces on the error for moderately separated particles and particles in close proximity.	
Two different types of geometries are considered: spheres and axisymmetric rods with smooth caps. The error in solutions to mobility problems is quantified for particles of varying inter-particle distances for systems containing a few particles, comparing to an accurate solution based on a second kind BIE-formulation where the quadrature error is controlled by employing quadrature by expansion (QBX).\\  

\end{abstract}
\\

\vspace*{1ex}

\noindent\textbf{Highlights}
%3-5 highlights
\begin{itemize}
	\itemsep0em
    \item Rigid rods and spheres are studied in Stokes flow in 3D free space.
    \item We improve on the accuracy of the multiblob method for the Stokes mobility problem.
    \item An optimal grid of blobs match the hydrodynamic interaction of a model particle.
    \item A self-interaction error dominant in the far-field is reduced with the optimal grid.
    \item Pair-corrections of Stokesian dynamics type reduce errors in the near-field. 
\end{itemize}
\clearpage
\tableofcontents
\clearpage
\section{Introduction}\label{sec:intro}
A fluid with immersed rigid particles on the micro scale can be modeled by the linear Stokes equations, along with no slip boundary conditions on all particle surfaces. The Stokes equations constitute the low Reynolds number limit of the Navier-Stokes equations, applicable under the assumption that the inertia of the particles is negligible compared to viscous effects. In such a fluid-particle system, every particle affects every other particle, due to the long range of the hydrodynamic interactions, meaning that the motion of all particles are coupled through the fluid. Examples of fluid-particle systems of this type are found both in biology and industry. Industrial applications are vast in materials science, with studies of flows of polymers, fibrils and fibers, and the forming of gels and crystalline phases \cite{Hakansson2014,Hakansson2016,KrishneGowda2019,Yan2019,Gowda2022}. Describing the dynamics on the micro level is key to understanding macro level properties of such processes to manufacture novel materials. 
%, phenomena studied for nano-cellulose crystals (CNCs) and nano-cellulose fibrils (CNFs) immersed in a fluid in 

The mobility problem for rigid particles in a Stokesian fluid is that of computing the translational and angular velocities of a set of non-deformable bodies, given assigned net forces and torques such as e.g.~gravity or electrostatic forces on every particle. We will focus on the free space problem in 3D, considering no confinements or periodicities for the particles.
The mobility problem for a system of $p$ particles can mathematically be stated as $\vec U=\vec M\vec F$, where $\vec F\in \mathbb R^{6p}$ denotes a vector of all (known) net forces and torques, and $\vec U\in \mathbb R^{6p}$ denotes a vector of all translational and angular velocities to be computed, i.e.~
\begin{equation}
\begin{aligned}
\vec F^T = \begin{bmatrix}
{\vec f^1}^T &  {\vec t^1}^T &\hdots &{\vec f^p}^T & {\vec t^p}^T
\end{bmatrix},
\vec U^T = \begin{bmatrix}
{\vec u^1}^T & {\vec{\omega}^1}^T &\hdots &{\vec u^p}^T & {\vec{\omega}^p}^T
\end{bmatrix},
\end{aligned}
\end{equation}
with $\vec f^i$ and $\vec t^i$ the net force and torque on particle $i$ and $\vec u^i$ and $\vec{\omega}^i$ the translational and rotational velocities of particle $i$. The mobility matrix $\vec M\in \mathbb R^{6p}\times \mathbb R^{6p}$ is dense, symmetric and depends on the position and orientation of all particles in the system \cite{Kim1991}. In addition, the mobility matrix is positive definite, which is a consequence of the dissipative nature of a Stokesian suspension \cite{Kim1991}. The inverse of the mobility matrix is termed the \emph{resistance} matrix, $\vec R$, and appears in the related resistance problem of computing forces and torques, given assigned particle velocities. We will return to the resistance matrix when discussing techniques for improving on the accuracy for a solution to the mobility problem, as presented in Section \ref{sec:pair}.

The molecules of the fluid collide with each other and with the immersed particles at a high rate, resulting in Brownian motion for small enough colloidal particles. This stochastic behaviour can be characterised by the overdamped Langevin equation, which is a stochastic differential equation incorporating not only the action of the mobility matrix, but also the action of its square root and divergence \cite{Ermak1978,Graham}. The last two quantities have to be approximated from matrix vector products of the form $\vec M\vec F$ for some ``force'' vector $\vec F$, using e.g.~a Krylov method for approximating the square root \cite{Ando2012} and a so called random finite difference quotient for the divergence term \cite{Delong2014}. The large number of such matrix-vector evaluations needed in either a dynamic simulation to determine a trajectory or for drawing statistical conclusions about some physical property of interest (such as e.g.~ the mean squared displacement, diffusion coefficients or equilibrium distributions) emphasizes the need of a method for which the matrix vector product is fast to evaluate, also for systems with many particles.  In a Brownian simulation, there are error contributions from several sources: modelling errors in the description of the geometry and in the physical assumptions for the studied particle type, a statistical error in determining physical quantities as averages of a large number of realisations or geometrical configurations, the time discretisation error and an error related to the numerical solution of the (deterministic) Stokes mobility problem. The latter is important to control also in non-Brownian simulations. The aim of this work is to understand the deterministic error related to solving the mobility problem and we assume no Brownian motion.

Specifically, this paper aims at studying the accuracy of a so called \emph{multiblob method}, where the surface of each rigid particle is discretized with spherical blobs and the blobs belonging to one particle are restricted to move as a rigid body, constrained by net forces located at the center of each blob. The method is simple to implement in its vanilla version and allows for a great flexibility in the particle shapes that can be studied without altering the method as such. In addition, particles of varying shapes and sizes can easily be handled within the same simulation and can be coupled to a fast method for evaluating the action of the blob-blob mobility matrix for different periodicities, which allows for a large number of particles to be studied at a low cost. In \cite{Sprinkle2017}, large systems of $\mathcal O(1000)$ \emph{Brownian} particles have been studied and in \cite{USABIAGA2016}, fast approximate solution techniques are discussed making the complexity close to linear in the number of blobs. The idea of forming larger particles from spheres was first introduced by Kirkwood in 1954 \cite{Kirkwood1954} and there is a large collection of work on methods of this type, including but not limited to  \cite{Cichocki1994,GarciaDeLaTorre2000,Fernandes2002,Zurita-Gotor2007,Kutteh2010,Ortega2011,Molina2013,Poblete2014,Dugosz2015,Pandey2016}. Multiblob methods have recently been employed for studying Brownian motion and active slip in a number of works by Usabiaga and coauthors \cite{Delong2015,USABIAGA2016,Sprinkle2017,Usabiaga2021}, and by Brosseau and coauthors \cite{Brosseau2019,Brosseau2021}, but also in connection to Stokesian dynamics, by the group of Swan \cite{Swan2011,Fiore2018,Fiore2019}. A similar technique is employed in \cite{Vazquez-Quesada2014}, where springs constrain the blobs to move as one body. Despite the multiblob method being approximate in its very nature, we present a strategy to understand, control and improve on its accuracy.  

Several accurate methods exist for solving the Stokes mobility problem, among which boundary integral equation (BIE) methods form an important class \cite{Bagge2021,AfKlinteberg2016,Corona2017}. Integral equations have successfully been employed for various particle shapes and various domains in 3D (including confinements \cite{Bagge2021} and/or periodicities \cite{AfKlinteberg2016}). A special quadrature method has to be used for accurate numerical treatment of the singularities and near-singularities appearing in any integral equation formulation for evaluation on, or close to, a boundary; one example of such quadrature for a double layer formulation is quadrature by expansion (QBX) developed by af Klinteberg \& Tornberg \cite{AfKlinteberg2016} and Bagge \& Tornberg \cite{Bagge2021}.  This solver, based on a second kind integral equation formulation, allows for accurate computations and has been used to evaluate the Stokes flow for $\mathcal O(100)$ closely interacting particles. However, to do so repeatedly for many time steps and/or statistical realisations would be unfeasibly slow, and in such situations, cheaper and hence less accurate methods must be used. In this paper, the QBX-based solver will serve two purposes: provide a reference solution when studying the accuracy of the multiblob method and be used to construct precomputable corrections to the multiblob method that are inexpensive to apply.

Multiple families of methods exist for computing the hydrodynamic interactions in a particle suspension, reviewed by Maxey in \cite{Maxey2017}. Another example of an approximate method is presented in the large collection of papers related to Stokesian Dynamics \cite{Brady1987,Brady1988,Brady2001,Swan2011,Swan2016,Fiore2019}, first introduced in 1987 by Brady, Durlofsky and Bossis in \cite{Brady1987}, where a near-field correction for each pair of spherical (or spheroidal) particles in close proximity is added to the global far-field resistance matrix, allowing for good approximations for very close particles and for widely separated particles. The near-field correction is constructed from lubrication expressions first presented by Jeffery \& Onishi \cite{Jeffrey1984a} and also listed in \cite{Kim1991}. However, the accuracy of Stokesian dynamics is worse for moderate particle separations than for very large or small separations \cite{Wilson2013}, a potential reason being that the lubrication expressions are dominant only for closely interacting pairs --  for moderately separated particles, it becomes evident that the additivity assumption of the resistance matrix does not hold \cite{Lefebvre-Lepot2015}. A potential cure to this problem is presented by Lefebvre-Lepot et al.~\cite{Lefebvre-Lepot2015}. In their paper, the hydrodynamic interaction is described using a multipole expansion with lubrication corrections, which however requires an evaluation of the lubrication field for all particles not in a closely interacting particle pair, relying on a multivariate interpolant that is not straight-forward to compute cheaply. Despite the possible drawbacks in Stokesian Dynamics, the multiblob method is coupled to that type of pair-corrections in \cite{Fiore2019}. We will here further develop this idea and study the accuracy of the coupling. In the results section, \ref{results}, we show numerically that in this setting, the additivity assumption is not limiting the accuracy of the method.
\begin{figure}[h!]
	\centering
	\begin{subfigure}[t]{0.35\textwidth}
		\centering	
		\includegraphics[trim = {3.0cm 19cm 10cm 2.2cm},clip,width=1.2\textwidth]{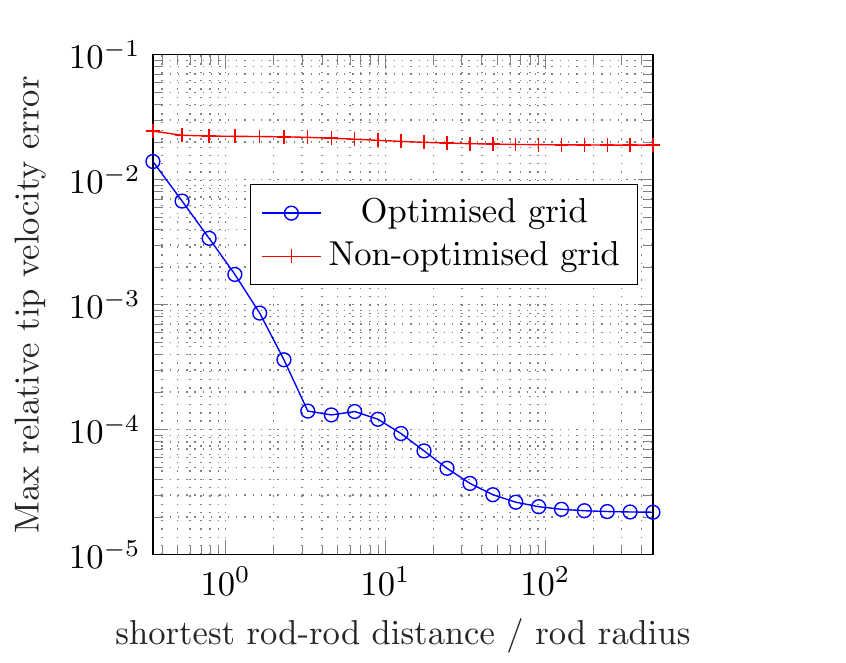}
		\caption{The maximum relative error in the tip velocity of the rods is plotted versus inter-particle distance. }
	\end{subfigure}~
	\begin{subfigure}[t]{0.33\textwidth}
		\centering
		\includegraphics[trim = {2.5cm 0cm 1.8cm 0.5cm},clip,width=0.95\textwidth]{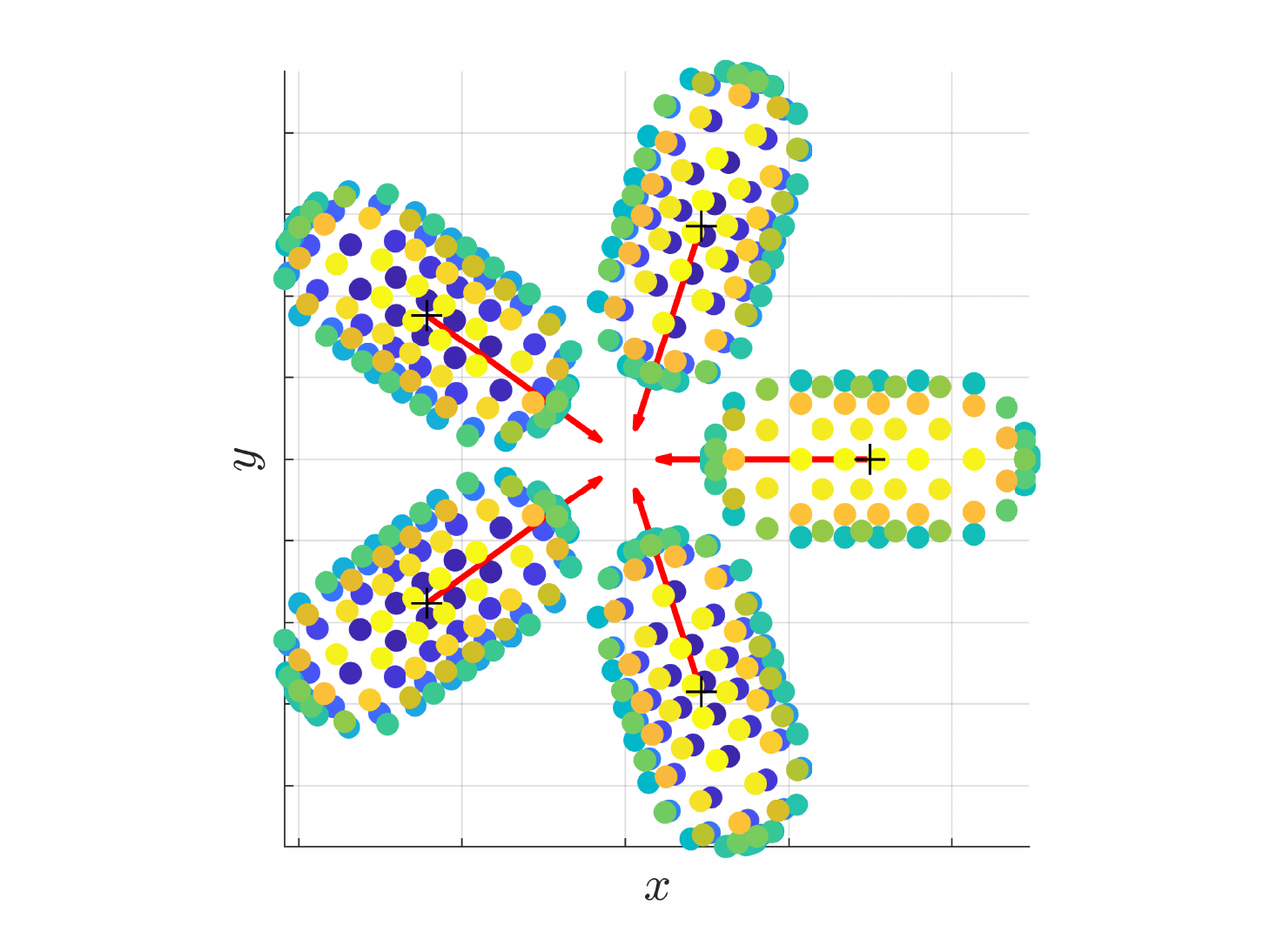}
		\caption{Geometry for a fixed shortest rod-rod distance, with indicated force and torque directions.}
		%\label{}
	\end{subfigure}~~	
	\caption{Numerical example with five rods of aspect ratio $L/R=4$ forming a circle. It is important to match the grid of the multiblob particle to the hydrodynamic response of the ideal particle that we would like to model. We gain up to three orders of magnitude in accuracy at larger distances by using the optimised grid. The non-optimised grid is obtained from the optimised grid parameters by adding a perturbation of $1\%$.}
	\label{important_opt}
\end{figure}

In Section \ref{sec:multiblob}, the details of the multiblob method are presented. In Section \ref{sec:match}, we describe how an optimisation problem can be solved to closely match the hydrodynamic response of a multiblob particle to that of a chosen particle type that we would like to model. We will talk about this hydrodynamic response as the self-interaction. The particle geometry and discretisations with blobs are presented, emphasizing that the multiblob particle is a model of an ideal particle with a certain geometric extension. By optimising the particle grid, we can obtain much smaller errors in the self-interaction, which will set the basic error level also in a multi-particle case. To display the importance of using such a matching technique, we consider a numerical example of rods in a circle centered at the origin in the xy-plane with rod tips pointing towards the center of the circle. Each rod is affected by a unit force towards the origin and a unit torque in the $z$-direction. We vary the circle radius and hence also the distance between rods. The relative error in the velocity at the tip of the rods, visualised in Figure \ref{important_opt}, is determined compared to a BIE reference solution. If an optimised grid is used, we gain up to three orders of magnitude in accuracy compared to using a grid with only slightly perturbed parameters. For closely interacting particles, note that there is a need to correct for interaction errors, also for the optimised grid. The specific parameters used in this example are presented in Table \ref{circle_params} in Section \ref{sec:opt_param} in the appendix, and in the manuscript we carefully describe why two seemingly similar discretisations of a particle can yield fundamentally different results in terms of accuracy. 

The optimisation technique is fit for use when the mobility matrix is known for a single particle from an accurate method or analytical expressions.  In Section \ref{sec:pair}, we introduce the technique to correct for pair-interactions, cheap to apply to large systems of multiblob particles. In Section \ref{results}, numerical results are presented for spheres and rods for systems consisting of a small number of particles at varying inter-particle distances, for which  comparisons can be made to results obtained with an accurate solver. The interplay between the self-interaction error and a pair-interaction error and how they can be improved will be discussed. We present the performance of what we refer to as the original multiblob method, with an optimised grid matching both translational and rotational properties of the ideal particle. This is done for particles of various resolutions and we also display the improved accuracy when applying pair-corrections. When a coarse discretisation is used, two solves based on two differently matched grids can also be combined for improved accuracy, with one grid optimised for translational properties and one for rotational. 
%%%%%%%%%%%%%%%%%%%%%%%%%%%%%%%%%%%%%%%%%%%%%%%%%%%%%%%%%%5
\subsection{The multiblob method}\label{sec:multiblob}
We will use the same description of the multiblob method as in the works by Usabiaga, Donev and coauthors \cite{Delong2015,Sprinkle2017,USABIAGA2016,Usabiaga2021}, by Brosseau et al.~in \cite{Brosseau2019,Brosseau2021} and by Swan and coauthors \cite{Swan2016,Fiore2018}. See especially the work by Usabiaga et al.~\cite{USABIAGA2016}, which presents the method in detail and also investigates the accuracy of the multiblob method in its deterministic setting for spheres in free space. A list of commonly used notation in this paper is collected for reference in Table \ref{symlist}.

\begin{figure}[h!]
	\centering	
	\begin{subfigure}[t]{0.3\textwidth}
		\centering
		\hspace*{-5ex}
		\includegraphics[trim = {3.9cm 1cm 3.3cm 1.7cm},clip,width=1.05\textwidth]{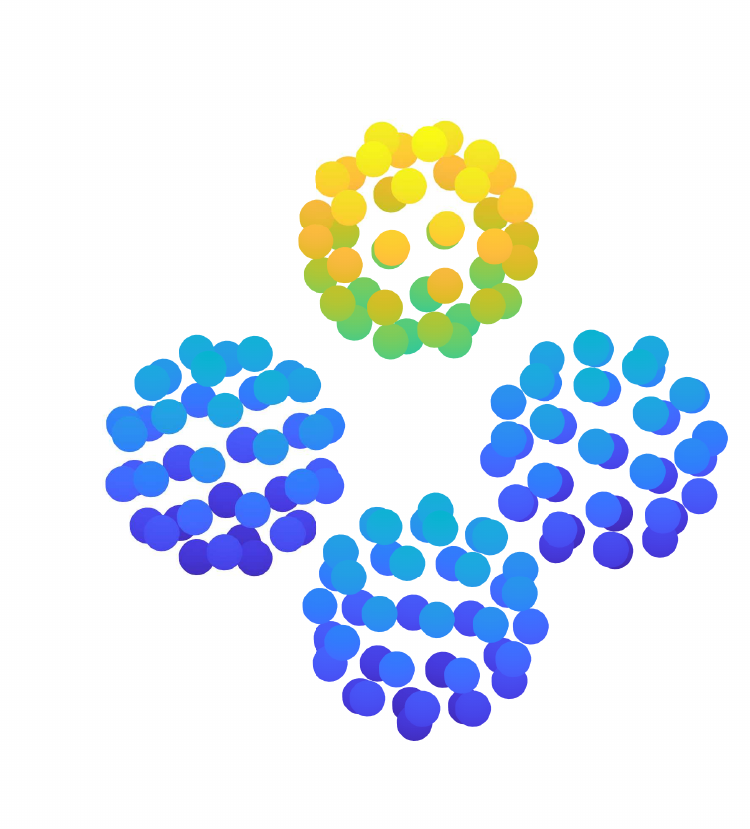}
		\caption{A tetrahedron of multiblob spheres.}
		\label{tetra_geom}
	\end{subfigure}~
	\begin{subfigure}[t]{0.3\textwidth}
		\centering
		\hspace*{-3ex}
		\includegraphics[trim = {2.5cm 1cm 1.8cm 1cm},clip,width=1.2\textwidth]{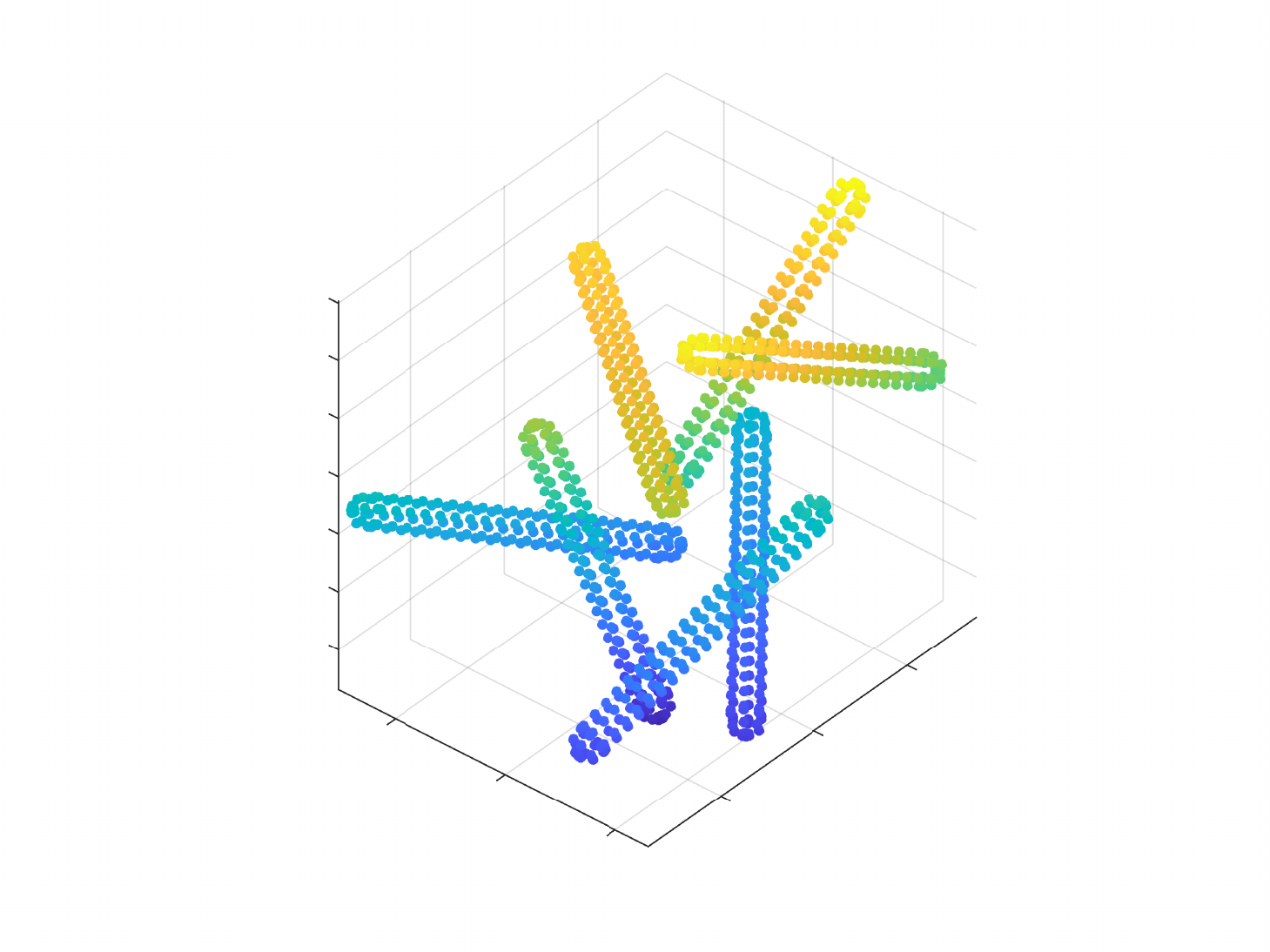}
		\caption{Slender rods of aspect ratio 20.}
		%\label{}
	\end{subfigure}~
	\begin{subfigure}[t]{0.3\textwidth}
		\centering
		\hspace*{2ex}
		\includegraphics[trim = {3.5cm 1cm 3cm 1.5cm},clip,width=0.95\textwidth]{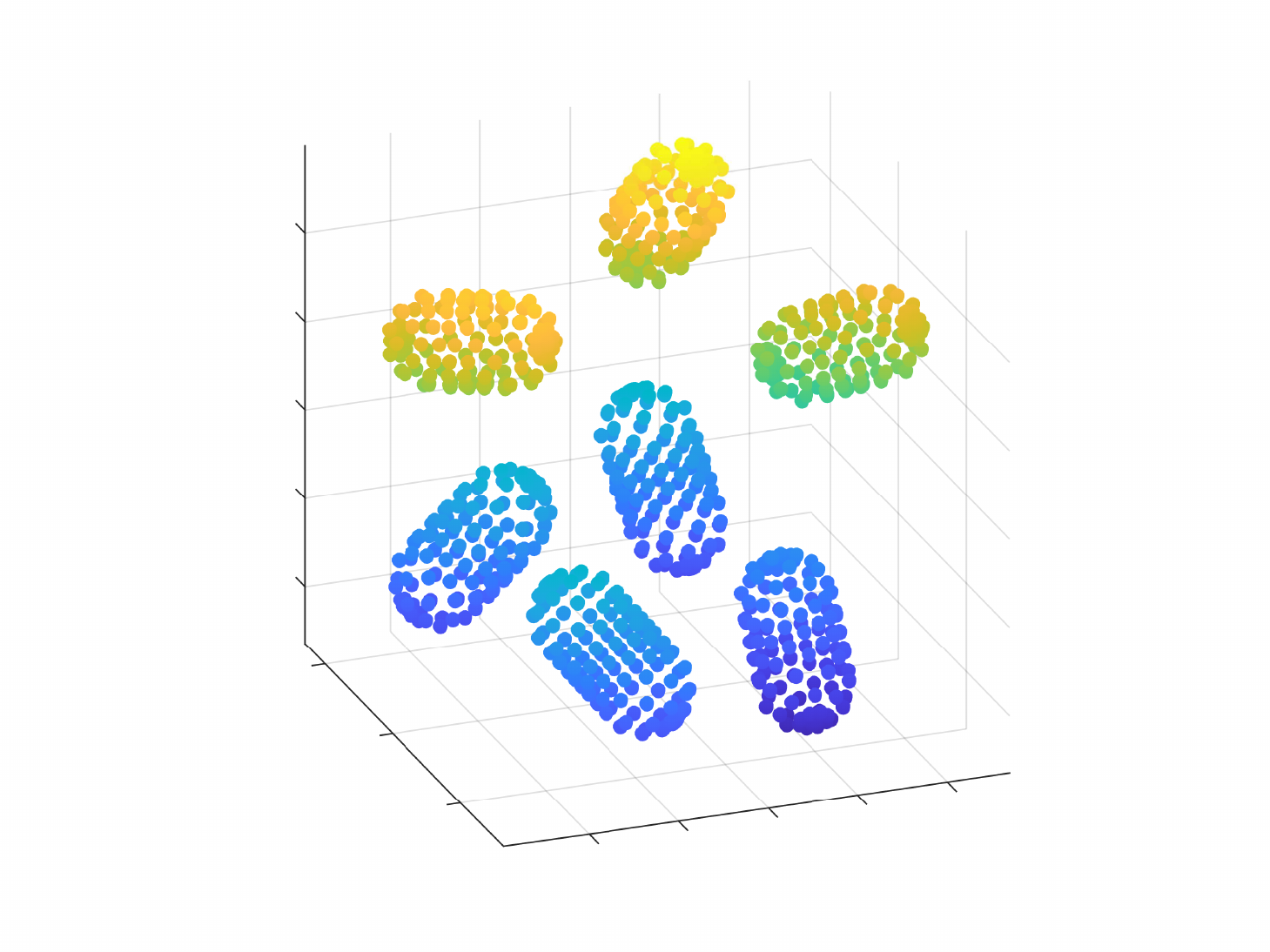}
		\caption{Fat rods of aspect ratio 4.}
		\label{short_rods}
	\end{subfigure}
	\caption{Example multiblob geometries.}
	\label{example_geom}
\end{figure}

Each particle in a suspension is described as a set of spheres or \emph{blobs} distributed on a surface. A few example geometries are visualised in Figure \ref{example_geom}. Blobs can be distributed on a surface in a magnitude of ways, as will be discussed in Section \ref{sec:match}. Given the discretisation points, at which blobs are to be centered, we define the characteristic grid spacing $s$ as the distance between discretisation points in the cross-section of the particle where the discretisation grid is the coarsest. To be more specific, for an axisymmetric particle, with cross-section perpendicular to the axis and with blob $i$ centered at $\vec b_i$, let
\begin{equation}\label{grid_s}
s = \max\limits_{\text{cross-section,\,} C}\left(\min\limits_{\lbrace b_j,b_i\rbrace\in C,i\neq j} \|\vec b_i-\vec b_j\|\right).
\end{equation}
The blobs are associated with their \emph{hydrodynamic} radius $a_h$, which serves as an effective model for how a blob interacts hydrodynamically with other blobs. Blobs with radius $a_h$ may overlap or be separated by a gap on the particle surface. The ratio between the hydrodynamic blob radius and the characteristic blob spacing is related to the extent of these overlaps or gaps.  This parameter, $a_h/s$, has to be chosen and is discussed in terms of the blob-blob spacing in Chapter 4.1 of \cite{USABIAGA2016}. In Section \ref{sec:match}, we solve an optimisation problem to determine $a_h$, given $s$. The effect of this optimal choice is discussed in Section \ref{results}.

The hydrodynamic interaction between particles is accounted for by first computing the mobility for all the blobs in the system, interacting in a \emph{pair-wise} manner only. The blob-mobilities are described in its most reduced and simplified form through the Rotne-Prager-Yamakawa tensor (RPY) \cite{USABIAGA2016,Ando2012}, where the block $\vec N_{ij}$ describes the motion on blob $i$ resulting from a given force on blob $j$, governed by the far field approximation 
\begin{equation}
\vec N_{ij} \approx \eta^{-1}\left(\vec I+\frac{1}{6}a_h^2\nabla^2_{\vec x}\right)\left(\vec I+\frac{1}{6}a_h^2\nabla^2_{\vec y}\right)\mathbb G(\vec r_{ij}),
\end{equation}
with $\eta$ the viscosity and 
\begin{equation}
\mathbb G(\vec r) = \frac{1}{8\pi r}\left(\vec I+\frac{\vec r\otimes \vec r}{r^2}\right)
\end{equation}
the Stokeslet, with $r=|\vec r|$ \cite{Wajnryb2013}. The RPY-tensor, corrected for overlapping blobs so that the resulting mobility  $\vec N_{ij}$ is positive-definite, takes the form \cite{USABIAGA2016,Wajnryb2013}
\begin{equation}\label{RPY}
\vec N_{ij} = \frac{1}{6\pi\eta a_h}\left\{
\begin{aligned}
C_1(r_{ij})\vec I+C_2(r_{ij})(\vec r_{ij}\otimes \vec r_{ij})/r_{ij}^2,\quad r_{ij}>2a_h,\\
C_3(r_{ij})\vec I+C_4(r_{ij})(\vec r_{ij}\otimes \vec r_{ij})/r_{ij}^2,\quad r_{ij}\leq 2a_h,\\
\end{aligned}
\right.
\end{equation}
with 
\begin{equation}
\begin{aligned}
C_1(r)& = \frac{3a_h}{4r}+\frac{a_h^3}{2r^3},\quad &C_2(r)= \frac{3a_h}{4r}-\frac{3a_h^3}{2r^3},&\\ \quad C_3(r) &= 1-\frac{9r}{32a_h},&C_4(r) = \frac{3r}{32a_h},&
\end{aligned}
\end{equation}
where $\vec r_{ij} = \vec x_i-\vec x_j$ is the center-center vector for two blobs $i$ and $j$. The diagonal blocks simply reduce to $\left(6\pi\eta a_h\right)^{-1}\vec I$, i.e.~the well-known translational part of the mobility matrix for a single sphere. Note that the RPY-tensor is a good approximation to the translation-translation coupling between two blobs when the separation between the blobs is sufficiently large.

The blobs belonging to one particle are restricted to move together as a rigid body, which is assured by applying a net force, $\vec{\lambda}_i^l\in\mathbb R^3$, to every blob center $\vec r_i^l$ on particle $l$. The forces on the blobs belonging to one particle are summed to yield the net force, $\vec f^l$, and torque, $\vec t^l$, on the particle. If we let $B^l$ be the indices of the set of $n_b$ blobs belonging to particle $l$ with center coordinate $\vec c^l$, let the blobs be centered in $\lbrace \vec b_i^l\rbrace_{i \in B^l}$ and impose no-slip boundary conditions on the multiblob particles, we obtain the set of equations for particle $l$
\begin{equation}\label{eq:multiblob}
\begin{aligned}
\sum_j \vec N^{ll}_{ij}\vec\lambda_j^l & = \vec u^l+\vec \omega^l\times(\vec b_i^l-\vec c^l) \quad \text{for all } i\in B^l,\\
\sum_{i\in B^l}\vec \lambda_i^l & =\vec f^l, \\
\sum_{i\in B^l}(\vec b_i^l-\vec c^l)\times \vec \lambda_i^l & = \vec t^l,
\end{aligned}
\end{equation}
to be solved for the unknown velocity pair $\vec u^l$ and $\vec \omega^l$ and blob force vector $\vec \lambda^l$. The formulation can be motivated as being the regularised discretisation of the first-kind integral equation \cite{USABIAGA2016}
\begin{equation}
\vec v(\vec x) = \vec u+\vec \omega\times \vec x = \eta^{-1}\int_{\Gamma}\mathbb G(\vec x -\vec y)\vec \psi(\vec y)\mathrm d\vec y\quad\text{for all }\vec x\in\Gamma,
\end{equation}
where $\Gamma$ is the particle boundary, and $\vec \psi$ is an unknown single layer density representing traction, being a continuous analogue of $\vec \lambda$.

Defining the matrix $\vec K^l\in \mathbb R^{3n_b}\times\mathbb R^6$ from the center coordinates of the blobs and the particle as
\begin{equation}
(\vec K^l \vec U^l)_I = \vec u^l + \vec\omega^l\times(\vec b_i-\vec c^l),
\end{equation}
with $I = \lbrace 3(i-1)+k \rbrace_{k=1}^3$,
the system in \eqref{eq:multiblob} can be written on the form
\begin{equation}\label{saddle}
\begin{bmatrix}
\vec N^{ll} & -\vec K^l \\ -\vec {K^l}^T & \vec 0 
\end{bmatrix}\begin{bmatrix}
\vec \lambda^l \\ \vec U^l
\end{bmatrix} = \begin{bmatrix}
\vec 0 \\ -\vec F^l
\end{bmatrix}
\end{equation}
with ${\vec F^l}^T = [{\vec f^l}^T {\vec t^l}^T]$ and ${\vec U^l}^T = [{\vec u^l}^T {\vec{\omega}^l}^T]$ (ignoring all other particles in the system). We can now eliminate $\vec{\lambda}^l$ and solve for $\vec U^l$ from
\begin{equation}
\left({\vec K^l}^T({\vec N^{ll}})^{-1} {\vec K^l}\right)\vec U^l =\vec F^l.
\end{equation}
Note that this allows us to define the single particle mobility matrix as $\vec M^l = \left({\vec K^l}^T({\vec N^{ll}})^{-1} \vec K^l\right)^{-1}$ (and its inverse, the resistance matrix $\vec R^l$).\footnote{See the discussion of p. 229 of \cite{USABIAGA2016} on the invertibility of $\vec N$ and $\vec M$. Special concerns are particles consisting of a single row of blobs or with infinitely many blobs covering its surface, but none of these settings will be considered here and $\vec M$ will always be invertible for the particle types studied in the paper.}
The formulation can be motivated as being the regularised discretisation of the first-kind integral equation \cite{USABIAGA2016}. For a suspension of multiple particles, we obtain a system of a similar form:
\begin{equation}\label{saddle_total}
\begin{bmatrix}
\vec N & -\vec K \\ -\vec K^T & \vec 0 
\end{bmatrix}\begin{bmatrix}
\vec \lambda \\ \vec U
\end{bmatrix} = \begin{bmatrix}
\vec 0 \\ -\vec F
\end{bmatrix},\text{ with } \vec K = \begin{bmatrix}
\vec K^1 & \vec 0 & \dots & \vec 0 \\
\vec 0 & \vec K^2 & \vec 0 & \dots \\
\vdots &\vdots & \ddots & \vec 0\\
\vec 0 &\hdots &\vec 0 & \vec K^p
\end{bmatrix}\in \mathbb R^{3n_bp}\times \mathbb R^{6p},
\end{equation}
where $\vec N^{ll}$ now is a block on the diagonal of the larger global blob-blob matrix $\vec N$ and $\vec \lambda$ contains the forces on all blobs. The mobility matrix for the system is then given by $\vec M = \left(\vec K^T\vec N^{-1}\vec K\right)^{-1}$.

In general, this is not how we would solve the mobility problem and there are efficient preconditioning strategies for solving the system in \eqref{saddle} \cite{USABIAGA2016,Swan2016}. However, when comparing the solution of the mobility problem for a small number of particles to that of a more accurate method, it can be motivated to actually compute mobility matrices if the computational cost for doing so is manageable, and apply a large number of right hand side force/torque vectors.

Note that if the particles are subject to a background flow, the upper block of the right hand side vector in \eqref{saddle_total} would be modified to contain a vector of flow velocities at the blob centers. This would hence not affect the form of the mobility matrix and in the remainder of this work, we assume that particles are immersed in a quiescent fluid. 
\clearpage
\section{Matching the multiblob grid}\label{sec:match}
Discretising a particle surface with blobs introduces an approximation on several levels. We will quantify the error in the mobility of a multiblob particle by comparing to a well-resolved model where a BIE-solver with QBX is employed for two different particle shapes: the sphere and the rod with smooth caps. Of importance is that a multiblob particle is viewed as a \emph{model} of what we refer to as an \emph{ideal} particle of a certain type. To match with the ideal particle, the centers of the blobs should for good results be placed on a surface interior to that of the ideal particles. We refer to this interior surface as the \emph{geometric} surface. There are several choices to be made, and the combination of these will determine how well the hydrodynamic ``response'' of the multiblob particle will match that of the ideal particle. Hence, given an ideal particle, we need to consider the questions
\begin{enumerate}
	\item How should the geometric surface be chosen?
	\item What is the optimal placement of blobs (given some restriction on the number of blobs)?
	\item What is the optimal hydrodynamic radius of the blobs (the parameter $a_h$ in the RPY-tensor in \eqref{RPY})?
\end{enumerate} 
If the blob discretisation of a particle surface is coarse, it is reasonable to believe that the hydrodynamic response will be different from that of the ideal particle. One option is to view the blobs as quadrature nodes on the particle surface, which intuitively is reasonable in the limit with many blobs covering the surface. If the blobs are large and the resolution is coarse however, the geometry of the multiblob particle is far from the geometry of the particle that we would like to model, and the blobs popping out from the particle surface will make the fluid ``interpret'' the particle as larger. By introducing an offset from the surface, such that blobs are placed centered on a geometric surface close to the boundary but in the interior of the ideal particle, the hydrodynamic response of a coarse multiblob model will be closer to that of the ideal particle. 

To see specifically how the geometry of the multiblob particle should be chosen, we start with a sphere in Section \ref{spheres} and later use a similar strategy also for the rod geometry, in Sections \ref{rods}-\ref{rod-rt}. The geometric surface for a sphere will be a sphere with the geometric radius $R_g$. For an axisymmetric particle, such as a rod, the geometric surface is defined using both a geometric radius $R_g$ and a geometric length $L_g$. 

Given a strategy for how to place $n_b$ blobs on a geometric surface, we have to select $R_g$, $L_g$ and the hydrodynamic radius $a_h$. This will be done through an optimisation procedure, where we seek to match the mobility matrix for a single particle, alone in a fluid. The reason is that in a multi-particle suspension where the particles are widely separated, this \emph{self-interaction} will be dominating. Said differently, independent of how dilute the suspension is, the error level will never be lower than the error in the self-interaction. When analytical formulas are not available, an accurate reference mobility matrix for a single particle can be obtained from an accurate numerical method such as the BIE-solver with QBX \cite{Bagge2021,AfKlinteberg2016}. We will in this paper consider axisymmetric particles, for which the resistance matrix of a single particle takes a particularly simple form. A resistance matrix can generally be written as 
\begin{equation}\label{resistance}
\vec R = \begin{bmatrix}
\vec R^{UF} & \vec R^{\Omega F} \\
\vec R^{UT} & \vec R^{\Omega T}
\end{bmatrix},
\end{equation}
where the four blocks represent the coupling between the assigned translational and angular velocities, indicated by $U$ and $\Omega$, and the induced net force and torque, denoted by $F$ and $T$.
For a single particle with symmetry axis described by the unit vector $\vec s$,  $\vec R^{\Omega F} = \vec R^{UT} = \vec 0$, $\vec R^{UF} = \xi_t^{\parallel}(\vec s\vec s^T)+ \xi_t^{\perp}(\vec I- \vec s\vec s^T)$ and $\vec R^{\Omega T} = \xi_r^{\parallel}(\vec s\vec s^T)+ \xi_r^{\perp}(\vec I- \vec s\vec s^T)$, with the subscript $t$ denoting a coefficient representing translation, the subscript $r$ denoting rotation and the superscripts $\parallel$ and $\perp$ representing motion parallel or perpendicular to the axis of symmetry  \cite{Graham}. The four positive coefficients $\xi_t^{\parallel}$, $\xi_t^{\perp}$, $\xi_r^{\parallel}$ and $\xi_r^{\perp}$ are specific for a given particle shape and are known analytically for some simple geometries, such as spheres, ellipsoids and infinitely long rods. The mobility matrix, $\vec M$, is the inverse of the resistance matrix, relating applied forces and torques to computed velocities. Its blocks are given by $\vec M^{F\Omega } = \vec M^{TU} = \vec 0$, $\vec M^{FU} = \left(1/\xi_t^{\parallel}\right)(\vec s\vec s^T)+ \left(1/\xi_t^{\perp}\right)(\vec I- \vec s\vec s^T)$ and $\vec M^{T\Omega} = \left(1/\xi_r^{\parallel}\right)(\vec s\vec s^T)+ \left(1/\xi_r^{\perp}\right)(\vec I- \vec s\vec s^T)$.

For what follows, let $\lbrace \xi_t^{\parallel}, \xi_t^{\perp},\xi_r^{\parallel},\xi_r^{\perp}\rbrace$ be the set of coefficients determined analytically or with an accurate method and let $\lbrace \hat{\xi}_t^{\parallel}, \hat{\xi}_t^{\perp},\hat{\xi}_r^{\parallel},\hat{\xi}_r^{\perp}\rbrace$ be another set of coefficients deduced from the approximate multiblob method.
%%%%%%%%%%%%%%%%%%%%%%%%%%%%%%%%%%%%%%%%%%5
\subsection{Spheres}\label{spheres}
A sphere constitutes a simple special case of an axisymmetric particle. It is well-known \cite{Kim1991} that the non-zero blocks of the resistance matrix for a single sphere of radius $R$ in a fluid with viscosity $\eta$ modelled in free space are given by $\vec R^{UF} = 6\pi\eta R\vec I$ and  $\vec R^{\Omega T} = 8\pi\eta R^3\vec I$; this is a consequence of the fact that the translational velocity of a  single sphere affected by  a net force $\vec f$  theoretically is given by $\vec f/(6\pi\eta R)$ and, similarly, that the angular velocity of a single sphere affected by a net torque $\vec t$ is $\vec t/(8\pi\eta R^3) $ (these relations are often referred to as Stokes' first and second law). The Stokes' laws can be used to numerically identify the hydrodynamic radius $\hat{R}_h$ of a spherical multiblob particle, given a computed translational and angular velocity, $\vec u$ and $\vec \omega$, corresponding to an assigned force and an assigned torque respectively. The effective hydrodynamic radius will however in general not automatically be the same for the translational and angular motion, and we denote these two computed hydrodynamic radii by $\hat R_h^{\text{trans}}$ and $\hat R_h^{\text{rot}}$.  Relating to the general resistance expressions, the exact coefficients are given by $\xi_t^{\parallel} =\xi_t^{\perp} = 6\pi\eta R$ and $\xi_r^{\parallel} =\xi_r^{\perp} = 8\pi\eta R^3$, while the computed coefficients for the multiblob particle are $\hat{\xi}_t^{\parallel} =\hat{\xi}_t^{\perp} = 6\pi\eta \hat R_h^{\text{trans}}$ and $\hat{\xi}_r^{\parallel} =\hat{\xi}_r^{\perp} = 8\pi\eta \left(\hat{R}_h^{\text{rot}}\right)^3$. %Note that typically, $\hat R_h^{\text{trans}}\neq \hat R_h^{\text{rot}}$ even though they both should equal $R$ for an exact representation.  

\begin{figure}[h!]
	\centering
	\includegraphics[trim = {2cm 3.8cm 2cm 0cm},clip,width=0.9\textwidth]{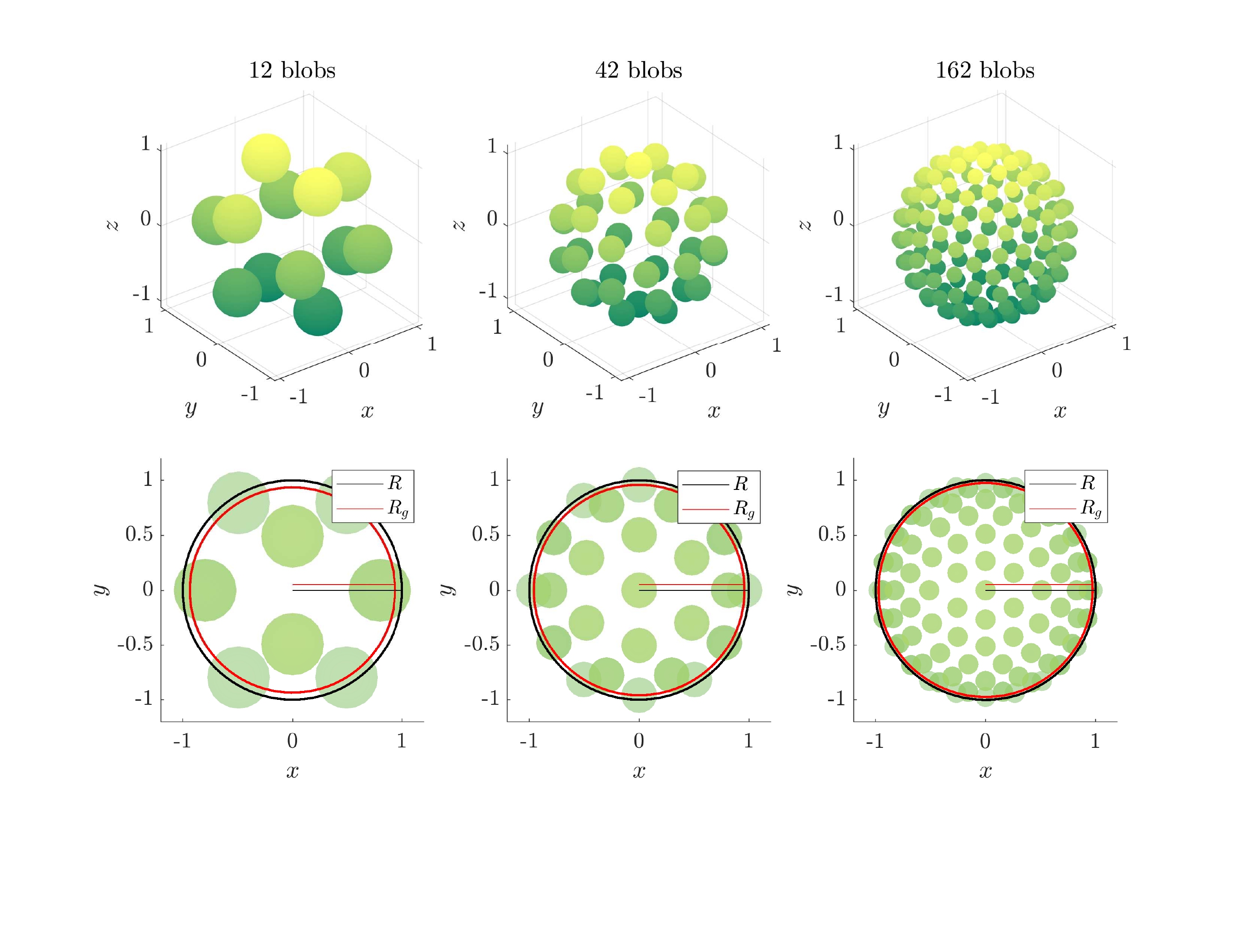}
	\caption{Spherical particles of unit hydrodynamic radius $R=1$ modelled with multiblobs. The geometric radius $R_g$ sets the surface on which to place the blob centers. A sphere is constructed from uniform subdivisions of an icosahedron (the first three subdivisions are depicted from left to right). In the bottom panel, a 2D projection is displayed for each discretisation, where the geometric and hydrodynamic radius are indicated. We have solved for the optimal $R_g$ and blob spacing $a_h/s$ as presented in Table \ref{opt_sphere}.}
	\label{icosa}
\end{figure}

The sphere geometry can be discretized with blobs from uniform subdivisions of an icosahedron projected onto the sphere geometry\footnote{Code for generating these subdivisions is taken from  (https://www.mathworks.com/matlabcentral/fileexchange/50105-icosphere), MATLAB Central File Exchange. Retrieved March 10, 2021.}, as is illustrated for three consecutive refinements in Figure \ref{icosa}, resulting in models with $10\cdot 4^{k-1}+2$ blobs in subdivision $k$. This is the same sphere geometry as used for multiblob spheres in \cite{USABIAGA2016,Sprinkle2020}. The blobs are placed on spheres of \emph{geometric} radius $R_g$ not necessarily equal to the hydrodynamic radius $R$.  We choose the characteristic grid spacing, $s$, to be the minimum spacing between grid nodes  for each sphere resolution (this corresponds to the definition given in \eqref{grid_s}) and define the hydrodynamic radius $a_h$ relative to $s$. An illustration of the parameters $R$, $R_g$ and $a_h$ is displayed in Figure \ref{circle}. We want to select $R_g$ and $a_h$ to match the hydrodynamic response of the multiblob particle to that of the ideal particles as closely as possible.

\begin{figure}[h!]
	\begin{minipage}[t]{0.6\textwidth}
	%	\begin{figure}[h]
			\centering
			\hspace*{-3ex}
			\includegraphics[trim = {2.5cm 13.7cm 2.5cm 6.7cm},clip,width = 1.45\textwidth]{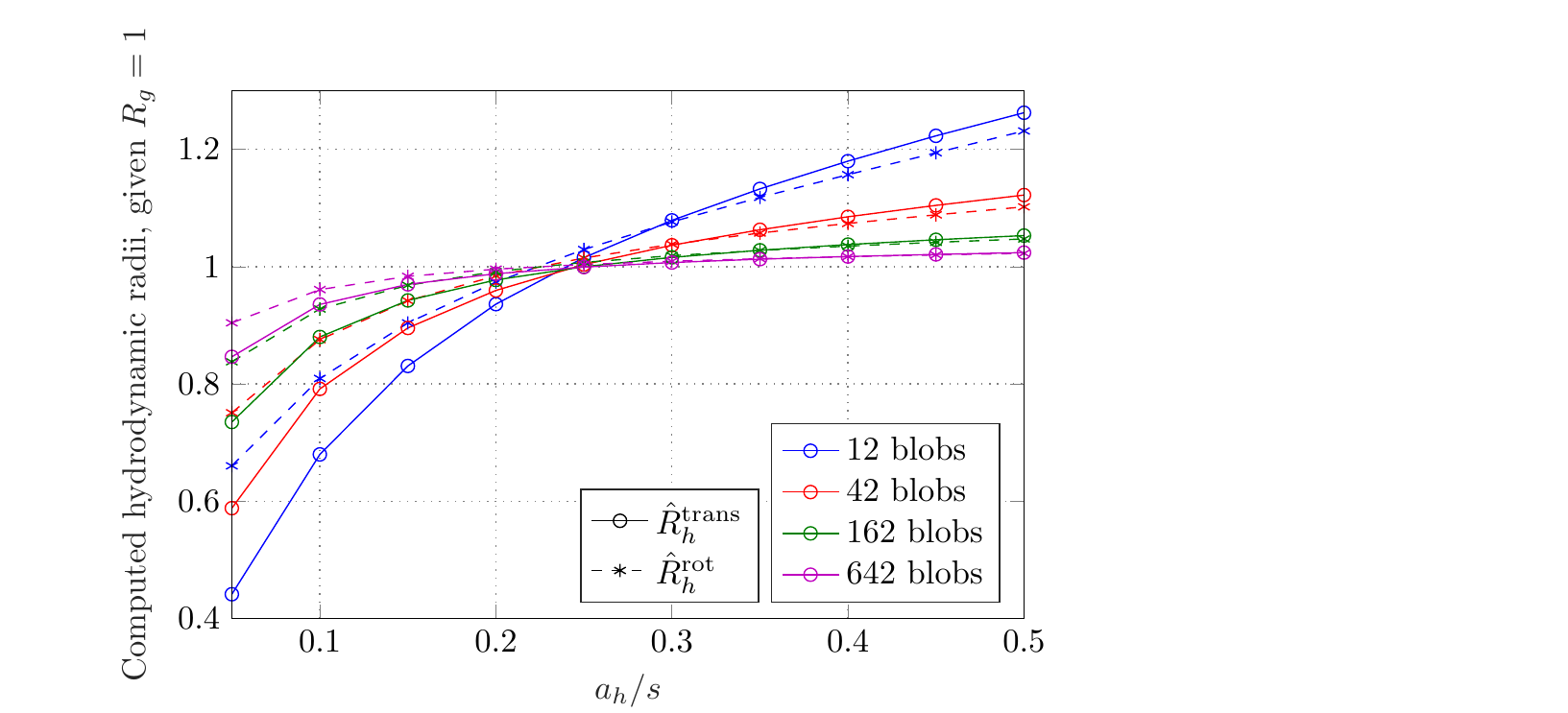}
			\caption{The computed effective translational and rotational radius for a spherical mulitblob particle of different resolutions plotted versus $a_h/s$, given $R_g = 1$. Note that in general $\hat R_h^{\text{trans}}\neq\hat R_h^{\text{rot}}$ and we need to solve for $R_g$ and $a_h/s$ such that the translational and rotational hydrodynamic radii agree. The optimal $R_g$ and $a_h/s$ given $\hat R_h^{\text{trans}}=\hat R_h^{\text{rot}}=1$ are presented in Table \ref{opt_sphere}.}
			\label{scaled_radius}
	%	\end{figure}	
	\end{minipage}\qquad
		\begin{minipage}[t]{0.35\textwidth}
			%\begin{figure}[h]
				\centering
				\hspace*{-6ex}
				\includegraphics[trim = {0cm 15.6cm 6cm 3.4cm},clip,width = 2.2\textwidth]{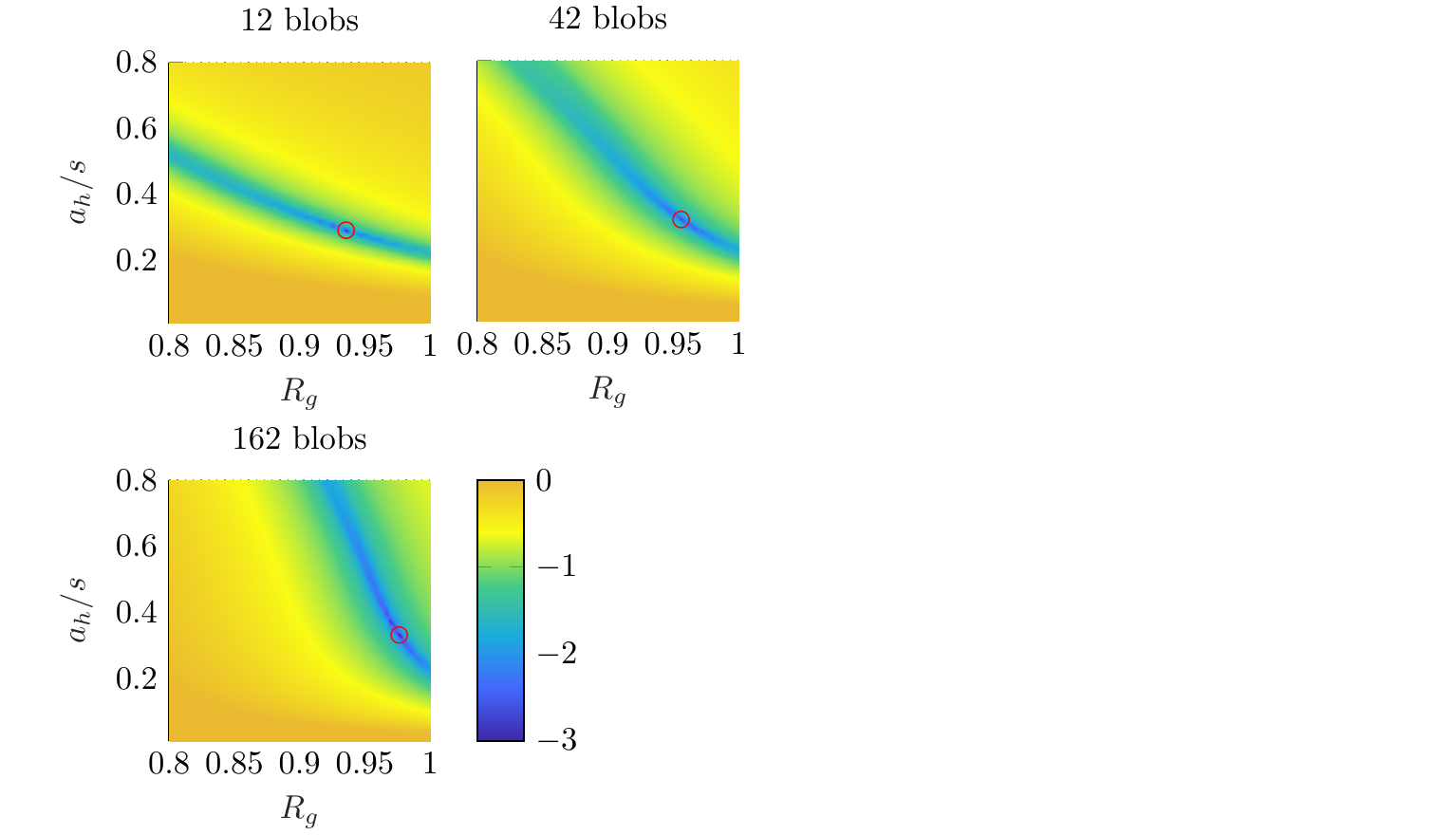}
				\caption{Optimisation landscape for the problem of determining the sphere grid in \eqref{eq2}. The maximum relative mobility coefficient error $(\log_{10})$ is displayed for each ($R_g,a_h/s)$ with the optimum marked with a red circle for three resolutions.}
				\label{sphere_landscape}
			%\end{figure}	
		\end{minipage}\qquad
\end{figure}

The hydrodynamic particle radii $\hat{R}_h^{\text{trans}}$ and $\hat{R}_h^{\text{rot}}$ depend both on $R_g$ and $a_h$. In general, $\hat{R}_h^{\text{trans}}\neq \hat{R}_h^{\text{rot}}$, if $R_g$ and $a_h$ are not chosen carefully. In Figures \ref{scaled_radius}-\ref{sphere_landscape}, the dependence on the parameters is visualised. We can however determine $R_g$ and $a_h$ such that $\hat{R}_h^{\text{trans}} \approx \hat{R}_h^{\text{rot}} \approx R$. For this purpose, we minimise the relative error in the mobility coefficients, that is
\begin{equation} \label{eq2}
\begin{aligned}
\min\limits_{R_g,a_h}  \quad&\max\left\{R\Bigg|\frac{1}{R}-\frac{1}{\hat{R}_h^{\text{trans}}\left(R_g,a_h\right)}\Bigg|,R^3\Bigg|\frac{1}{R^3}-\frac{1}{\left(\hat{R}_h^{\text{rot}}\left(R_g,a_h\right)\right)^3}\Bigg|\right\}. \\
%\text{s.t. }
%\quad 
%&\left(\frac{1}{6\pi}-\frac{1}{6\pi\hat{R}_h^{\text{trans}}\left(R_g(a_h)\right)}\right)^2 \leq \epsilon, \\
%%\min\limits_{R_g^{\text{rot}}(a_h)}
%& \left(\frac{1}{8\pi}-\frac{1}{8\pi\left(\hat{R}_h^{\text{rot}}\left(R_g(a_h)\right)\right)^3}\right)^2 \leq \epsilon.
\end{aligned}
\end{equation}
The problem in \eqref{eq2} might have multiple local minima and we choose the minimizing $(R_g,a_h)$ such that $R_g\approx R$. The reason for this is that the boundary conditions are imposed at the blob centers, as presented in \eqref{eq:multiblob} in the description of the multiblob method, which is physically reasonable if $R_g\approx R$. The optimal pairs $(R_g,a_h)$ are presented for four resolutions of the sphere in Table \ref{opt_sphere}, given $R = 1$. We would like to emphasise the importance of this optimisation procedure by inspecting the optimisation landscape in Figure \ref{sphere_landscape}. The relative error in the mobility coefficients is very sensitive to the values of $R_g$ and $a_h$. 
%with 
%\begin{equation}\label{eq3}
%\begin{aligned}
%\quad R_g^{\text{trans}}(a_h) & = \underset{R_g}{\arg\min} 
%\left(\frac{1}{6\pi}-\frac{1}{6\pi\hat{R}_h^{\text{trans}}\left(R_g(a_h)\right)}\right)^2, \\
% %\min\limits_{R_g^{\text{rot}}(a_h)}
%\quad R_g^{\text{rot}}(a_h) & = \underset{R_g}{\arg\min} \left(\frac{1}{8\pi}-\frac{1}{8\pi\left(\hat{R}_h^{\text{rot}}\left(R_g(a_h)\right)\right)^3}\right)^2.
%\end{aligned}
%%\end{equation}
%\begin{equation} 
%\begin{align}
%&&\min\limits_{a_h} & \quad\max\left\{\Big|\frac{1}{6\pi}-\frac{1}{6\pi\hat{R}_h^{\text{trans}}\left(R_g^{\text{rot}}(a_h)\right)}\Big|,\Big|\frac{1}{8\pi}-\frac{1}{8\pi\left(\hat{R}_h^{\text{rot}}\left(R_g^{\text{trans}}(a_h)\right)\right)^3}\Big|\right\}, \label{eq2} \\
%&&\text{with} & %\min\limits_{R_g^{\text{trans}}(a_h)}
%\quad R_g^{\text{trans}}(a_h) = \underset{R_g}{\arg\min} 
%\left(\frac{1}{6\pi}-\frac{1}{6\pi\hat{R}_h^{\text{trans}}\left(R_g(a_h)\right)}\right)^2, \label{eq3}\\
%&& & %\min\limits_{R_g^{\text{rot}}(a_h)}
%  \quad R_g^{\text{rot}}(a_h) = \underset{R_g}{\arg\min} \left(\frac{1}{8\pi}-\frac{1}{8\pi\left(\hat{R}_h^{\text{rot}}\left(R_g(a_h)\right)\right)^3}\right)^2. \label{eq4}
%\end{align}
%\end{equation}
%with $\epsilon$ a small tolerance set to $\epsilon = 10^{-7}$.

\begin{table}[h!]
	\centering	
	\begin{tabular}{c|c| c|c|c}
		Resolution & $R_g$ & $a_h/s$ & $|1-\left(\hat{R}_h^{\text{trans}}(R_g,a_h)\right)^{-1}|$ &  $|1-\left(\hat{R}_h^{\text{rot}}\left(R_g,a_h\right)\right)^{-3} |$\\ \hline
		12 blobs & 0.936 & 0.291 & $4.94\times 10^{-9}$ & $1.96\times 10^{-8}$ \\
		42 blobs & 0.959 & 0.311 & $9.23\times 10^{-9}$ & $1.22\times 10^{-8}$  \\
		162 blobs & 0.974 & 0.345 & $1.97\times 10^{-8}$ &$1.96\times 10^{-8}$ \\
		642 blobs & 0.984 & 0.388 & $5.19\times 10^{-9}$ & $9.15\times 10^{-9}$ \\
	\end{tabular}
	\caption{Optimal $R_g$ and $a_h/s$ computed as solutions to \eqref{eq2} for spherical multiblobs of varying resolutions with $R=1$, presented together with the relative error in each of the mobility coefficients.}
	\label{opt_sphere}	
\end{table}

In the work by Usabiaga et al., \cite{USABIAGA2016}, $R_g$ is chosen so that $R = \hat{R}_h^{\text{trans}}$, with the blob radius ratio $a_h/s = 0.5$. The motivation is that the translational coupling is the most long ranged in the fluid. This is a good choice if we would only care about translational motion. However, if rotational velocities of spheres are to be considered, we have large errors in $\hat{R}_h^{\text{rot}}$ as presented in Table \ref{rot_error_radius}, resulting in large relative errors in the rotational mobility coefficient $1/\left(8\pi(\hat{R}_h^{\text{rot}})^3\right).$ We will show the consequence of such an error in multi-particle simulations in Section \ref{sec:spheres}. The blob radius ratio $a_h/s = 0.5$ has previously been the choice in e.g.~\cite{Molina2013,Swan2011}, motivated by a view where blobs are seen as tightly packed spheres covering the surface of a particle, but not overlapping.

%\begin{figure}[h!]
%	\centering
%	\hspace*{9ex}
%	\includegraphics[trim = {2.5cm 13.7cm 2.5cm 6.7cm},clip,width = \textwidth]{figures/sphere/sphere_optland_1.pdf}
%	\caption{Optimisation landscapes in the two parameters $(R_g,a_h)$ for multiblob spheres of varying resolutions.}
%	\label{opt_landscape}	
%\end{figure}

\begin{table}[h!]
	\centering
	\begin{tabular}{c|c| c| c|c}
		Resolution & $R_g$ &$\Big|1-\left(\hat{R}_h^{\text{rot}}(R_g,a_h)\right)\Big|$ & $\Big|1-\left(\hat{R}_h^{\text{rot}}\left(R_g,a_h\right)\right)^{-3}\Big|$ \\ \hline
		12 blobs & 0.792  &$2.46\times 10^{-2}$ & $7.77\times 10^{-2}$\\
		42 blobs & 0.891  &$1.79\times 10^{-2}$ & $5.58\times 10^{-2}$\\
		162 blobs & 0.950  &$5.48\times 10^{-3}$ & $1.66\times 10^{-2}$\\
		642 blobs & 0.977  &$1.25\times 10^{-3}$ & $3.77\times 10^{-3}$
	\end{tabular}
	\caption{Relative errors in the rotational radius and rotational mobility coefficient using $a_h/s = 0.5$ and $R_g$ chosen so that $\hat{R}_h^{\text{trans}}(R_g,a_h) =1$. This is the relation chosen between the geometric and hydrodynamic radius of the sphere in \cite{USABIAGA2016}. Compare the error levels in the rightmost column to those presented in Table \ref{opt_sphere}, where we have optimised for $R_g$ and $a_h/s$ to minimise the errors in both mobility coefficients (matching both $\hat{R}_h^{\text{trans}}$ and $\hat{R}_h^{\text{rot}}$). The error in the rotational mobility coefficient is much larger if the blob radius ratio $a_h/s = 0.5$ is fixed.}
\label{rot_error_radius}
\end{table}

\begin{remark}
	We could also compute an effective stress radius for a sphere in shear flow, similarly as we do for the rotational and translational radii, and use also this radius as a parameter that we choose to match for in the optimisation problem. Computing the stress radius is done in \cite{USABIAGA2016}. Along a similar note, there might be other effective quantities of the multiblob particle that affects e.g.~the  rheological properties of a multi-particle suspension. In an application of multiblobs to model physical particles with specific properties, the objective function in \eqref{eq2} can potentially be modified or possible additional constraints added to account for such properties.
\end{remark}

%%%%%%%%%%%%%%%%%%%%%%%%%%%%%%%%%%%%%%%%%%%%55%%%%%%%%%%%%%%%%%%%%%%%%%
\subsection{Rods: The rt-grid}\label{rods}
For any axisymmetric particle, there are four mobility coefficients that need to be matched: two for translation, $1/\xi_t^{\parallel}$ and $1/\xi_t^{\perp}$, and two for rotation, $1/\xi_r^{\parallel}$ and $1/\xi_r^{\perp}$. For a rod-like particle with length $L$ and radius $R$, there are approximate expressions from slender body theory for the translational resistance coefficients, where \cite{Tao2005,Dhont1996}
\begin{equation}
\xi_t^{\parallel} = \frac{2\pi\eta L}{\ln(L/2R)},\quad \xi_t^{\perp} = 2\xi_t^{\parallel},
\end{equation}
however only valid in the limit $L\to\infty$. For the rotational resistance coefficients, theoretical results  mainly focus on rotation perpendicular to the axis of symmetry, as the rotation around the axis of symmetry is assumed negligible for an axisymmetric particle. From slender body theory, the friction coefficient $\xi_r^{\perp}$ for an infinitely thin rod can be approximated by, \cite{Tao2005,Dhont1996}
\begin{equation}
\xi_r^{\perp} = \frac{\pi\eta L^3}{3\ln(L/2R)}.
\end{equation}
As these resistance coefficients are inaccurate for any rod of finite aspect ratio, we will use the resistance coefficients $\lbrace \xi_t^{\parallel}, \xi_t^{\perp},\xi_r^{\parallel},\xi_t^{\perp}\rbrace$ computed with the BIE-method to determine how the blobs should be placed on the surface of a given length and radius rod with smooth caps. In \cite{USABIAGA2016}, mobility coefficients are extrapolated for a cylinder, using that the convergence to the true mobility coefficients is linear in $a_h$. In this paper, we employ different matching techniques, one of which is presented in this section.

Rods  of two different aspect ratios are studied: a fat rod with $L/R = 4$ and a slender rod with $L/R = 20$, with the accurate mobility coefficients reported in Table \ref{coeffs} in Appendix \ref{sec:BIE}. Given the length and radius determining the shape of the ideal particle, a geometric radius, $R_g$, and a geometric length, $L_g$, are determined for each rod size and discretisation so that the hydrodynamic response of the multiblob particle closely matches that of the BIE-rod. More specifically, the mobility matrix for a single multiblob rod is matched as closely as possible to the mobility matrix for the corresponding BIE-rod. We seek $L_g$ and $R_g$, along with the hydrodynamic radius of the blobs, $a_h$, that minimises the maximum relative error in the mobility coefficients. We can phrase this as an optimisation problem  on the form
\begin{equation}\label{opt}
\begin{aligned}
\min\limits_{L_g,R_g,a_h}&\max\left\{\xi_r^{\perp}\Bigg|\frac{1}{\xi_r^{\perp}}-\frac{1}{\hat{\xi}_r^{\perp}\left(L_g,R_g,a_h\right)}\Bigg|,
\xi_r^{\parallel}\Bigg|\frac{1}{\xi_r^{\parallel}}-\frac{1}{\hat{\xi}_r^{\parallel}\left(L_g,R_g,a_h\right)}\Bigg|,\right.\\
& \quad\qquad\left.\xi_t^{\perp}\Bigg|\frac{1}{\xi_t^{\perp}}-\frac{1}{\hat{\xi}_t^{\perp}\left(L_g,R_g,a_h\right)}\Bigg|,
\xi_t^{\parallel}\Bigg|\frac{1}{\xi_t^{\parallel}}-\frac{1}{\hat{\xi}_t^{\parallel}\left(L_g,R_g,a_h\right)}\Bigg|\right\}. 
\end{aligned}
\end{equation}
To find a reasonable local minimum to this problem, the three variables $R_g$, $L_g$ and $a_h$ are related to the ideal particle and constrained so that
\begin{equation}
\begin{gathered}
L/2\leq L_g\leq 1.5L,\\
R/2\leq R_g\leq 2R,\\
0.2\leq a_h/s \leq 1.
\end{gathered}
\end{equation}
The problem is solved with \texttt{fminimax} in \textsc{Matlab} with a solver tolerance set to $10^{-11}$. The minimisation of the maximum error in \eqref{opt} ensures that the error level is kept small and approximately equal in all four mobility coefficients. We term the optimised grid the \emph{rt-grid}, as both rotational and translational mobility coefficients are matched.

\begin{figure}[h!]
	\centering
\begin{subfigure}[t]{0.33\textwidth}
		\centering
			\includegraphics[trim = {0cm 19.2cm 15cm 3.2cm},clip,width=0.85\textwidth]{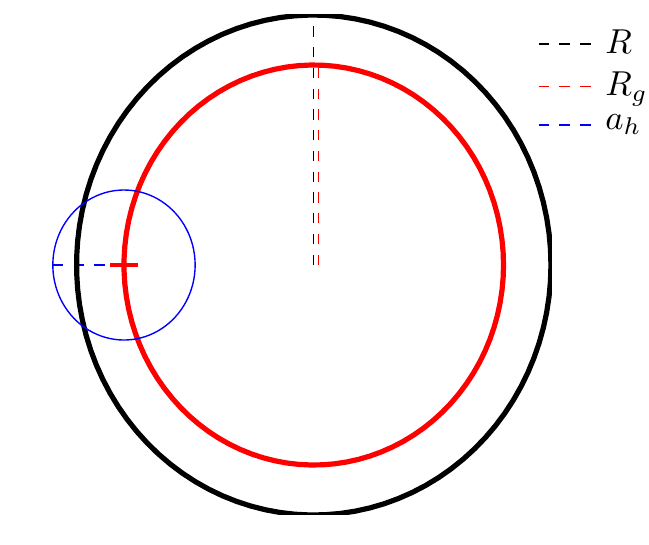}
		\caption{Sketch of the cross-section of a sphere with geometric radius $R_g$ and hydrodynamic radius $R$. The position and hydrodynamic radius $a_h$ of a single blob placed on the geometric surface is indicated.}
		\label{circle}
	\end{subfigure}~~
	\begin{subfigure}[t]{0.33\textwidth}
		\centering
			\includegraphics[trim = {5cm 19.5cm 11cm 2.55cm},clip,width=0.7\textwidth]{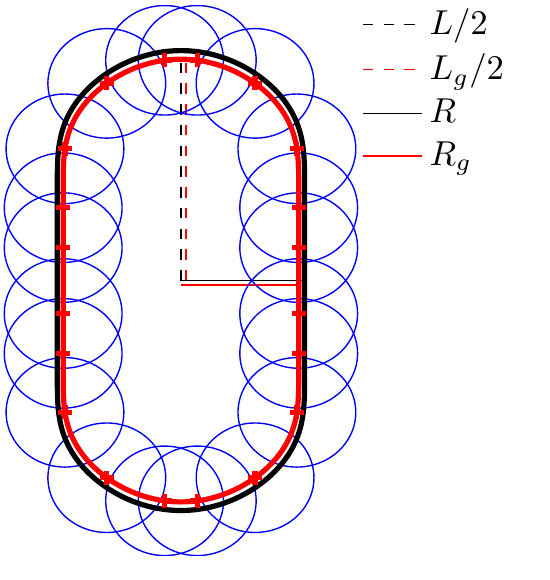}
		\caption{Cross-section of a rod with $L/R = 4$ displaying the geometric surface and the surface of the ideal particle, with the geometric surface in the interior determined by $R_g$ and $L_g$.}
		\label{rods4}
	\end{subfigure}~~
	\begin{subfigure}[t]{0.33\textwidth}
	\centering
	%\hspace*{3ex}
	\includegraphics[trim = {7cm 11cm 2cm 9cm},clip,width=1.15\textwidth]{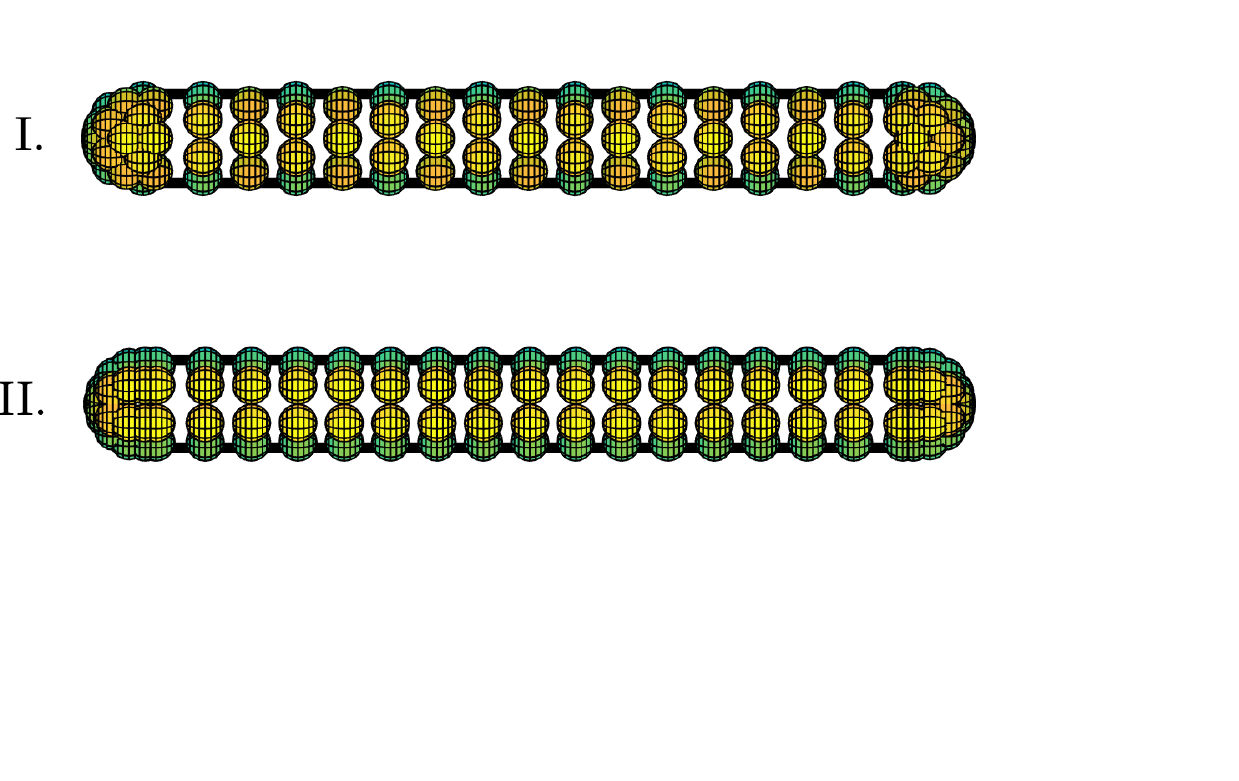}
	\caption{Rod with  aspect ratio $L/R = 20$, where I. the position of the blobs in the layers in the axial direction are shifted and II. aligned.}
	\label{rods20}
\end{subfigure}
	\caption{Multiblob particles discretised with blobs placed on a geometric surface chosen with an optimisation procedure to hydrodynamically match the corresponding ideal particle.}
	\label{geom}
\end{figure}

The geometry and parameterisation of the surface of rods with smooth caps, designed for the BIE-method, is described in detail by Bagge \& Tornberg in the appendix of \cite{Bagge2021}. For the multiblob rods, we use the same parameterisation of the rod geometry. For a multiblob rod of length $L_g$ and radius $R_g$, three parameters determine its discretisation: The top and bottom cap each occupy a length corresponding to $1.5R_g$ (a choice made for smooth caps for the BIE-rods -- the caps are hence not half-spheres) and are discretised in the axial direction by $n_{\text{cap}}$  nodes. The middle cylindrical part of the rod is discretised with $n_{\text{cyl}}$ equally spaced nodes (in the BIE-method, these are chosen as Gauss-Legendre nodes). Both the cap and the cylindrical middle part of the rod is discretised with $n_{\varphi}$ equally spaced points in the cross section of the rod. Thus, a total of $n_b = (2n_{\text{cap}}+n_{\text{cyl}})n_{\varphi}$ blobs discretise the rod. We have experimented with different ways of sampling these parameters: \emph{aligning} the different layers along the axial direction of the rod or \emph{shifting} the layers so that every second layer is aligned and sampling the cap in the axial direction with either equally spaced nodes or Gauss-Legendre nodes in the arc length parameter. Two example rods are displayed in Figure \ref{geom}, where the grid is shifted on one rod and aligned on the other. Different strategies of placing the blobs do not result in very large differences in neither the computed geometries $(L_g,R_g,a_h)$ nor the resulting error levels upon solving \eqref{opt}. Hence, only a subset of the results are presented. We have also tried to approximate the rod with a cylinder without caps, to quantify the importance of the caps in the hydrodynamic response (results not reported, but the errors in the mobility coefficients are larger). 

\begin{figure}[h!]
	\centering
	%\hspace*{-5ex}
	\includegraphics[trim = {0.5cm 20.5cm 4cm 3.5cm},clip,width=0.9\textwidth]{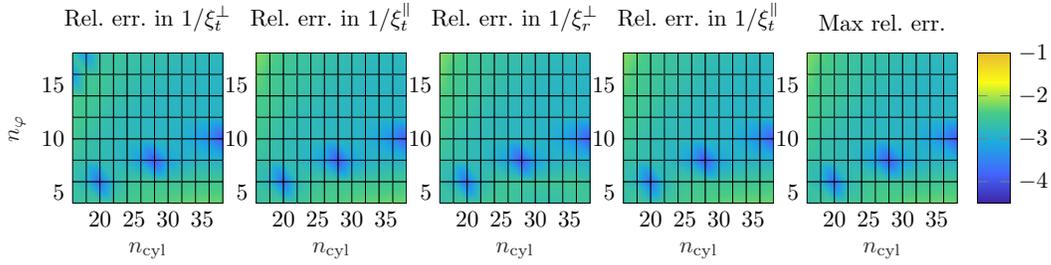}
	\caption{Relative error $(\log_{10})$ in the four mobility coefficients, $1/\xi_t^{\parallel}$, $1/\xi_t^{\perp}$, $1/\xi_r^{\parallel}$ and $1/\xi_r^{\perp}$,  for a rod of aspect ratio $L/R = 20$ discretised with $n_{\text{cap}} = 3$ and varying $n_{\text{cyl}}$ and $n_{\varphi}$ at the optimal value of the problem in \eqref{opt}.  Here, the cap is sampled with Gauss Legendre nodes in the axial direction and on the cylindrical part of the rod, layers of blobs in the axial direction are aligned.}
	\label{all_err_ar20}	
\end{figure}

\begin{figure}[h!]
	\centering
	%\hspace*{-17ex}
	\includegraphics[trim = {0.5cm 19.6cm 2cm 3.3cm},clip,width=1\textwidth]{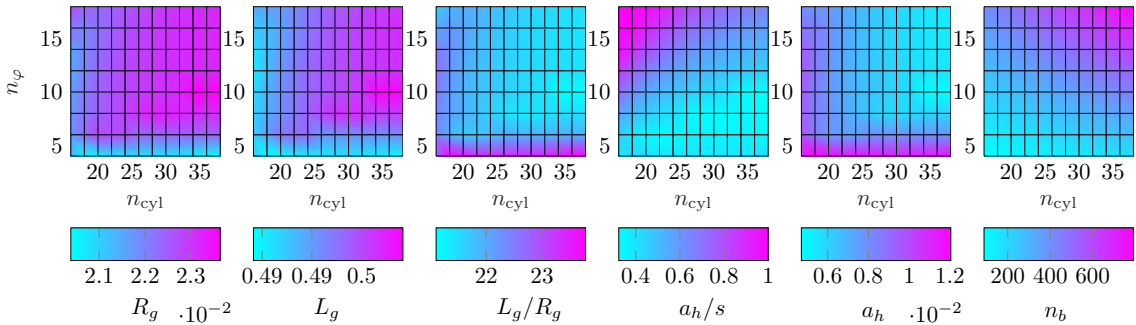}
	\caption{Same as Figure \ref{all_err_ar20}, but showing computed geometric radii, $R_g$, geometric lengths, $L_g$, aspect ratio $L_g/R_g$, blob scalings, $a_h/s$, hydrodynamic blob radii, $a_h$, and number of blobs, $n_b$, for each grid. The length and radius of this slender rod is $L=0.5$ and $R = 0.025$. With finer grid discretisation in the axial direction of the particle (larger $n_{\text{cyl}}$), $R_g$ closely matches $R$ and $L_g$ closely matches $L$.}	
	\label{param_20}
\end{figure}

\begin{figure}[h!]
	\centering
	%\hspace*{-8ex}
	\includegraphics[trim = {0.5cm 20.6cm 4cm 3.2cm},clip,width=0.9\textwidth]{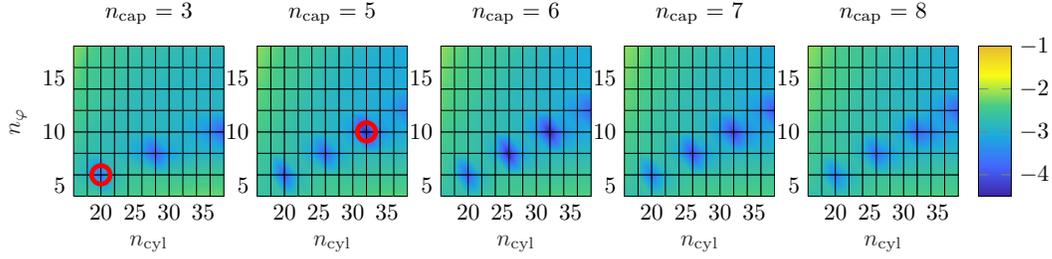}
	\caption{Maximum relative error $(\log_{10})$ in the four mobility coefficients for rods of aspect ratio $L/R = 20$ with increasing $n_{\text{cap}}$ from left to right, corresponding to the rightmost panel in Figure \ref{all_err_ar20}. The grid is aligned along the axial direction of the rod and Gauss-Legendre nodes are used for the caps. For some relation between the grid spacings in the vertical and horisontal direction along the rod, errors are significantly smaller. 
%		These ``best'' choices of grid parameters are not dependent on $n_{\text{cap}}$ for $L/R = 20$ (errors in blue). For $L/R = 4$, these ``best'' choices will depend on $n_{\text{cap}}$ - a reason being that the caps occupy a much larger part of the rods. \note{The optimal ratio between the grid spacing in the axial and horisontal direction for this particular type of discretisation seems to be $\approx0.88$, i.e. the blobs are then more clustered in the axial direction.} 
		Two sets of grid parameters $\lbrace n_{\text{cap}},n_{\text{cyl}},n_{\varphi}\rbrace$ are marked in red, to be considered for further numerical investigations in Section \ref{sec:num}. }	
	\label{max_err_ar20} 
\end{figure}

\begin{figure}[h!]
	\centering
	\hspace*{-3ex}
	\includegraphics[trim = {0.5cm 20.5cm 2cm 3.3cm},clip,width=1.01\textwidth]{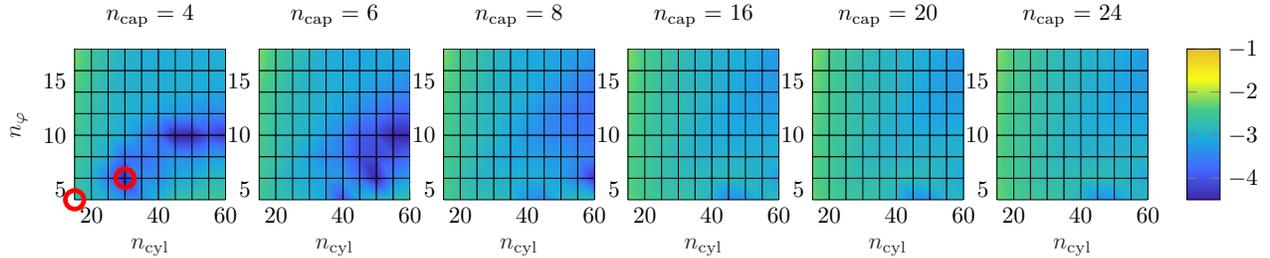}
	\caption{The same type of investigation as in Figure \ref{max_err_ar20}, but with the grid rotated between each layer so that not all the blobs of all layers are aligned, but only the blobs of every second layer. Errors are generally slightly lower than if the grid is not rotated (compare to Figure \ref{max_err_ar20}). Gauss-Legendre nodes are used for the caps. This choice was found superior to using equally spaced points along the axial direction of the caps or chopping the caps of the rod to avoid clustered blobs near the rod endpoints (results not displayed).}	
	\label{max_err_ar20_type8} 
\end{figure}

\begin{figure}[h]
	\centering
	\hspace*{2ex}
	\includegraphics[trim = {0.5cm 19.3cm 2cm 3.4cm},clip,width=1\textwidth]{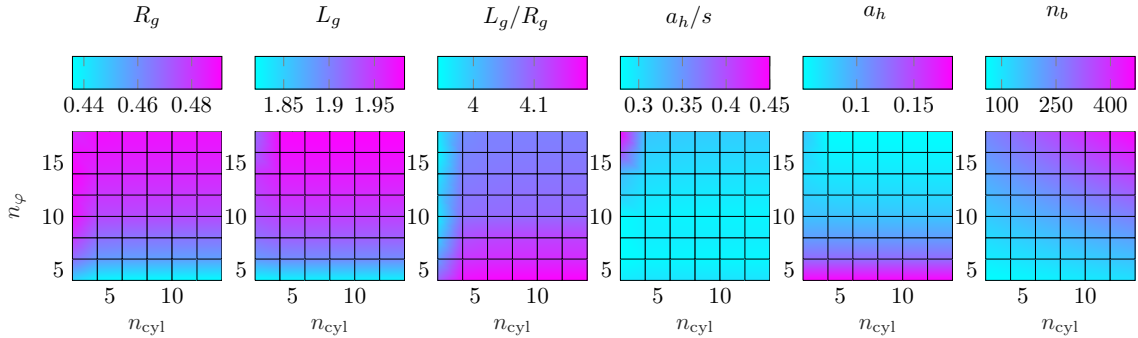}
	\caption{Solving the optimisation problem \eqref{opt} for a rod of aspect ratio $L/R = 4$, for which $L=2$ and $R=1/2$. The particle is discretised with $n_{\text{cap}} = 6$ and varying $n_{\text{cyl}}$ and $n_{\varphi}$. Displayed is the computed geometric radii, $R_g$, geometric lengths, $L_g$, computed aspect ratio $L_g/R_g$, blob scalings, $a_h/s$, hydrodynamic blob radii, $a_h$, and the number of blobs for each grid. With finer grid discretisation in the cross-section of the particle (larger $n_{\varphi}$), $R_g$ closely matches $R$ and $L_g$ closely matches $L$. }	
	\label{param_4}	
\end{figure}

\begin{figure}[h]
	\centering
	%\hspace*{-4ex}
	\includegraphics[trim = {0.5cm 20.5cm 2cm 3.4cm},clip,width=1\textwidth]{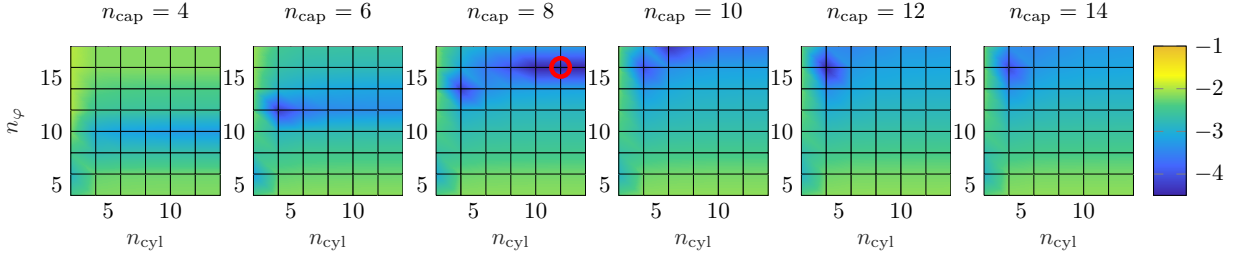}
	\caption{Maximum relative error in the four mobility coefficients $(\log_{10})$ for rods of aspect ratio $L/R = 4$ with increasing $n_{\text{cap}}$ from left to right, computed as optimal values in \eqref{opt}. Here, all layers of blobs along the axial direction of the rod are aligned and caps are discretised with Gauss-Legendre nodes in the axial direction. The choice $\lbrace n_{\text{cap}},n_{\text{cyl}},n_{\varphi}\rbrace$ marked in red is considered in numerical experiments in Section \ref{results}. The error and corresponding optimised parameters are presented in Tables \ref{ar20_tab} and \ref{tab:compare}.}	
	\label{max_err_ar4} 
\end{figure}

\begin{figure}[h]
	\centering
	%\hspace*{-5ex}
	\includegraphics[trim = {0.5cm 20.5cm 2cm 3.4cm},clip,width=1\textwidth]{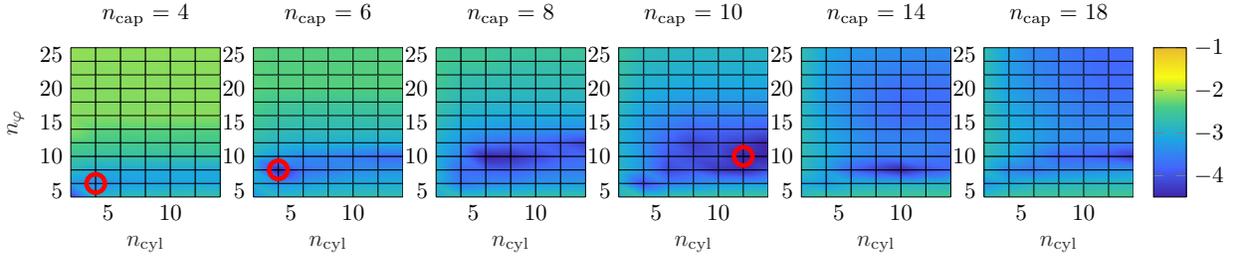}
	\caption{The same type of investigation as in Figure \ref{max_err_ar4}, but with the grid shifted between each layer of the rod. Results are improved with larger $n_{\text{cap}}$. Three sets of discretisation triplets $\lbrace n_{\text{cap}},n_{\text{cyl}},n_{\varphi}\rbrace$ are marked in red and are considered in numerical experiments in Section \ref{results}.}	
	\label{max_err_ar4_type8} 
\end{figure}

The optimisation problem in \eqref{opt} is solved for a large set of different discretisation triplets $\lbrace n_{\text{cap}},n_{\text{cyl}},n_{\varphi}\rbrace$  for each of the two aspect ratios $L/R = 20$ and $L/R = 4$. The parameter $n_{\text{cap}}$ is kept moderate to avoid excessive clustering of blobs near the endpoints of the rods, which might cause ill-conditioning of the matrix $\vec N$. Figures \ref{max_err_ar20}-\ref{max_err_ar20_type8} and \ref{max_err_ar4}-\ref{max_err_ar4_type8} display the maximum relative error in the mobility coefficients depending on the number of blobs used to discretise the particle, presented in terms of $\lbrace n_{\text{cap}},n_{\text{cyl}},n_{\varphi}\rbrace$, for shifted and aligned grids. Note that some certain choices of $\lbrace n_{\text{cap}},n_{\text{cyl}},n_{\varphi}\rbrace$ will be considerably more favorable than others and it is not only the total number of blobs that is important nor an increased refinement in a certain direction. The distribution of blobs with shifted layers generally constitutes the best choice for both aspect ratios, but for the aligned grid (Figures \ref{max_err_ar20} and \ref{max_err_ar4}), some choices of $\lbrace n_{\text{cap}},n_{\text{cyl}},n_{\varphi}\rbrace$ are better than the shifted grid (Figures \ref{max_err_ar20_type8} and \ref{max_err_ar4_type8}).
The relative error in the four different mobility coefficients are visualised for a specific choice of $n_{\text{cap}}$ for the aspect ratio $L/R = 20$ in Figure \ref{all_err_ar20}. Note that the relative error is equal in all four mobility components. The computed length, radius, aspect ratio, blob radius ratio $a_h/s$, blob radius and total number of blobs for varying $n_{\text{cyl}}$ and $n_{\varphi}$ is visualised in Figures \ref{param_20} and \ref{param_4} for two specific choices of the parameter $n_{\text{cap}}$ for the slender and fat rod. These Figures illustrate that $L_g\to L$ and $R_g\to R$ for well-resolved models of the rod.\\

%\begin{figure}[h!]
%	\centering
%	\hspace*{-8ex}
%	\includegraphics[trim = {4cm 0.5cm 4cm 1cm},clip,width=1.15\textwidth]{figures/ar20_one_maxerr_type9_twist.pdf}
%	\caption{Maximum error in the four mobility coefficients for rods of aspect ratio $L/R = 20$ with increasing $n_{\text{cap}}$ from left to right.  \note{Here, the grid is rotated between each layer so that not all the blobs of all layers are aligned, but only the blobs of every second layer. A uniform discretisation is used for the caps. The red markers can be disregarded for now.}}	
%	\label{max_err_ar20_type9} 
%\end{figure}
%\begin{figure}[h!]
%	\centering
%	\hspace*{-8ex}
%	\includegraphics[trim = {4cm 0.5cm 4cm 1cm},clip,width=1.15\textwidth]{figures/ar20_one_maxerr_type7_twist.pdf}
%	\caption{Maximum error in the four mobility coefficients for rods of aspect ratio $L/R = 20$ with increasing $n_{\text{cap}}$ from left to right.  Here, the grid is rotated between each layer so that not all the blobs of all layers are aligned, but only the blobs of every second layer. Errors are generally lower than if the grid is not rotated. Gauss-Legendre nodes are used for the caps, but the caps are chopped so that very tips of the rods are cut off. The resulting rod is of the same aspect ratio as with the other grids and errors seem to be reasonably small. The reason for trying also this discretisation is that blobs are not as clustered towards the endpoint of the rods.}	
%	\label{max_err_ar20_type7} 
%\end{figure}

\begin{figure}[h]
	\centering
	%\hspace*{2ex}
	\includegraphics[trim = {0.5cm 20.0cm 3.2cm 3.4cm},clip,width=0.84\textwidth]{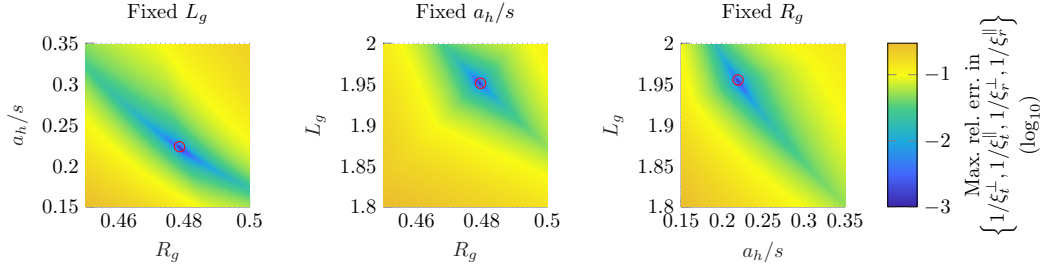}
	\caption{Optimisation landscape for the problem in \eqref{opt} for a rod of aspect ratio $L/R=4$ and $n_{\text{cap}}=6$, $n_{\text{cyl}}=4$, $n_{\varphi}=8$, with one parameter at the time fixed at its optimal value. The optimal value in all three variables $(R_g,L_g,a_h/s)$ is marked with a red circle in each of the three plots. The minimum of the maximum relative mobility coefficient error is highly dependent on the precise choice of $(L_g,R_g,a_h)$. This error level will set a baseline for the error also in a multi-particle simulation. }	
	\label{opt_rod}	
\end{figure}

The error in the solution of a mobility problem with multiple particles will be correlated with the error levels of the mobility coefficients for the rt-grid, as we will see in the numerical results section for rods, \ref{sec:num}, at least if the particles are sufficiently separated. If the error is in one of the translational coefficients, then, the translational velocity error would be limited from below on a level correlated with that coefficient error and similarly for the rotational coefficients and rotational velocity errors. We can expect no errors in the multi-particle tests for rods to be smaller than the errors seen for the single particle mobility coefficients. Reasonable error levels for multi-particle mobility problems are expected if a set of discretisation parameters are chosen so that the maximum relative mobility coefficient error is low for the rt-grid. Solving a minmax problem ensures that the error is weighted equally for rotation and translation, parallel and perpendicular to the axis of symmetry. To illustrate the importance of optimising for $(L_g,R_g,a_h)$, the optimisation landscape is visualised in Figure \ref{opt_rod}, where one parameter at the time is fixed at its optimum and the other two are varied. From the figure, it is clear that the error levels in the four mobility coefficients are highly sensitive to the choice of $(L_g,R_g,a_h)$.

For illustration purposes, we pick a number of discretisation triplets $\lbrace n_{\text{cap}},n_{\text{cyl}},n_{\varphi}\rbrace$ with varying number of total blobs and varying error levels for each aspect ratio to be used in numerical experiments. These sets are marked in red in Figures \ref{max_err_ar20} and \ref{max_err_ar20_type8} for the slender rod  and in Figures \ref{max_err_ar4} and \ref{max_err_ar4_type8} for the fat rod. For these particular choices, the single particle error levels are presented in Table \ref{ar20_tab}. Note that the error level is the same in all four mobility coefficients. The corresponding optimised parameters $L_g$, $R_g$ and $a_h/s$ are displayed in Table \ref{tab:compare}.

\begin{table}[h]
	\centering	
	\begin{tabular}{c|c|c|c|c|c|c|c|c}
		$n_{\text{cap}}$ & $n_{\text{cyl}}$ & $n_{\varphi}$ & $n_b$ & aligned & $|1-\xi_t^{\parallel}/\hat{\xi}_t^{\parallel}|$ & $|1-\xi_t^{\perp}/\hat{\xi}_t^{\perp}|$  & $|1-\xi_r^{\parallel}/\hat{\xi}_r^{\parallel}|$  & $|1-\xi_r^{\perp}/\hat{\xi}_r^{\perp}|$  \\ \hline \hline
		\multicolumn{9}{c}{$L/R = 20$} \\ \hline
		%3 & 18 & 6 & 144 &$1.44\times 10^{-3}$ & $1.44\times 10^{-3}$ & $1.44\times 10^{-3}$ & $1.44\times 10^{-3}$ \\
		4 & 15 & 4 & 92 & No & $1.73\times 10^{-3}$ & $1.73\times 10^{-3}$ & $1.73\times 10^{-3}$ & $1.73\times 10^{-3}$\\ 
		3 & 20 & 6 & 156 & Yes & $1.92\times 10^{-4}$ & $1.92\times 10^{-4}$ &$1.92\times 10^{-4}$ & $1.92\times 10^{-4}$ \\
		4 & 30 & 6 & 228 & No & $2.61\times 10^{-5}$ & $2.61\times 10^{-5}$& $2.61\times 10^{-5}$& $2.61\times 10^{-5}$\\ 
		5 & 32 & 10 & 420 & Yes &$2.21\times 10^{-6}$ & $2.21\times 10^{-6}$ &$2.21\times 10^{-6}$ & $2.21\times 10^{-6}$ \\ \hline \hline
		\multicolumn{9}{c}{$L/R = 4$} \\ \hline
		4 & 4 & 6 & 72 & No & $3.66\times 10^{-4}$ & $3.66\times 10^{-4}$& $3.66\times 10^{-4}$ & $3.66\times 10^{-4}$\\ 
		6 & 4 & 8 & 128 & No & $2.17\times 10^{-5}$ & $2.17\times 10^{-5}$ & $2.17\times 10^{-5}$ & $2.17\times 10^{-5}$ \\ 
		10 & 12 & 10 & 320 & No & $3.10\times 10^{-5}$ &$3.10\times 10^{-5}$ &$3.10\times 10^{-5}$ & $3.10\times 10^{-5}$\\
		8 & 12 & 16 & 448 & Yes & $3.46\times 10^{-6}$  & $3.47\times 10^{-6}$  &$3.47\times 10^{-6}$ & $3.46\times 10^{-6}$ 
		
	\end{tabular}
\caption{Relative error levels in the mobility coefficients for a single particle for four chosen discretisations, marked with red circles in Figures \ref{max_err_ar20}, \ref{max_err_ar20_type8}, \ref{max_err_ar4} and \ref{max_err_ar4_type8}. The presented errors are the computed optimal values in \eqref{opt}. The ``aligned'' column indicates if all blobs along the axial direction of the rod is aligned or shifted for every second layer.}
\label{ar20_tab}
\end{table}

\begin{table}[h]
	\centering	
	\begin{tabular}{c|c|c|c|c|c|c|c|c|c}
		$n_{\text{cap}}$ & $n_{\text{cyl}}$ & $n_{\varphi}$ & $n_b$ & aligned & $L_g$ & $L_g^*$ & $R_g$ & $R_g^*$ & $a_h/s$  \\ \hline \hline
		\multicolumn{10}{c}{$L/R = 20$} \\ \hline
		4 & 4 & 15 & 92 & No & 0.485 & 0.478 & 0.0213 & 0.0191 & 0.359\\ 
		3 & 20 & 6 & 156 & Yes & 0.493 & 0.484 & 0.0228 & 0.0209 &0.334\\ 
		4 & 30 & 6 & 228 & No & 0.495 & 0.483 & 0.0232 & 0.0208 &0.267\\ 
		5 & 32 & 10 & 420 & Yes & 0.496 & 0.491 & 0.0235 & 0.0225 & 0.341 \\ \hline \hline
		\multicolumn{10}{c}{$L/R = 4$} \\ \hline
		4 & 4 & 6 & 72 & No & 1.933 & 1.762 & 0.476 & 0.412 & 0.222\\ 
		6 & 4 & 8 & 128 & No & 1.953 & 1.820 & 0.479 & 0.433 & 0.221\\ 
		10 & 12 & 10 & 320 & No & 1.966 & 1.858 & 0.483 &  0.446& 0.201\\
		8 & 12 & 16 & 448 & Yes & 1.966 & 1.921 & 0.483 & 0.467 & 0.304\\ 
	\end{tabular}
\caption{Optimised geometry $(L_g,R_g,a_h/s)$ upon solving the minmax problem \eqref{opt}, compared to the resulting geometry $(L_g^*,R_g^*)$ from solving \eqref{opt1}, with fixed blob radius ratio $a_h^*/s = 0.5$. The geometries correspond to the error levels presented in Table \ref{compare_ar}. For reference, the slender ideal particle has length $L = 0.5$ and radius $R=0.025$ and the fat particle has length $L=2$ and radius $R = 0.5$.}
\label{tab:compare}
\end{table}

%\begin{figure}[h!]
%	\centering
%	\hspace*{-9ex}
%	\includegraphics[trim = {4cm 0.5cm 4cm 0cm},clip,width=1.15\textwidth]{figures/ar4_one_ncap6_err.pdf}
%	\caption{Relative error in the four mobility coefficients, $1/\xi_t^{\parallel}$, $1/\xi_t^{\perp}$, $1/\xi_r^{\parallel}$ and $1/\xi_r^{\perp}$ for a rod of aspect ratio $L/R = 4$ discretised with $n_{\text{cap}} = 6$ and varying $n_{\text{cyl}}$ and $n_{\varphi}$ upon solving the optimisation problem \eqref{opt}. The cap is sampled with Gauss Legendre nodes in the axial direction and layers of blobs in the axial direction of the cylindrical part of the rod are aligned. }	
%	\label{all_err_ar4}
%\end{figure}
%\begin{figure}[h!]
%	\centering
%	\hspace*{-10ex}
%	\includegraphics[trim = {4cm 0.5cm 4cm 1cm},clip,width=1.2\textwidth]{figures/ar4_one_maxerr_type9.pdf}
%	\caption{Maximum relative error in the four mobility coefficients for rods of aspect ratio $L/R = 4$ with increasing $n_{\text{cap}}$ from left to right, computed as optimal values in \eqref{opt}. \note{The grid is rotated between each layer of the rod and results are improved with larger $n_{\text{cap}}$. Caps are discretised uniformly in the axial direction.} }	
%	\label{max_err_ar4_type9} 
%\end{figure}
%\begin{figure}[h!]
%	\centering
%	\hspace*{-10ex}
%	\includegraphics[trim = {4cm 0.5cm 4cm 1cm},clip,width=1.2\textwidth]{figures/ar4_one_maxerr_type7.pdf}
%	\caption{Maximum relative error in the four mobility coefficients for rods of aspect ratio $L/R = 4$ with increasing $n_{\text{cap}}$ from left to right, computed as optimal values in \eqref{opt}. \note{The grid is rotated between each layer of the rod and results are improved with larger $n_{\text{cap}}$. Caps are discretised with Gauss-Legendre nodes in the axial direction but the outer part of the caps are chopped (type 7).} }	
%	\label{max_err_ar4_type7} 
%\end{figure}

\begin{figure}[h]
	\centering	
\begin{subfigure}[t]{\textwidth}	
\hspace*{12ex}
	\includegraphics[trim = {0.5cm 20.5cm 5cm 3.4cm},clip,width=0.85\textwidth]{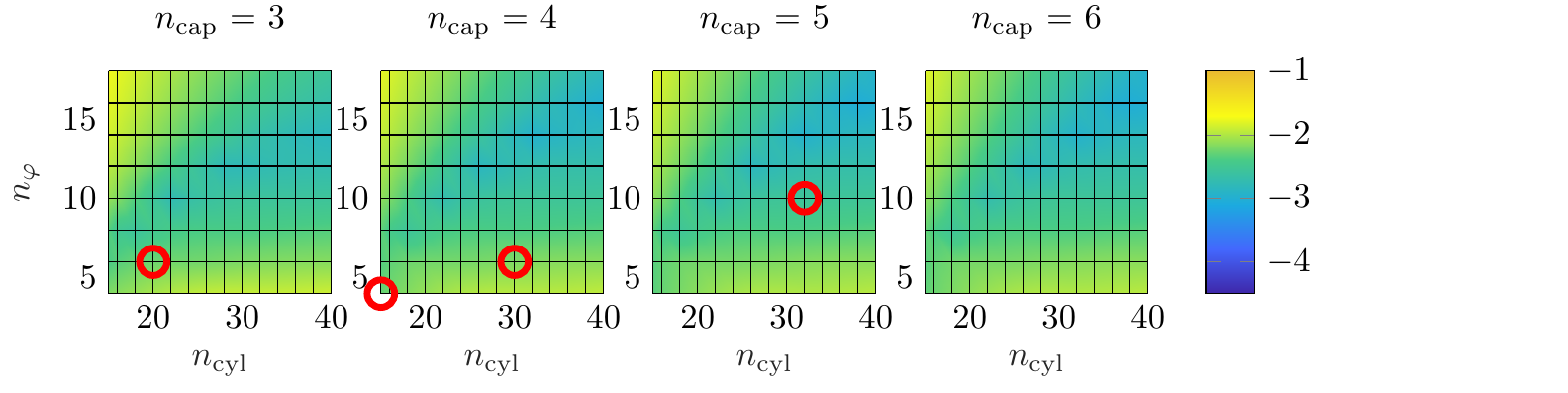}
\caption{$L/R = 20$}	
\end{subfigure}
\begin{subfigure}[t]{\textwidth}
	\centering
	%\hspace*{-14ex}
	\includegraphics[trim = {0.5cm 20.5cm 2cm 3.3cm},clip,width=1\textwidth]{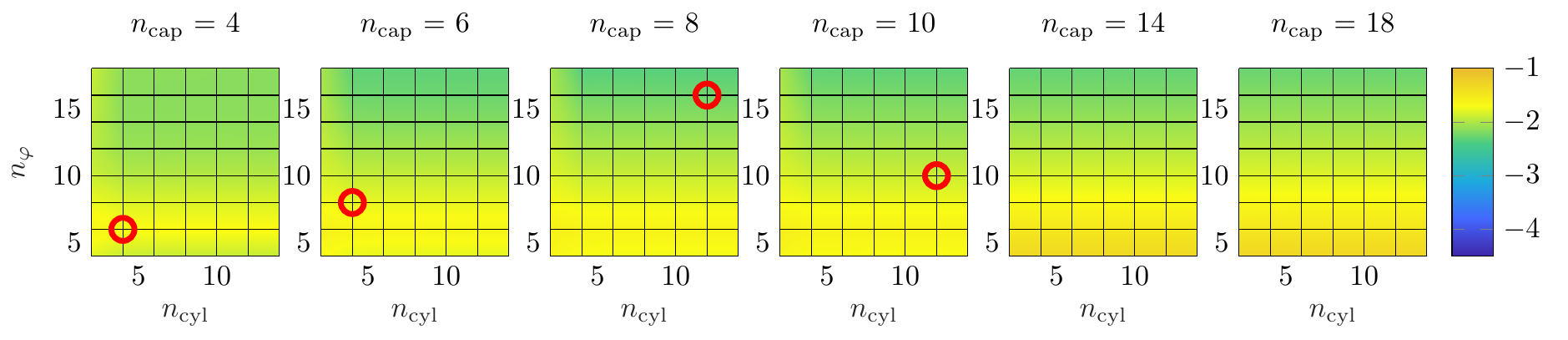}
	\caption{$L/R = 4$}	
	\end{subfigure}
		\caption{Maximum relative error in the four mobility coefficients $(\log_{10})$ while keeping the ratio $a_h/s = 0.5$ fixed and solving the problem \eqref{opt1} for $L_g$ and $R_g$. The same four sets of grid parameters $\lbrace n_{\text{cap}},n_{\text{cyl}},n_{\varphi}\rbrace$ are marked in red as in Figures \ref{max_err_ar20} and \ref{max_err_ar20_type8} for the slender rod and in Figures \ref{max_err_ar4} and \ref{max_err_ar4_type8} for the fat rod, where we optimise also for $a_h/s$. In Table \ref{compare_ar}, the error levels are compared and it can be concluded that the errors are considerably larger if we do not optimise for $a_h/s$. Only results for grids shifted along the axis of the rod are displayed.}	
			\label{ar_blobs_1} 
\end{figure}
\clearpage
To stress the importance of optimising also for $a_h/s$, and not only for the geometric surface in terms of $(L_g,R_g)$, we solve the related problem
\begin{equation}\label{opt1}
\begin{aligned}
\min\limits_{L_g,R_g}&\max\left\{\xi_r^{\perp}\Bigg|\frac{1}{\xi_r^{\perp}}-\frac{1}{\hat{\xi}_r^{\perp}\left(L_g,R_g\right)}\Bigg|,
\xi_r^{\parallel}\Bigg|\frac{1}{\xi_r^{\parallel}}-\frac{1}{\hat{\xi}_r^{\parallel}\left(L_g,R_g\right)}\Bigg|,\right.\\
& \quad\qquad\left.\xi_t^{\perp}\Bigg|\frac{1}{\xi_t^{\perp}}-\frac{1}{\hat{\xi}_t^{\perp}\left(L_g,R_g\right)}\Bigg|,
\xi_t^{\parallel}\Bigg|\frac{1}{\xi_t^{\parallel}}-\frac{1}{\hat{\xi}_t^{\parallel}\left(L_g,R_g\right)}\Bigg|\right\}, 
\end{aligned}
\end{equation}
with fixed $a_h/s = 0.5$, using shifted grids along the particle axis. The corresponding maximum relative error levels for different grids are presented in Figure \ref{ar_blobs_1} and Table \ref{compare_ar} for the two rod sizes. The errors in the self-interaction is much larger if we fix $a_h/s$ instead of optimising for this ratio. The corresponding $(L_g,R_g)$ is reported in Table \ref{tab:compare}.

\begin{table}[h]
	\centering	
	\begin{tabular}{c|c|c|c|c|c}
		$n_{\text{cap}}$ & $n_{\text{cyl}}$ & $n_{\varphi}$ & $n_b$ & Coeff.~err.,~optimal $a_h/s$ & Coeff.~err.,~$a_h/s= 0.5$  \\ \hline\hline
		\multicolumn{6}{c}{$L/R = 20$} \\ \hline
		4 & 15 & 4 & 92 & $1.73\times 10^{-3}$ & $5.23\times 10^{-3}$\\
		3 & 20 & 6 & 156 &$1.92\times 10^{-4}$ & $5.65\times 10^{-3}$ \\
		4 & 30 & 6 & 228 & $2.61\times 10^{-5}$ & $6.67\times 10^{-3}$\\
		5 & 32 & 10 & 420 & $2.21\times 10^{-6}$ & $2.78\times 10^{-3}$ \\ \hline \hline
		\multicolumn{6}{c}{$L/R = 4$} \\ \hline
		4 & 4 & 6 & 72 & $3.66\times 10^{-4}$ &  $2.11\times 10^{-2}$\\ 	
		6 & 4 & 8 & 128 & $2.17\times 10^{-5}$ & $1.71\times 10^{-2}$\\
		10 & 12 & 10 & 320 &$3.10\times 10^{-5}$  &  $1.32\times 10^{-2}$\\
		8 & 12 & 16 & 448 & $3.46\times 10^{-6}$  &  $7.51\times 10^{-3}$
	\end{tabular}
\caption{Comparing relative errors in the mobility coefficients when optimising for $L_g$ and $R_g$, using either an optimal blob radius ratio that is solved for in the optimisation problem, or choosing $a_h/s = 0.5$. The relative error levels in the mobility coefficients for a single particle indicate that errors are considerably larger if $a_h/s = 0.5$ is chosen. Data is presented for four chosen discretisations marked with red circles in Figure \ref{ar_blobs_1}, with the optimal parameters presented in Table \ref{tab:compare}. Note that the relative errors are equal for all of the four mobility coefficients (see Table \ref{ar20_tab} for the case with optimal $a_h/s$) and therefore only one number is reported per discretisation.}
\label{compare_ar}
\end{table}
%\clearpage
We have experimented with rotations of the multiblob grid about the particle axis when determining the parameters in the rt-grids in this section and found that even if the multiblob rods are not truly axisymmetric, the magnitude of the mobility coefficients for a single particle are equal to 14 digits with different rotations. In an unbounded fluid with no other particles present, this is not surprising. We can however conclude from these experiments that the known block diagonal form of the mobility matrix for axisymmetric particles hold also for our multiblobs. To ensure that our solutions of the minmax problem \eqref{opt} are not local minimas, we have also perturbed the initial guesses in all three variables $(L_g,R_g,a_h)$. These perturbations have not been done for all grids, but we are confident to conclude that the convergence to a solution in \eqref{opt} is not dependent on the initial guesses.

\clearpage

%%%%%%%%%%%%%%%%%%%%%%%%%%%%%%%%%%%%%%%%%%%%%%%%%%%%%%%%%
\subsection{Rods: The r- and t-grids}\label{rod-rt}
Even if we solve the minimisation problem \eqref{opt} (which we hereafter refer to as the self-interaction) to determine $R_g$, $L_g$ and $a_h$ in the rt-grid, the relative errors in the mobility coefficients may not be as small as desired. For a given number of blobs on the particle, the question is if there is any way to obtain smaller errors in the mobility coefficients. In this section, we explore the idea of combining two different discretisations to reduce the errors further.

The idea is to match the grid of the multiblob particle twice: \emph{either} so that translational coefficients are matched, with \begin{equation}\label{tgrid}
\xi_t^{\perp}\Bigg|\frac{1}{\xi_t^{\perp}}-\frac{1}{\hat{\xi}_t^{\perp}\left(L_g^t,R_g^t,a_h^t\right)}\Bigg| \leq \epsilon\text{ and }\xi_t^{\parallel}\Bigg|\frac{1}{\xi_t^{\parallel}}-\frac{1}{\hat{\xi}_t^{\parallel}\left(L_g^t,R_g^t,a_h^t\right)}\Bigg|\leq \epsilon
\end{equation}
  \emph{or} so that rotational coefficients are matched, with 
  \begin{equation}\label{rgrid}
\xi_r^{\perp}\Bigg|\frac{1}{\xi_r^{\perp}}-\frac{1}{\hat{\xi}_r^{\perp}\left(L_g^r,R_g^r,a_h^r\right)}\Bigg| \leq \epsilon\text{ and }\xi_r^{\parallel}\Bigg|\frac{1}{\xi_r^{\parallel}}-\frac{1}{\hat{\xi}_r^{\parallel}\left(L_g^r,R_g^r,a_h^r\right)}\Bigg|\leq \epsilon,
 \end{equation} 
for some small $\epsilon$ (note that in general, it is not possible to satisfy \eqref{tgrid} and \eqref{rgrid} with $\epsilon=0$). In other words, we match the self-interaction for the particle for translation and rotation separately. We denote the grid where the translational components are matched, fulfilling \eqref{tgrid}, as the \emph{t-grid} and the grid where rotational components are matched as the \emph{r-grid}, \eqref{rgrid}. If the t-grid is used to discretise the particle, we obtain errors of size $\epsilon$ in the translational velocity for a single particle. On the other hand, if the r-grid is used, errors of size $\epsilon$ are instead obtained in the rotational velocity. The errors in the rotational coefficients for the t-grid, and vice versa, can however be expected to be larger. We would like to utilize the good properties of the r- and t-grids also in a multi-particle setting. For increasing particle-particle distances, we have already mentioned that the particles behave more and more like isolated entities and the self-interaction becomes increasingly dominant. We can then expect to capture the translational velocities accurately with the t-grid and the rotational velocities accurately with the r-grid, also in the multi-particle case. The idea is therefore to solve a multi-particle mobility problem twice, once for each of the grids, keeping the translational velocities from the solution stemming from the t-grid and the rotational velocities computed with the r-grid. We term this solution strategy a \emph{combined solve}.

Let us for a moment consider the structure of the mobility matrix $\vec M$. If we were to explicitly compute $\vec M$, the errors in the diagonal blocks would be small by doing a combined solve. These blocks represent the force-translation and torque-rotation coupling in the self-interaction for all particles in the suspension.  A consequence is that we will have small relative velocity errors with the combined solve if the force and torque magnitudes are approximately equal -- then, the diagonal blocks of the mobility matrix are dominating the matrix vector product $\vec M\vec F$. However, for a very large torque in relation to the force on a particle, or vice versa, off-diagonal blocks of $\vec M$, representing force-rotation and torque-translation couplings through the fluid, will have a larger impact. The four mobility coefficients $\xi_t^{\perp}$, $\xi_t^{\parallel}$, $\xi_r^{\perp}$ and $\xi_t^{\parallel}$ will be important in the representation of these blocks too. The matrix $\vec M$ is symmetric and the r-grid will affect not only rotational components of the velocity, but also translational, and similarly for the t-grid. Hence, despite the fact that we seek two different grids, we would like the relative error in the translational mobility coefficients to be small when using the r-grid and vice versa. We refer to these errors as the \emph{cross-errors}.

If we do not match the r- and t-grids carefully, there is a risk of large cross-errors. The idea is to try to minimize the cross-errors by varying the blob radius $a_h$ such that the errors in the translational mobility coefficients are small when matching for the r-grid, and similarly, that the errors in the rotational mobility coefficients are small when matching for the t-grid.  We solve two different  optimisation problems: For the t-grid, we find a blob radius $a_h^t$ and geometry pair $(L_g^t,R_g^t)$ such that the error in the rotational mobility coefficients is minimized. For the r-grid, we solve the opposite problem: we find a blob radius $a_h^r$ and geometry pair $(L_g^r,R_g^r)$ such that the error in the translational coefficients is minimized. Mathematically, we can write the two problems as
%\begin{align}
%\min\limits_{a_h} \quad &\left(\frac{1}{\xi_t^{\perp}}-\frac{1}{\hat{\xi}_t^{\perp}\left(L_g^r(a_h^r),R_g^r(a_h^r)\right)}\right)^2+
%\left(\frac{1}{\xi_t^{\parallel}}-\frac{1}{\hat{\xi}_t^{\parallel}\left(L_g^r(a_h^r),R_g^r(a_h^r)\right)}\right)^2,\label{eq8} \\
%\text{s.t.}\qquad & \min\limits_{L_g^r(a_h^r),R_g^r(a_h^r)}
%\quad \left(\frac{1}{\xi_r^{\perp}}-\frac{1}{\hat{\xi}_r^{\perp}\left(L_g^r(a_h^r),R_g^r(a_h^r)\right)}\right)^2 +
%\left(\frac{1}{\xi_r^{\parallel}}-\frac{1}{\hat{\xi}_r^{\parallel}\left(L_g^r(a_h^r),R_g^r(a_h^r)\right)}\right)^2\label{eq9}
%\end{align}
\begin{equation}
\begin{aligned}
\min\limits_{L_g^r,R_g^r,a_h^r} \quad\max &\left\{\xi_t^{\perp}\Bigg|\frac{1}{\xi_t^{\perp}}-\frac{1}{\hat{\xi}_t^{\perp}\left(L_g^r,R_g^r,a_h^r\right)}\Bigg|,
\xi_t^{\parallel}\Bigg|\frac{1}{\xi_t^{\parallel}}-\frac{1}{\hat{\xi}_t^{\parallel}\left(L_g^r,R_g^r,a_h^r\right)}\Bigg|\right\},\label{eq8} \\
\text{s.t.}\qquad & 
\xi_r^{\perp}\Bigg|\frac{1}{\xi_r^{\perp}}-\frac{1}{\hat{\xi}_r^{\perp}\left(L_g^r,R_g^r,a_h^r\right)}\Bigg| \leq \epsilon, \\
&\xi_r^{\parallel}\Bigg|\frac{1}{\xi_r^{\parallel}}-\frac{1}{\hat{\xi}_r^{\parallel}\left(L_g^r,R_g^r,a_h^r\right)}\Bigg|\leq \epsilon. 
\end{aligned}
\end{equation}
and
%\begin{align}
%\min\limits_{a_h} \quad &\left(\frac{1}{\xi_r^{\perp}}-\frac{1}{\hat{\xi}_r^{\perp}\left(L_g^t(a_h^t),R_g^t(a_h^t)\right)}\right)^2+
%\left(\frac{1}{\xi_r^{\parallel}}-\frac{1}{\hat{\xi}_r^{\parallel}\left(L_g^t(a_h^t),R_g^t(a_h^t)\right)}\right)^2,\label{eq10} \\
%\text{s.t.}\qquad & \min\limits_{L_g^t(a_h^t),R_g^t(a_h^t)}
%\quad \left(\frac{1}{\xi_t^{\perp}}-\frac{1}{\hat{\xi}_t^{\perp}\left(L_g^t(a_h^t),R_g^t(a_h^t)\right)}\right)^2 +
%\left(\frac{1}{\xi_t^{\parallel}}-\frac{1}{\hat{\xi}_t^{\parallel}\left(L_g^t(a_h^t),R_g^t(a_h^t)\right)}\right)^2\label{eq11}.
%\end{align}
\begin{equation}
\begin{aligned}
\min\limits_{L_g^t,R_g^t,a_h^t} \quad\max &\left\{\xi_r^{\perp}\Bigg|\frac{1}{\xi_r^{\perp}}-\frac{1}{\hat{\xi}_r^{\perp}\left(L_g^t,R_g^t,a_h^t\right)}\Bigg|,
\xi_r^{\parallel}\Bigg|\frac{1}{\xi_r^{\parallel}}-\frac{1}{\hat{\xi}_r^{\parallel}\left(L_g^t,R_g^t,a_h^t\right)}\Bigg|\right\},\label{eq9} \\
\text{s.t.}\qquad & 
\xi_t^{\perp}\Bigg|\frac{1}{\xi_t^{\perp}}-\frac{1}{\hat{\xi}_t^{\perp}\left(L_g^t,R_g^t,a_h^t\right)}\Bigg| \leq \epsilon, \\
&\xi_t^{\parallel}\Bigg|\frac{1}{\xi_t^{\parallel}}-\frac{1}{\hat{\xi}_t^{\parallel}\left(L_g^t,R_g^t,a_h^t\right)}\Bigg|\leq \epsilon,
\end{aligned}
\end{equation}
with $\epsilon$ a small tolerance to be chosen. With a larger $\epsilon$, we allow for some slackness in the constraints and a larger feasibility region, potentially leading to smaller cross-errors, while with smaller $\epsilon$, smaller errors are obtained in the translational coefficients for the t-grid and in the rotational coefficients for the r-grid. These errors appear in the $3\times3$ diagonal blocks of the mobility matrix and since these diagonal blocks are dominating the matrix vector product $\vec M\vec F$, we would like to pick a small $\epsilon$. From empirical studies of the optimal values in \eqref{eq8} and \eqref{eq9}, we pick $\epsilon = 10^{-7}$ for rods with $L/R = 4$, while for $L/R = 10$, we pick $\epsilon = 10^{-6}$. As a rule of thumb, we allow for cross-errors a few orders larger in magnitude than the self-interaction error. The problems in \eqref{eq8} and \eqref{eq9} are solved with \texttt{fminimax} in \textsc{Matlab}.

\begin{figure}[h!]
\begin{subfigure}[t]{1\textwidth}
		\centering
		%\hspace*{-10ex}
		\includegraphics[trim = {0.5cm 20.5cm 2cm 3.3cm},clip,width=0.96\textwidth]{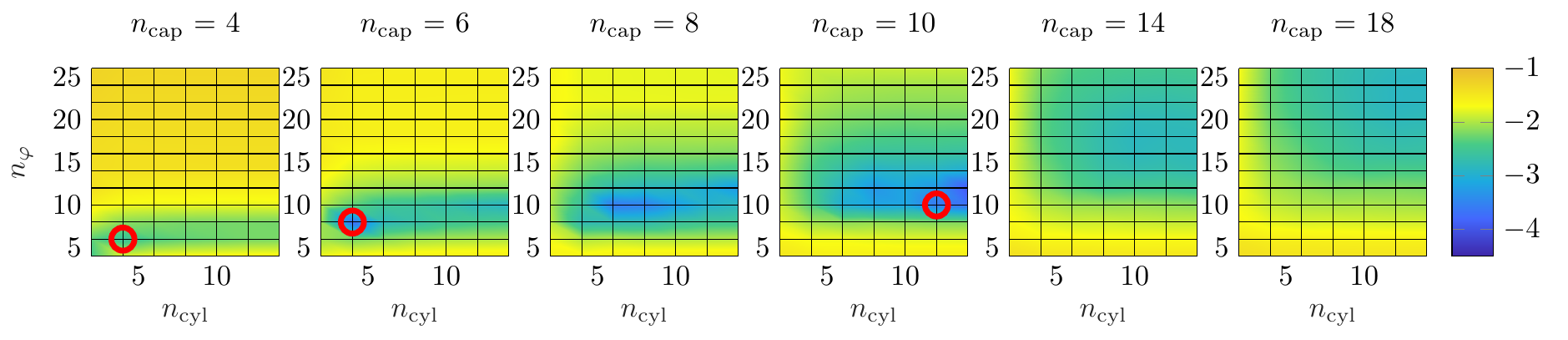}
		\caption{$L/R = 4$, $\epsilon = 10^{-7}$}
		\label{max_err_ar4_rt} 
		\end{subfigure}
\begin{subfigure}[t]{1\textwidth}
	\centering
%	\hspace*{-10ex}
	\includegraphics[trim = {0.5cm 20.5cm 7.5cm 3.3cm},clip,width=0.7\textwidth]{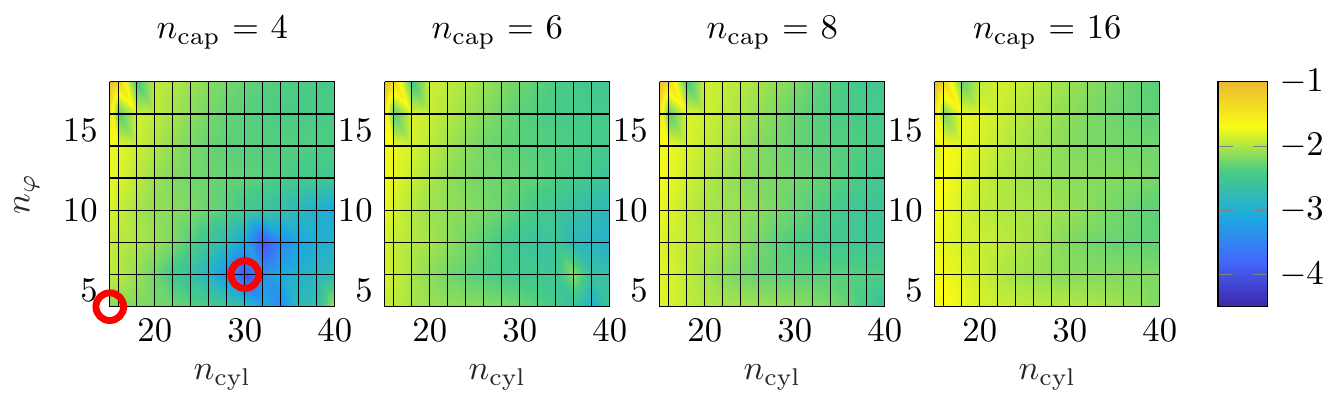}
\caption{$L/R = 20$, $\epsilon = 10^{-6}$}
	\label{max_err_ar20_rt} 
	\end{subfigure}
			\caption{Maximum cross-errors for the r- and t-grids used in the combined solve, with blob geometry computed as solution to \eqref{eq8}-\eqref{eq9}.  Cross-errors affect the off-diagonal blocks of the mobility matrix. In the diagonal blocks, the relative error is limited by the set tolerance $\epsilon$ in \eqref{eq8} and \eqref{eq9} for sufficiently separated particles. With a combined solve, the error level in the diagonal blocks are smaller than when using the rt-grid. Marked in red are discretisation parameters $\lbrace n_{\text{cap}},n_{\text{cyl}},n_{\varphi}\rbrace$ for each rod size that are considered for further numerical investigations in Section \ref{sec:num}. Results for grids with shifted layers of blobs in the axial direction of the rod are displayed.}
			\label{max_err_ar_rt}	
\end{figure}

Note that for spheres, we can find an optimal blob radius, such that all mobility coefficients are matched simultaneously, meaning that it suffices to use a \emph{single} optimised grid. 
%The optimisation problem becomes
%\begin{align}
%\min\limits_{a_h} \quad \max&\left\{\left(\frac{1}{\xi_t^{\perp}}-\frac{1}{\hat{\xi}_t^{\perp}\left(L_g^r(a_h),R_g^r(a_h)\right)}\right)^2+
%\left(\frac{1}{\xi_t^{\parallel}}-\frac{1}{\hat{\xi}_t^{\parallel}\left(L_g^r(a_h),R_g^r(a_h)\right)}\right)^2,\right. \label{eq5} \\
% &  \left.\quad\left(\frac{1}{\xi_r^{\perp}}-\frac{1}{\hat{\xi}_r^{\perp}\left(L_g^t(a_h),R_g^t(a_h)\right)}\right)^2+
%\left(\frac{1}{\xi_r^{\parallel}}-\frac{1}{\hat{\xi}_r^{\parallel}\left(L_g^t(a_h),R_g^t(a_h)\right)}\right)^2
%\right\}, \notag \\
%\text{s.t.}\qquad & \min\limits_{L_g^t(a_h),R_g^t(a_h)}
%\quad \left(\frac{1}{\xi_t^{\perp}}-\frac{1}{\hat{\xi}_t^{\perp}\left(L_g^t(a_h),R_g^t(a_h)\right)}\right)^2 +
%\left(\frac{1}{\xi_t^{\parallel}}-\frac{1}{\hat{\xi}_t^{\parallel}\left(L_g^t(a_h),R_g^t(a_h)\right)}\right)^2\label{eq6}\\
% & \min\limits_{L_g^r(a_h),R_g^r(a_h)}
%\quad \left(\frac{1}{\xi_r^{\perp}}-\frac{1}{\hat{\xi}_r^{\perp}\left(L_g^r(a_h),R_g^r(a_h)\right)}\right)^2 +
%\left(\frac{1}{\xi_r^{\parallel}}-\frac{1}{\hat{\xi}_r^{\parallel}\left(L_g^r(a_h),R_g^r(a_h)\right)}\right)^2\label{eq7}.
%\end{align}
%\question{Remove the formulation above}
%The solution strategy is the same as for the spheres.
%\subsubsection{Two different $a_h$}
For rods, the relative error in all four mobility coefficients cannot be forced to a prescribed tolerance $\epsilon$ simultaneously, i.e.~we cannot determine $(R_g,L_g,a_h)$ fulfilling both \eqref{tgrid} and \eqref{rgrid}, and we therefore consider this idea where the mobility problem is determined with a combined solve from the r- and t-grids.  The size of the cross-errors are displayed for r- and t-grids minimizing \eqref{eq8} and \eqref{eq9} for a few chosen discretisation triplets $\lbrace n_{\text{cap}},n_{\text{cyl}},n_{\varphi}\rbrace$ in Table \ref{cross_ar}. The corresponding blob geometries $(R_g^t,L_g^t,a_h^t)$ and  $(R_g^r,L_g^r,a_h^r)$ are reported in Table \ref{rt_geometry} in Appendix \ref{sec:opt_param}. Cross-errors are visualised for a larger number of discretisation triplets in Figure \ref{max_err_ar_rt}. Note that the presented cross-errors are larger than the errors in the self-interaction obtained when all mobility coefficients are matched simultaneously for the rt-grid. The cross-errors however do not affect the largest contribution to the hydrodynamical interaction of the particles, related to diagonal blocks of the mobility matrix, and are hence less severe.  In a multi-particle simulation where force and torque magnitudes are not equal, we expect the relative velocity errors to plateau at a level correlated with the cross-errors for well-separated particles, if the combined solve is applied. We investigate the accuracy using a combined solve in numerical experiments in Section \ref{sec:num}.
\begin{table}[h!]
	\centering
	\begin{tabular}{c| c| c | c| c|c | c|c}

		$n_{\text{cap}}$	& $n_{\text{cyl}}$ & $n_{\varphi}$ & aligned & r-trans  & t-rot & t-trans & r-rot  \\	 \hline \hline
			\multicolumn{8}{c}{$L/R = 20$} \\ \hline 	
		4 & 15 & 4 & No & $2.31 \times 10^{-3}$& $6.96\times 10^{-3}$ & $1.00\times 10^{-6}$& $1.00\times 10^{-6}$\\ 
		3 & 20 & 6 & Yes & $2.51\times 10^{-4}$  & $8.09\times 10^{-4}$ & $1.00\times 10^{-6}$ & $8.45\times 10^{-7}$ 	\\ 	
		4 & 30 & 6 & No & $3.45\times 10^{-5}$& $1.01\times 10^{-4}$ & $1.00\times 10^{-6}$ & $1.00\times 10^{-6}$\\ 		
		5 & 32 & 10 & Yes & $2.20\times 10^{-6}$    & $8.09\times 10^{-6}$ & $1.00\times 10^{-6}$ &	$4.97\times 10^{-7}$\\	 \hline \hline
			\multicolumn{8}{c}{$L/R = 4$} \\ \hline
		4  & 4 & 6 & No &$4.20\times 10^{-4}$  &  $2.84\times 10^{-3}$ & $1.00\times 10^{-7}$& $1.00\times 10^{-7}$\\ 			
		6 & 4 & 8 & No &$2.46\times 10^{-5}$ &   $1.82\times 10^{-4}$ & $1.00\times 10^{-7}$& $1.00\times 10^{-7}$\\ 		
		10 & 12 & 10 & No& $3.53\times10^{-5}$ & $2.54\times10^{-4}$ & $1.00\times 10^{-7}$& $1.00\times 10^{-7}$ \\	
		8 & 12 & 16 & Yes & $3.93\times 10^{-6}$ &   $2.77\times 10^{-5}$ & $1.00\times 10^{-7}$ & $1.00\times 10^{-7}$\\		
	\end{tabular}
\caption{Relative coefficient errors in matching the r-grid and t-grid for both types of rods, upon solving \eqref{eq8}-\eqref{eq9}. The cross-errors ``r-trans'', $|1-\xi_t/\hat{\xi}_t(R_g^r,L_g^r,a_h^r)| $, and ``t-rot'', $|1-\xi_r/\hat{\xi}_r(R_g^t,L_g^t,a_h^t)| $, are displayed, i.e.~the relative error in the translational coefficients using the r-grid and vice versa. Note that errors in parallel and perpendicular coefficients are of the same magnitude and therefore not distinguished. For rods with $L/R = 4$, $\epsilon = 10^{-7}$ in the problems \eqref{eq8}-\eqref{eq9}, while for $L/R = 20$, $\epsilon = 10^{-6}$. These levels agree well with the obtained error levels ``t-trans'' and ``r-rot'', corresponding to $|1-\xi_t/\hat{\xi}_t(R_g^t,L_g^t,a_h^t)| $ and $|1-\xi_r/\hat{\xi}_r(R_g^r,L_g^r,a_h^r)| $, meaning that the constraints in \eqref{eq8}-\eqref{eq9} are met. The resulting optimal geometry for the two grids is reported in Table \ref{rt_geometry} in Appendix \ref{sec:opt_param}.}
\label{cross_ar}
\end{table}

%\begin{table}[h!]
%	\centering
%	\caption{Relative coefficient error in matching the r-grid and t-grid for rods of aspect ratio $L/R=20$, where the relative error in a translational coefficients using the r-grid and vice versa is displayed.}
%	\label{cross_ar20}
%	\begin{tabular}{c| c| c | c| c}
%			$n_{\text{cap}}$	& $n_{\text{cyl}}$ & $n_{\varphi}$ & $|1-\xi_t/\hat{\xi}_t(R_g^r,L_g^r,a_h^r)| $  & $|1-\xi_r/\hat{\xi}_r(R_g^t,L_g^t,a_h^t)| $  \\	 \hline	
%		3  & 18 & 6 &  $1.92\times 10^{-3}$   & $5.75\times 10^{-3}$	\\ \hline
%		3 & 20 & 6 & $2.48\times 10^{-4}$  & $7.80\times 10^{-4}$	\\ \hline				
%		5 & 32 & 10 &  $5.64\times 10^{-7}$    & $5.99\times 10^{-7}$
%	\end{tabular}
%\end{table}
%\clearpage

To be physically sensible, the mobility matrix $\vec{M}$ needs to be symmetric and positive definite. Introduce the notation 
\begin{equation}
\vec B_u = \begin{bmatrix}
\vec I & \vec 0\\ \vec 0 & \vec 0 
\end{bmatrix}\text{ and }
\vec B_{\omega} = \begin{bmatrix} \vec 0 & \vec 0\\ \vec 0 & \vec I \end{bmatrix},
\end{equation} 
Then, we can write the mobility matrix where the t-grid and r-grid are combined as 
\begin{equation}\label{eq1} 
\vec{M}_{\text{combined solve}} = \left(\vec I \otimes \vec B_{\omega}\right)\vec M_r+\left(\vec I\otimes\vec B_u\right)\vec M_t\\
\end{equation}
with $\vec M_r$ the mobility matrix computed with the r-grid and $\vec M_t$ the mobility matrix computed with the t-grid. Theoretically, we want to ensure that the mobility matrix is symmetric, which could be done by computing $\bar{\vec M} =\left(\vec{M}_{\text{combined solve}}+\vec{M}_{\text{combined solve}}^T\right)/2 $.  Numerically however, the symmetry error is small and it is therefore sufficient to solve the problem directly as in \eqref{eq1}, or equivalently, extracting translational velocities from the t-grid and rotational velocities from the r-grid.

\begin{remark}
	A considered option to using a combine solve is the application a self-correction, with
	$\vec R_{\text{corr}}^l = \vec R^l_{\text{exact}}-\vec R^l_{\text{blob}}$ introduced for  particle $l$, such that the new resistance matrix is given by 
	\begin{equation}\label{selfcorr}
	\vec R = \vec M^{-1}_{\text{blob}}+\vec R_{\text{corr}}^{\text{self}},
	\end{equation}
	with $\vec R_{\text{corr}}^{\text{self}}$ a block diagonal matrix with 
	\begin{equation}\label{self-corr}
	\vec R_{\text{corr}}^{\text{self}}=
	\begin{bmatrix}
	\vec R_{\text{corr}}^1 & & &\\
	& \vec R_{\text{corr}}^2 & &\\
	& & \ddots & \\
	& & & \vec R_{\text{corr}}^p
	\end{bmatrix}.
	\end{equation}
	The correction matrix $\vec R_{\text{corr}}^l$ only depends on the geometry of particle $l$ and could be constructed from an accurate representation of the resistance matrix for a single particle, ignoring all other particles, and subtracting the corresponding contribution from the multiblob method, not to count the self-interaction twice. The technique allows for an accurate self-interaction in the diagonal blocks of $\vec M$ in dilute suspensions, however at a high cost: large pollution in off-diagonal blocks of the mobility matrix as a result of the global operation of inverting the resistance matrix in \eqref{selfcorr}.  Such pollution is not obtained with the combined solve and therefore that technique is favoured.
\end{remark}

\section{Pair-corrections}\label{sec:pair}
With optimised blob grids as presented in Section \ref{sec:match}, we aim at good accuracy for well-separated particles. In this section, we consider a pair-correction strongly inspired by Stokesian dynamics, aiming to improve the accuracy also for closely interacting and moderately separated particles. Denoting the correction of the resistance matrix by $\vec R_{\text{corr}}$,  the corrected system can be written on a saddle-point form as
\begin{equation}\label{eq12}
\begin{bmatrix}
\vec N & -\vec K \\ -\vec K^T & -\vec R_{\text{corr}} 
\end{bmatrix}\begin{bmatrix}
\vec \lambda \\ \vec U
\end{bmatrix} = \begin{bmatrix}
\vec 0 \\ -\vec F
\end{bmatrix},
\end{equation}
as identified by Fiore \& Swan \cite{Fiore2019} (note that this is only a modification of the system in \eqref{saddle} to be solved in the non-corrected case). Solution methods for saddle-point problems are well-studied and reviewed in \cite{Benzi2005} by Benzi, Golub and Liesen. For this particular system, fast solution techniques can be applied as summarized in \cite{USABIAGA2016}, utilizing a fast implementation of an approximation of the matrix vector product $\vec{N\lambda}$. Fiore \& Swan also stress that the resistance matrix $\vec R$ (such that $\vec R\vec U=\vec F$) can be identified as the negative Schur complement to the matrix in \eqref{eq12} and that the system in \eqref{eq12} is block diagonisable using the Schur complement.  This means that the system can be solved using GMRES (with efficient preconditioning outlined in \cite{USABIAGA2016,Fiore2019}) to compute the matrix vector product $\vec M\vec F$ at low cost. Moreover, the correction matrix $\vec R_{\text{corr}}$ only induces an extra cost in setting up the system, but not in solving it. It is of course possible to track the mobility matrix explicitly, solving \eqref{eq12} for $\vec U$ straightforwardly to obtain $\vec U = \left(\vec K^T\vec N^{-1}\vec K+\vec R_{\text{corr}}\right)^{-1}\vec F$ and identifying the corrected mobility matrix as 
\begin{equation}
\vec M = \left(\vec K^T\vec N^{-1}\vec K+\vec R_{\text{corr}}\right)^{-1}.
\end{equation}

Accounting for lubrication effects for particles in close proximity  is expensive in any grid-based method for which the fluid domain or the particle boundaries are discretised; a fine grid of the particle surfaces must be used to properly resolve the physics of the fluid. A pair-correction, as introduced in this section, can be included to avoid resolving the particle surface, but still obtain reasonable accuracy.

In Stokesian dynamics \cite{Brady1988,Brady1993,Brady2001,Swan2011,Fiore2019}, a correction is added to the resistance matrix accounting for lubrication forces, so that 
\begin{equation}
\vec R = \vec M^{-1}_{\text{coarse}}+\vec R_{\text{corr}}^{\text{pair}},
\end{equation}
with $\vec R_{\text{corr}}^{\text{pair}}$ built pair-wise by blocks on the form $\vec R_{\text{corr}}^{ij} =\vec R^{ij}_{\text{exact}}-\vec R^{ij}_{\text{coarse}}$,
for all particles $(i,j)$ closer to each other than some set cut-off, ignoring all other particles in the system. 
The correction is constructed from accurate analytical lubrication expressions for the particle pair, subtracting off the corresponding approximate resistance matrix constructed from a multipole expansion, making sure not to count the same contribution twice in the coarse representation. The small $12\times 12$ correction matrix for the pair takes the form
\begin{equation}\label{submatrix}
\vec R_{\text{corr}}^{ij} = \begin{bmatrix}
{\vec R_{\text{corr}}^{ij}}_{11} &   {\vec R_{\text{corr}}^{ij}}_{12} \\
{\vec R_{\text{corr}}^{ij}}_{12} &  {\vec R_{\text{corr}}^{ij}}_{11}
\end{bmatrix},
\end{equation}
where the blocks on the diagonal represent self-interaction within the pair and the blocks off the diagonal represent interaction with the other particle in the pair. The contribution from every pair is added, such that the correction to the resistance matrix takes the form 
\begin{equation}\label{eq14}
\vec R_{\text{corr}}^{\text{pair}} = \begin{bmatrix}
\sum_{j\neq 1}{\vec R^{1j}_{\text{corr}}}_{11} & {\vec R^{12}_{\text{corr}}}_{12} & \dots & {\vec R^{1p}_{\text{corr}}}_{11} \\
{\vec R^{21}_{\text{corr}}}_{12} & \sum_{j\neq 2}{\vec R^{2j}_{\text{corr}}}_{11} & & {\vec R^{2p}_{\text{corr}}}_{12} \\
\vdots & & \ddots & \vdots \\
{\vec R^{p1}_{\text{corr}}}_{12}  & {\vec R^{p2}_{\text{corr}}}_{12} & \dots & \sum_{j\neq p}{\vec R^{pj}_{\text{corr}}}_{11}
\end{bmatrix}.
\end{equation}
The submatrix ${\vec R^{ij}_{\text{corr}}}$ as defined in \eqref{submatrix} is set to zero for well-separated particles. Corrections on the particle level is a standard idea in Stokesian dynamics. In contrast to treating the interaction between particles  directly as in Stokesian dynamics with an RPY or higher order multipole expansion (treating spherical particles as a single blob), the coarse mobility and resistance matrices are in this paper computed using the multiblob method.  The strategy to introduce corrections for multiblob particles was presented and favoured in \cite{Fiore2019}. Here, we do not think about this as a correction to account for lubrication forces only, but as a means of encoding accurate information for pairs of particles to improve on the total mobility description of the particle system. We therefore term this correction the \emph{pair-correction}. It is however still natural to introduce a cut-off distance so that the correction is set to zero for well-separated particles.

Analytical results for the close interaction of a particle pair are however known only for spherical and spheroidal particle geometries. In this paper, we investigate how the technique can be generalised to other particle shapes, but instead of using known analytical results for the pair interaction, the pair-wise resistance matrix is precomputed with the BIE-solver equipped with QBX (one could use any accurate method at hand). Hence, we are able to generalise also to distances outside of where any lubrication approximation would be accurate. For spheres, the mobility matrix for a particle pair depends on the distance only and we can compute an accurate interpolant for the pair mobility, which allows for rapid evaluations of the corrections. For rods, pre-computations  can still be done, but creating a multi-variate interpolant is more cumbersome and is outside the scope of this work. In Section \ref{results}, pair-corrections are employed both for rods and spheres. To the knowledge of the authors, a pair-correction of this type has previously not been adopted for other particle shapes than spheres and spheroids. For spheroidal geometries, a new advantage is that the pair-corrections are applicable also for larger particle separations. This is in contrast to Stokesian dynamics, where it  numerically has been shown that the technique is sensitive to the approximation introduced in an additive correction of this type, when the correction is not  dominating the behaviour of the resistance matrix (which is only the case when treating lubrication for close to touching bodies) \cite{Lefebvre-Lepot2015}.

When pair-corrections are applied to the rt-grid, we solve a system of the form \eqref{eq12}. With pair-corrections instead applied to the combined solve introduced in Section \ref{rod-rt}, we solve two corrected mobility problems on the form \eqref{eq12}, one for the t-grid and one for the r-grid. This means that pair-corrections infer no additional solving cost, compared to using the combined solve as is. The small pair-correction matrix for the isolated particle pair is formed explicitly for all pairs sufficiently close to each other using the combined grid. Forming this matrix is however a cheap operation compared to working with the mobility matrix for a large system of particles.  Note that we, with a pair-correction  (on the rt-grid or to the combined solve), cannot guarantee that the correction is symmetric positive definite nor prove that the the corrected mobility matrix has the same property. We will return to this question in the numerical results section for rods, \ref{sec:num}, where we check for positive definiteness. 

An alternative discussed in \cite{Fiore2019} is to correct for lubrication on the \emph{blob} level, applicable mainly for particle shapes with an unknown analytical pair-wise mobility matrix. Interpreting each blob as a rigid sphere individually affected by lubrication forces from any close blobs belonging to \emph{other} particles, the lubrication effects on the blob level could be incorporated by constructing the resistance matrix with corrections included for any \emph{blobs} sufficiently close to each other \cite{Fiore2019,Wajnryb2013,Reichert2006}.
%as 
%\begin{equation}
%\vec R = \vec{\Sigma}^T\left(\vec{\tilde{N}}+\mathbb R_{\text{corr}}^{\text{lub}}\right)\vec{\Sigma},
%\end{equation}
%with $\mathbb R_{\text{corr}}^{\text{lub}}$ formed identically as $\vec R_{\text{corr}}^{\text{pair}}$ in \eqref{eq14}, but now for any \emph{blobs} sufficiently close to each other. The matrix $\vec{\tilde{N}}$ is the \emph{extended} RPY-tensor of size $\mathbb R^{6n_bp}\times\mathbb R^{6n_bp}$ including not only the translational coupling between blobs, as in $\vec N$, but also the rotational coupling, between angular velocities and torques, and the translational-rotational coupling, relating translational velocities and torques \cite{Wajnryb2013,Reichert2006}, with a positive definite extended RPY-tensor for all distances presented (also for overlapping blobs) in \cite{Wajnryb2013}. The matrix $\vec{\Sigma}$ is an extension of the matrix $\vec K$ that relates the translational and angular velocities of the blobs to the rigid body motion of the particles; its transpose couples the forces and torques applied on the blobs to the net forces and torques on the particles, i.e.~for each block $\vec \Sigma^l$ in the matrix $\vec\Sigma$,
%\begin{equation}
%(\vec \Sigma^l \vec U)_I = \begin{bmatrix}\vec u^l + \vec{\omega}^l\times(\vec b_i-\vec c^l)\\
%\vec{\omega}^l
%\end{bmatrix},
%\end{equation}
%with $I = \lbrace 6(i-1)+k\rbrace_{k=1}^6$.
As multiblob particles have a rough surface by construction, they are not allowed to come too close to each other, i.e blobs on different particles should physically not be allowed to overlap (even if this is not mathematically hindered). Therefore,  a pair-correction would only be of interest for particles of some minimum separation. On the scale of the blobs, this would correspond to large blob-blob separations relative to the blob radius between pairs of blobs for which lubrication effects are to be extracted. Therefore, each contribution to such a correction is expected to be too small in magnitude to be able to correct the error inherent in the multiblob method. Numerical tests have confirmed this hypothesis. As a consequence, we will choose to only present results for corrections on the particle level.\\

%%%%%%%%%%%%%%%%%%%%%%%%%%%%%%%%%%%%%%%%%%%%%%%%%%%%%%%%%%%%%%%%%%%%%%%%%

\begin{table}[h!]
	\centering
	\begin{tabular}{l l} \hline\hline
		s & Characteristic grid spacing, p. \pageref{sec:multiblob} \\%Section \ref{sec:multiblob}\\
		$a_h$ & Hydrodynamic radius of a blob, p. \pageref{sec:multiblob} \\
		$R$ & Radius of an ideal particle, p. \pageref{spheres}\\ 
		$L$ & Length of an ideal particle, p. \pageref{rods}\\ 
		$R_g$ & Geometric radius of a multiblob sphere or rod, p. \pageref{sec:match}\\
		$L_g$ & Geometric length of a multiblob  rod, p. \pageref{sec:match}\\
		$\hat{R}_h^{\text{trans}}$ & Computed translational radius of a multiblob sphere, p. \pageref{spheres}\\
		$\hat{R}_h^{\text{rot}}$ & Computed rotational radius of a multiblob sphere, p. \pageref{spheres}\\
		$\xi^{\parallel}_t$, $\xi^{\perp}_t$, $\xi^{\parallel}_r$, $\xi^{\perp}_r$ & Resistance coefficients, p. \pageref{sec:match}\\ 
		$\hat{\xi}^{\parallel}_t$, $\hat{\xi}^{\perp}_t$, $\hat{\xi}^{\parallel}_r$, $\hat{\xi}^{\perp}_r$ & Approximate resistance coefficients computed with the multiblob method,  p. \pageref{sec:match}\\ 
		$n_b$ & The number of blobs on one particle, p. \pageref{sec:multiblob} \\
		$n_{\text{cyl}}$, $n_{\text{cap}}$, $n_{\varphi}$ & Discretisation parameters for a rod-like particle, p. \pageref{rods} \\
		$\vec N$ & The RPY-tensor, p.	\pageref{sec:multiblob} \\
		$\vec M$ & System mobility matrix relating given particle forces and torques and the computed particle \\ & velocities, p.	\pageref{sec:intro}\\
		$\vec R$ & System resistance matrix, $\vec R = \vec M^{-1}$, p. \pageref{sec:intro} \\
		$\vec\lambda$ & Vector of forces on blobs, p.	\pageref{sec:multiblob}\\
		$\vec u^i$ & Translational velocity of particle $i$, p. \pageref{sec:intro} \\
		$\vec \omega^i$ & Angular velocity of particle $i$, p. \pageref{sec:intro} \\
		$\vec v^i$ & Velocity at the tip of particle $i$, p. \pageref{results} \\
		$\vec f^i$ & Net force on particle $i$,  p. \pageref{sec:intro} \\
		$\vec t^i$ & Net torque on particle $i$,  p. \pageref{sec:intro} \\ 
		$\delta$ & Smallest particle-particle distance, p. \pageref{results} \\
		$\alpha$ & Relation between particle force and torque magnitudes in numerical experiments, p. \pageref{alpha}\\
rt-grid & Optimised blob discretisation where errors in translational and rotational mobility coefficients \\ & are minimised simultaneously, p. \pageref{rods} \\
t-grid & Optimised blob discretisation where translational mobility coefficients are matched, p. \pageref{rod-rt} \\
r-grid & Optimised blob discretisation where rotational mobility coefficients are matched, p. \pageref{rod-rt} \\	
combined solve 	& Solving for $\vec u^i$ with the r-grid and $\vec\omega^i$ with the t-grid, p. \pageref{rod-rt} \\
		\hline\hline
	\end{tabular}
	\caption{List of commonly used symbols and notation and page references to where they are first introduced.}
\label{symlist}

\end{table}

%\clearpage
\section{Numerical results}\label{results}
We would like to quantify the error in the mobility matrix for each particle configuration, aiming for a general result for the worst possible error in solving mobility problems with particles of a certain type. One option could then be to compute the relative error in the mobility matrix, $\|\vec M_{\text{BIE}}-\vec M\|_{2}/\|\vec M_{\text{BIE}}\|_{2}$, as this metric sheds light on the appearance of the error not only for a specific right hand side $\vec F$, but also for a general force/torque vector. The mobility matrix however contains elements of largely varying magnitude and is dominated by its diagonal blocks representing the  force-translation and torque-rotation couplings in the self-interaction for all particles. Quantifying the relative error in the norm of the mobility matrix would hence mainly capture errors in these diagonal blocks. The interaction between particles is nevertheless important and so is different force-rotation and torque-translation couplings through the fluid. We therefore instead quantify the error for a large number of force/torque vectors $\vec F$ in each test, with the components of the particle forces and torques drawn independently from some distribution. In many of the simulations, we let 
\begin{equation}\label{alpha}
\| \vec t^i\|=\alpha \|\vec f^i\|,
\end{equation}
 with the direction of $\vec t^i$ and $\vec f^i$ drawn from the unit sphere. The parameter $\alpha$ is varied to account for three important scenarios: a dominating torque, a dominating force or approximately equal magnitudes of the force and torque. With these cases, different blocks of the mobility matrix will be important and we can in this way better understand the error level in different parts of the mobility matrix and the worst error level in the matrix as a whole.

One strategy for quantifying the error in the matrix vector product $\vec U = \vec M \vec F$ is to extract the relative errors in the translational and rotational velocities, $\vec u^i$ and $\vec \omega^i$, separately. This is for instance the choice made for spheres. For the rods, remember that we are ultimately interested in particle dynamics. A large relative error in rotational velocities does not necessarily have a large impact in cases where the translational velocities are large, and vice versa. Hence, we then choose to compute the relative error in the velocity at the tip of the particle, given by
$\vec v^{\text{tip}^i} = \vec u^i +\vec{\omega^i}\times\left((L/2)\vec s_i\right)$, with $\vec s_i$ the unit direction of the symmetry axis of the particle.

The error in the solutions to the mobility problems will mainly depend on the shortest distance between particles, hereafter referred to as $\delta$ or the ``gap''. For rods, there is also a dependence on the relative orientation of the particles. For large gaps, we will see that the error in the self-interaction in the rt-grid will set the lowest error level attainable for dilute suspensions, in accordance with our previous hypothesis. 

For reference, a list of commonly used notation throughout the paper is presented in Table \ref{symlist}. Parameters and settings used for the computations of the accurate reference solutions using the BIE-solver with QBX (see \cite{AfKlinteberg2016,Bagge2021}) are reported in Appendix \ref{sec:BIE}.

\subsection{Spheres}\label{sec:spheres}
The accuracy of the multiblob method is presented with and without pair-corrections for a few spheres in different configurations with different degrees of symmetry, as summarised in Table \ref{sum_spheres}. We observe that the mobility errors  decrease with particle separations and that error levels obtained with the optimised multiblob grids in $R_g$ and $a_h/s$ can be further improved with pair-corrections. We also study the dependence on the particle resolution for the accuracy and describe the error in the rotational and translational velocities for different relations between the force and torque magnitudes (different $\alpha$ in \eqref{alpha}).  The optimised grid where we have solved for $R_g$ and $a_h/s$ is compared to choosing $R_g$ such that $\hat{R}_h^{\text{trans}}=1$, with $a_h/s = 0.5$, as in \cite{USABIAGA2016}. We illustrate that it is important to match all mobility coefficients, as long as we do not only care about the translational velocity. As we have at hand an accurate interpolant for the mobility matrix for a pair of particles at any separation distance, the pair-correction can easily be computed for varying separation distances.
\clearpage
\begin{table}[h!]
	\centering
	\begin{tabular}{c|l|c}
		\textbf{Geometry} & \textbf{Studied properties} & \textbf{Subsection} \\ \hline \hline
		%	&\multicolumn{1}{c}{\textbf{Spheres}} & \\ \hline 
		Two spheres & \textbullet\quad Comparing optimised grids to the choice $\hat{R}_h^{\text{trans}}=1$ with $a_h/s = 0.5$.  & \ref{two_spheres}\\ 
		&	\textbullet\quad Grid orientation and resolution dependence for the relative error in $\vec M$. & \\ \hline
		A tetrahedron & \textbullet\quad Accuracy in $\vec u^i$ and $\vec \omega^i$ with the optimised grid with and without pair- & \ref{tetrahedron} \\of spheres &\medspace\medspace\quad  correction for varying resolutions, depending on the gap $\delta$. & \\
		 & \textbullet\quad Accuracy for different $\alpha$, with $\|\vec t^i\| = \alpha\|\vec f^i\|$ and different gaps $\delta$. & \\ \hline
		Five random  & \textbullet\quad Accuracy in $\vec u^i$ and $\vec \omega^i$ for asymmetric particle configurations with $\alpha=1$ & \ref{five_spheres} \\ spheres &  \medspace\medspace\quad  and the optimised grid with and without pair-correction for varying $n_b$. & \\  \hline
		Twisted chain & \textbullet\quad Generalisation of the tetrahedron test, but for a larger number of particles   & \ref{Ex:twist_spheres} \\ 
	of spheres.	& \medspace\medspace\quad and a fixed $\alpha$. & \\
		 & \textbullet\quad Pair-corrections applied only for neighbouring particles. &  \\ \hline\hline 
	\end{tabular}
	\caption{Overview of numerical tests presented for spheres.}
\label{sum_spheres}
\end{table}

\subsubsection{Two spheres}\label{two_spheres}
\begin{figure}[h!]
	\centering
	\includegraphics[trim = {2.5cm 15.2cm 2cm 2.5cm},clip,width=\textwidth]{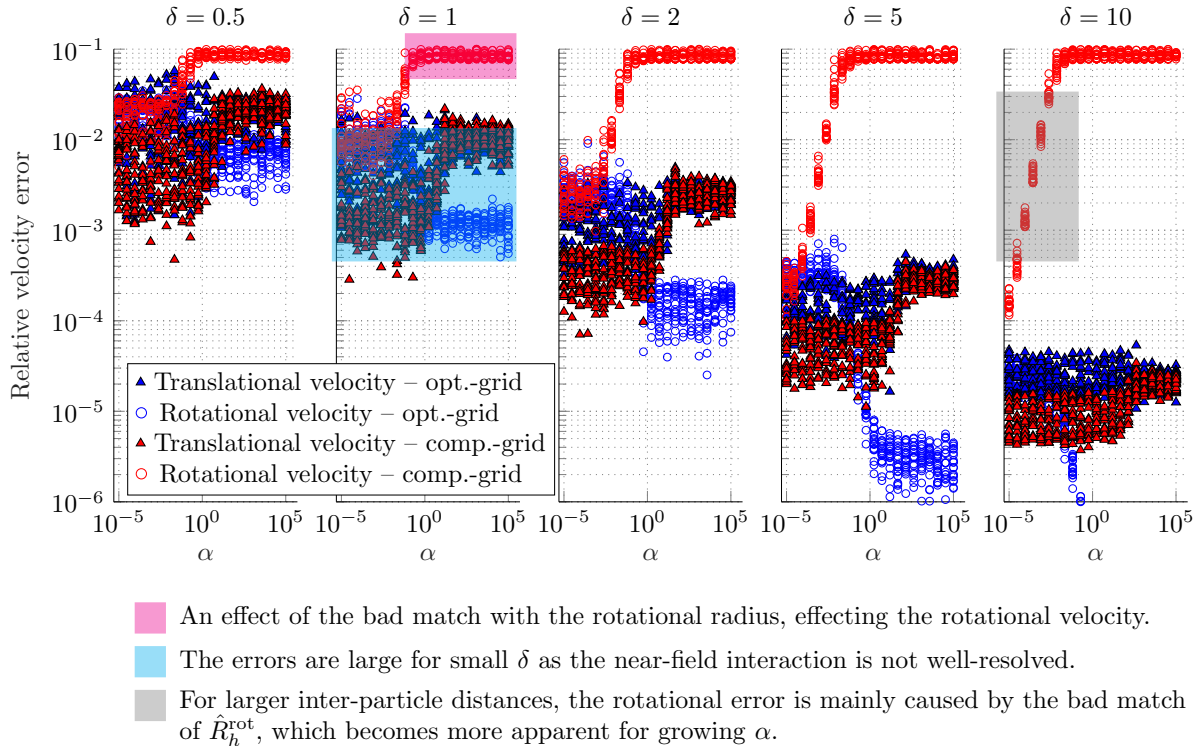}
	\caption{Velocity errors for translation and rotation for two multiblob spheres  with two different grids of 12 blobs: the comparative grid chosen so that $\hat{R}_h^{\text{trans}} = 1$ (to match a unit sphere) and $a_h/s =0.5$, and the optimised grid given from the matching problem in \eqref{opt_sphere}. Both spheres are affected by a force and torque with $\|\vec t^i\| = \alpha\|\vec f^i\|$ and directions uniformly sampled from the unit sphere.  For each $\alpha$, 20 force/torque pairs are assigned to the particles. The depicted error for each $\alpha$ is the relative error for each realisation of forces and torques, displaying the maximum taken over the two spheres. The error in the rotational velocity plateaus at a high level for the comparative grid, due to the large error in the rotational hydrodynamic radius, see Table \ref{rot_error_radius}. The particle-particle distance $\delta$ is varied between the different panels. }
	\label{two_spheres_trans12}
\end{figure}

Consider a setting with two spheres at the set separation distances $\delta\in\lbrace 0.5,1,2,5,10\rbrace$. Each particle is assigned a force and torque with $\|\vec t^i\| = \alpha \|\vec f^i\|$ and varying $\alpha$. For each $\alpha$, 20 different force/torque sets are assigned to the particles, with directions uniformly sampled from the unit sphere. We solve the mobility problem with the optimised grid and with the \emph{comparative} grid, with $\hat{R}_h^{\text{trans}}=1$ and $a_h/s = 0.5$, as in \cite{USABIAGA2016}\footnote{In that work, the effective hydrodynamic radius is chosen as $R = \hat{R}_h^{\text{trans}}$ for each resolution, given $R_g = 1$.}. In Figure \ref{two_spheres_trans12}, 12 blobs discretise each particle. If we only care about the translational motion of spheres, the error is low at large $\delta$ if the comparative grid is used. This is reasonable to do in a dynamic simulation, as the rotational velocity has no effect on the position of the spheres. On the other hand, if the optimised grid is used, we can obtain good accuracy at large $\delta$ also for the rotational velocity. For the comparative gird, errors in the rotational velocity plateau at the level of the relative error in the rotational mobility coefficient, as presented in Table \ref{rot_error_radius}. For the optimised grid, on the other hand, only the inter-particle error has an effect.

%\begin{figure}[h!]
%	\centering
%	\includegraphics[trim = {2.5cm 15.5cm 2cm 2cm},clip,width=\textwidth]{figures/nocorr_rot.pdf}
%	\caption{Velocity errors with the r-grid for two spheres of three different resolutions affected by a force and torque with $\|\vec t^i\| = \alpha\|\vec f^i\|$. The directions of the forces and torques are uniformly sampled from the unit sphere. The error visualised is the maximum relative error among the two spheres. For each $\alpha$, ten force/torque pairs are assigned to the particles. There is a rapid decay in the error for the rotational velocity with $\delta$, most notably for large $\alpha$, where the torque is dominating.}
%	\label{two_spheres_match_rot}
%\end{figure}

\begin{figure}[h!]
	\centering
	\includegraphics[trim = {1.5cm 16.9cm 10cm 3.2cm},clip,width=0.55\textwidth]{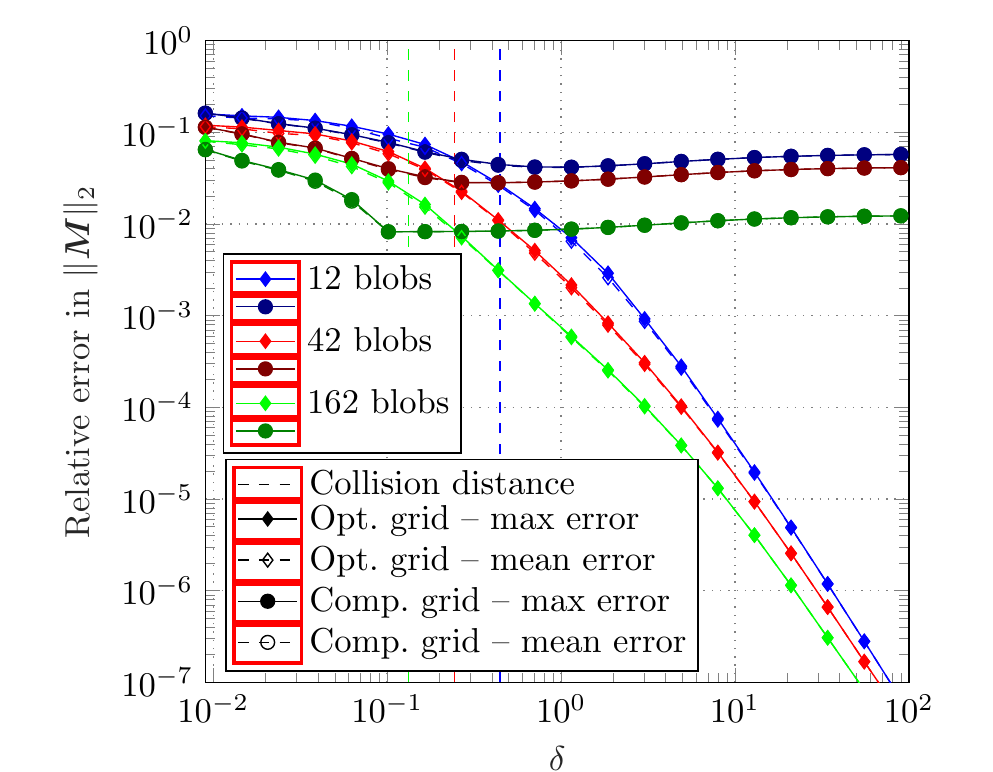}
	\caption{Relative error in the mobility matrix for two spheres at varying gaps $\delta$. Close to coinciding maximum and mean errors using 10 different rotations of the multiblob particles are displayed for spheres of three different resolutions. The errors with the comparative grids plateau at the level of the self-interaction error (for the rotational mobility coefficient), while the accuracy for the optimised grid is improved with $\delta$.}
	\label{two_spheres_match}
\end{figure}

The relative error in the mobility matrix is visualised for varying $\delta$ in Figure \ref{two_spheres_match}. Spheres of three different resolutions are investigated and it is clear that the error in the rotational mobility coefficient sets the error level for the comparative grid, whereas the velocity errors decrease with increasing $\delta$ for the optimised grid. For each $\delta$, 10 different orientations of the grid are considered and it can be concluded that the grid orientation has a very small impact on the error level. For the smallest gaps, the comparative grid captures the lubrication effects between close to touching spheres better than with the optimised grid. Note however that this is for distances closer than where blobs on adjacent particles start to overlap (marked with vertical lines in the figure). Due to the dominance of the self-interaction error using the comparative grid, we will only use the optimised grid in the remaining tests.
%\clearpage

\subsubsection{A tetrahedron of spheres}\label{tetrahedron}
\begin{figure}[h]
	\centering
	\includegraphics[trim = {2.5cm 17.7cm 1cm 2cm},clip,width=\textwidth]{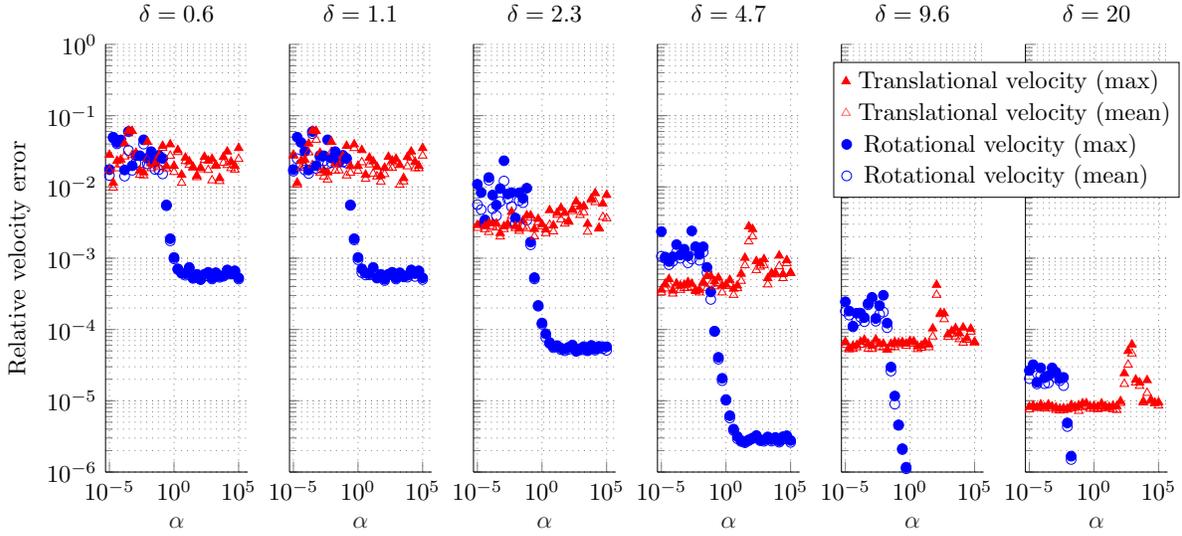}
	\caption{Velocity errors for a tetrahedron of four equally spaced spheres (see Figure \ref{tetra_geom}) discretised with 42 blobs each, affected by a force and torque with $\|\vec t^i\| = \alpha\|\vec f^i\|$ and directions uniformly sampled from the unit sphere. Note that the same $\alpha$ is used for all four spheres. For each $\alpha$, 200 different force/torque sets are assigned to the particles. The total error level counting both translation and rotation varies with varying $\alpha$ and the smallest total error is obtained for $\alpha\approx 1$. Here and in Figures \ref{tetra_all}--\ref{random_corr_12}, the optimised grid is used. }
	\label{tetra}
\end{figure}

Four multiblob spheres are placed at the vertices of a tetrahedron as in Figure \ref{tetra_geom} and the distance $\delta$ between the particles (equal among all pairs) is varied. We also vary the forces and torques by setting $\|\vec t^i\| = \alpha \|\vec f^i\|$ and sampling the direction of the force and torque uniformly from the unit sphere. The constant $\alpha$ is equal for all particles and for each $\alpha$, 200 different force/torque sets are considered. In Figure \ref{tetra}, the mean and maximum error is displayed for the translational and rotational velocity over all particles resulting from all such sets of forces and torques, using 42 blobs to discretise each sphere. Note that the behaviour of the error is the same regardless of resolution, but that the number of blobs used in the discretisation sets the error level (not displayed). The mirrored S-shape in the error for the rotational velocity indicates small errors for large $\alpha$, that is, in simulations where the torque is dominant relative to the force. The smallest total error taken over all velocity components is obtained for $\alpha\approx 1$.
%\begin{figure}[h!]
%	\centering
%	\includegraphics[trim = {1.5cm 13.6cm 8cm 7cm},clip,width=0.65\textwidth]{figures/tetra_mb_matrix12.pdf}
%	\caption{For comparison: Error in the mobility matrix when applying self- and pair-corrections in the tetrahedron test where the spheres consist of 12 blobs. The combined tr-grid solve in Figure \ref{tetra} is comparable in accuracy to the corrected solutions.}
%\end{figure}

%\begin{figure}[h!]
%	\centering
%	\includegraphics[trim = {2.5cm 15.5cm 1.5cm 2.5cm},clip,width=\textwidth]{figures/rt_grid_with_corr_12.pdf}
%	\caption{Applying a pair-correction to the tetrahedron test, with spheres consisting of 12 blobs. For each $\alpha$, 200 force/torque sets are assigned to the particles. The mean and max errors for each method are displayed for each $\alpha$. See caption of Figure \ref{tetra} for further details. }
%	\label{tetra_pair12}
%\end{figure}

\begin{figure}[h!]
	\centering
	\includegraphics[trim = {2.5cm 17.5cm 3cm 2.5cm},clip,width=0.88\textwidth]{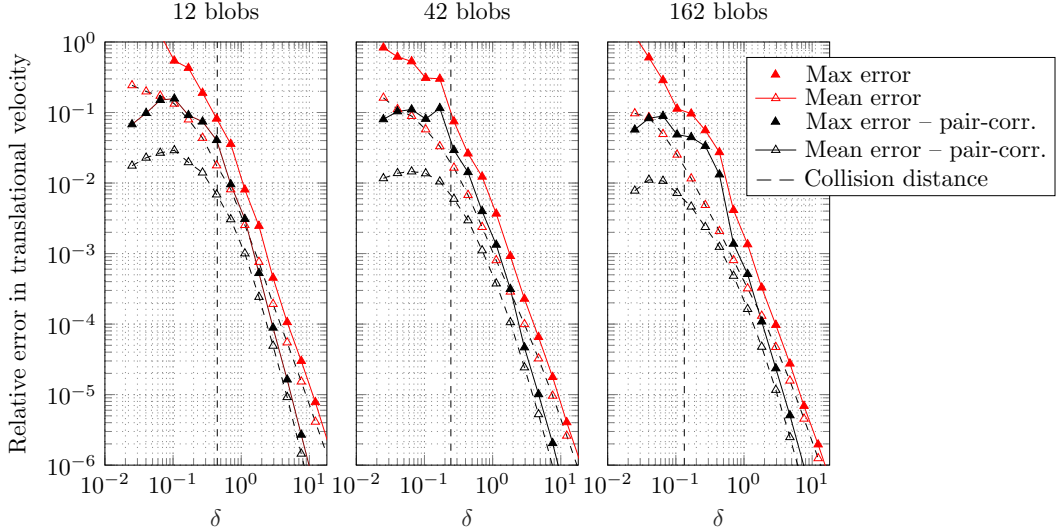}
	\caption{Applying a pair-correction to the tetrahedron test, with spheres of varying resolutions. Only the error in translational velocity for the case $\alpha = 1$ is displayed. A slight improvement can be noted with resolution. The distance for which collision with blobs on adjescent particles might occur is marked for each resolution. See caption of Figure \ref{tetra} for further details. }
	\label{tetra_all}
\end{figure}
\begin{figure}[h]
	\centering
	\includegraphics[trim = {1.7cm 18.9cm 0.5cm 2.5cm},clip,width=1.02\textwidth]{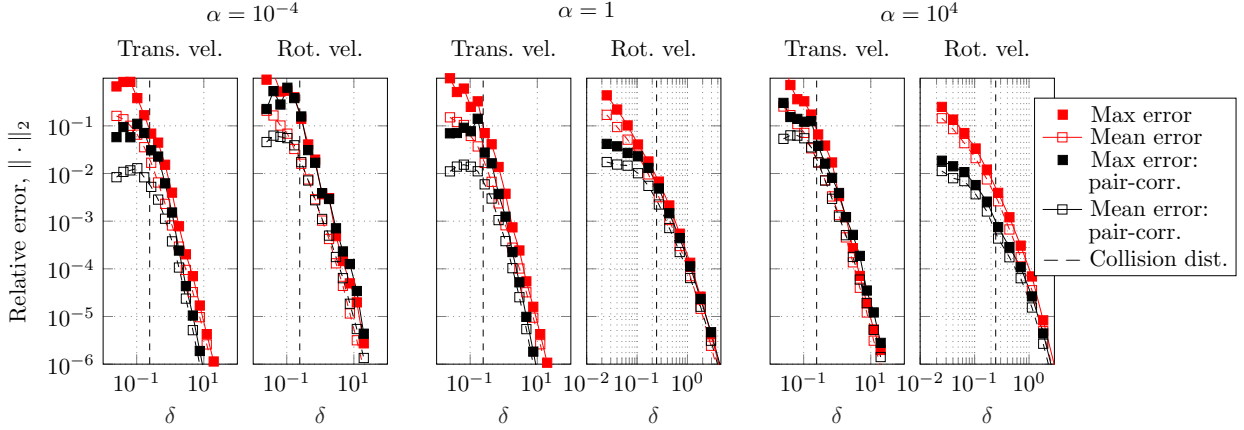}
	\caption{Applying a pair-correction to the tetrahedron test, with spheres consisting of 42 blobs.  See caption of Figure \ref{tetra} for further details. The error in the rotational velocities decays faster than the error for translational velocities when the torque is non-negligible (seen for $\alpha\geq 1$). Pair-corrections have the largest impact on the dominating component of the velocity -- translational velocities for small $\alpha$ and rotational velocities for large $\alpha$.}
	\label{tetra_pair42_delta}
\end{figure}
We now try to improve on the accuracy further by applying pair-corrections. The test is first done with $\alpha\in\lbrace 10^{-4}, 1, 10^4\rbrace$ and $n_b = 42$ in Figure \ref{tetra_pair42_delta}.  Again for each $\alpha$, 200 sets of randomly directed forces and torques are considered and the mean and max velocity error is visualised for each $\alpha$ and velocity type. The pair-correction decreases the error with approximately one order in magnitude for sufficiently close particles. The improvement is notable for $\delta<1$ and affects the dominating components of the velocity the most, i.e.~translation or rotation depending on if $\alpha$ is small or large. Note however that the maximum error for the smallest gaps is relatively large. In Figure \ref{tetra_all}, the number of blobs used to discretise the particles is varied and only the error in the translational velocity is considered for $\alpha=1$. A small improvement in accuracy can be noted with larger $n_b$.
%\begin{figure}[h!]
%	\centering
%	\includegraphics[trim = {2.5cm 15.5cm 1.5cm 2.5cm},clip,width=\textwidth]{figures/rt_grid_with_corr_42.pdf}
%		\caption{Applying a pair-correction to the tetrahedron test, with spheres consisting of 42 blobs. For each $\alpha$, 200 force/torque set are assigned to the particles. See caption of Figure \ref{tetra} for further details. }
%		\label{tetra_pair42}
%\end{figure}
%\clearpage

%\clearpage
%%%%%%%%%%%%%%%%%%%%%%%%%%%%%%%%%%%%%%%%%%%%%%%%%%%%%%%%%%%

\subsubsection{Five random spheres}\label{five_spheres}
\begin{figure}[h!]
	\centering
	\includegraphics[trim = {2.1cm 16.3cm 1.8cm 3.4cm},clip,width=\textwidth]{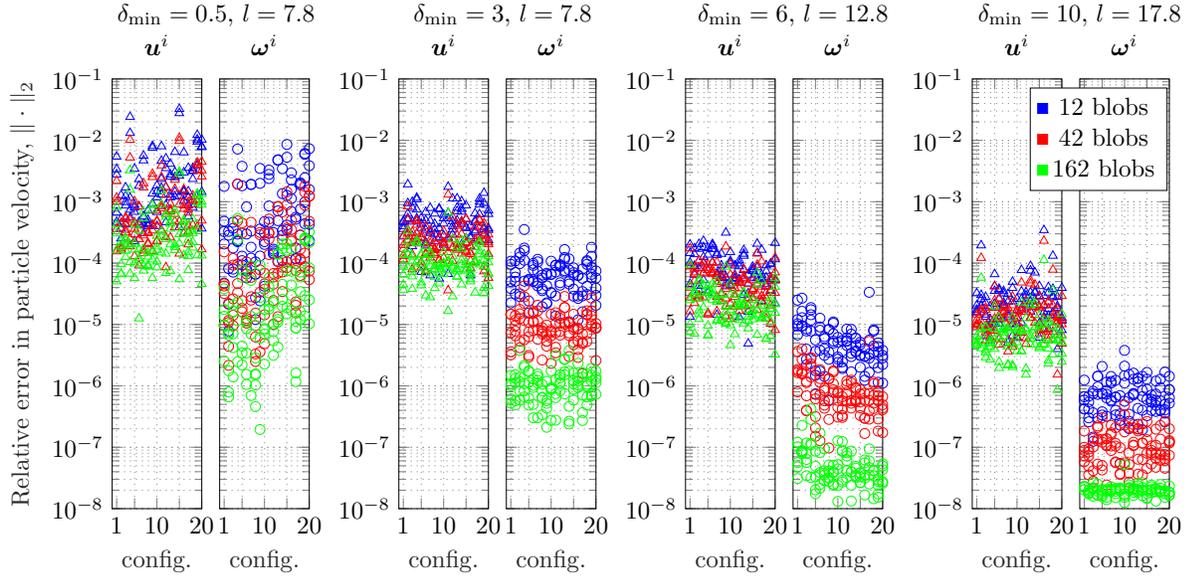}
	\caption{The example in Section \ref{five_spheres}: Random configurations of five unit spheres of different resolutions, affected by forces and torques with each component sampled from  $\mathcal U(-1,1)$. The spheres are contained in a box with minimum allowed separation distance $\delta_{\min}$ (see details in main text). The relative errors in the translational velocity, $\vec u^i$, and angular velocity, $\vec\omega^i$, are displayed for each particle in 20 different randomized configurations, yielding five marks for each configuration for each blob resolution. }
	%\caption{Results for a chain of rods of aspect ratio 20.}
		\label{random_spheres}
\end{figure}

Random configurations of five unit spheres are considered of different particle densities. A minimum allowed particle-particle distance $\delta_{\min}$ is set, with $\delta_{\min}$ in the ordered list $\lbrace 0.5, 3, 6,10 \rbrace$, meaning that no particle-pair is closer to each other than $\delta_{\min} $ for each setting.
 Particles are positioned at random in a cube of side length $l$ such that the distance for each sphere to any other sphere is at least $\delta_{\min}$, with $l=\lbrace 7.8,7.8,12.8,17.8 \rbrace$. Each pair of $\delta_{\min}$ and $l$ is tested with 20 different random configurations, with each sphere in each configuration assigned a random force and torque, with every component independently drawn from $\mathcal U(-1,1)$. This means that also here, $\|\vec f^i\| \approx \|\vec t^i\|$. The relative error in the translational velocity, $\vec u^i$, and angular velocity, $\vec\omega^i$, is computed for each sphere $i$ for the original multiblob method with the optimised grid. Results are depicted in Figure \ref{random_spheres}. Note the difference in accuracy using a coarse or a fine grid.

%	\begin{figure}[h!]
%	\centering
%	\hspace*{-10ex}
%	\includegraphics[trim = {2.1cm 16.4cm 1.8cm 3.4cm},clip,width=\textwidth]{figures/random_sub0.pdf}
%	%\input{figures/random_sub0.pdf}
%	\caption{For comparison, with self- and pair-corrections: Random suspensions of five unit spheres consisting of 12 blobs. The relative error in the translational velocity, $\vec u^i$, and angular velocity, $\vec\omega^i$, is displayed for each particle in 20 different randomized configurations. Self-and pair-corrections improve on the worst velocity error in the original multiblob method (compare the error level of the corrections to the original error level for $\vec\omega^i$), but not on the error level in $\vec u^i$.}
%	\label{random_spheres}
%\end{figure}

\begin{figure}[h!]
	\centering
	\includegraphics[trim = {2.1cm 16.2cm 1.8cm 3.4cm},clip,width=\textwidth]{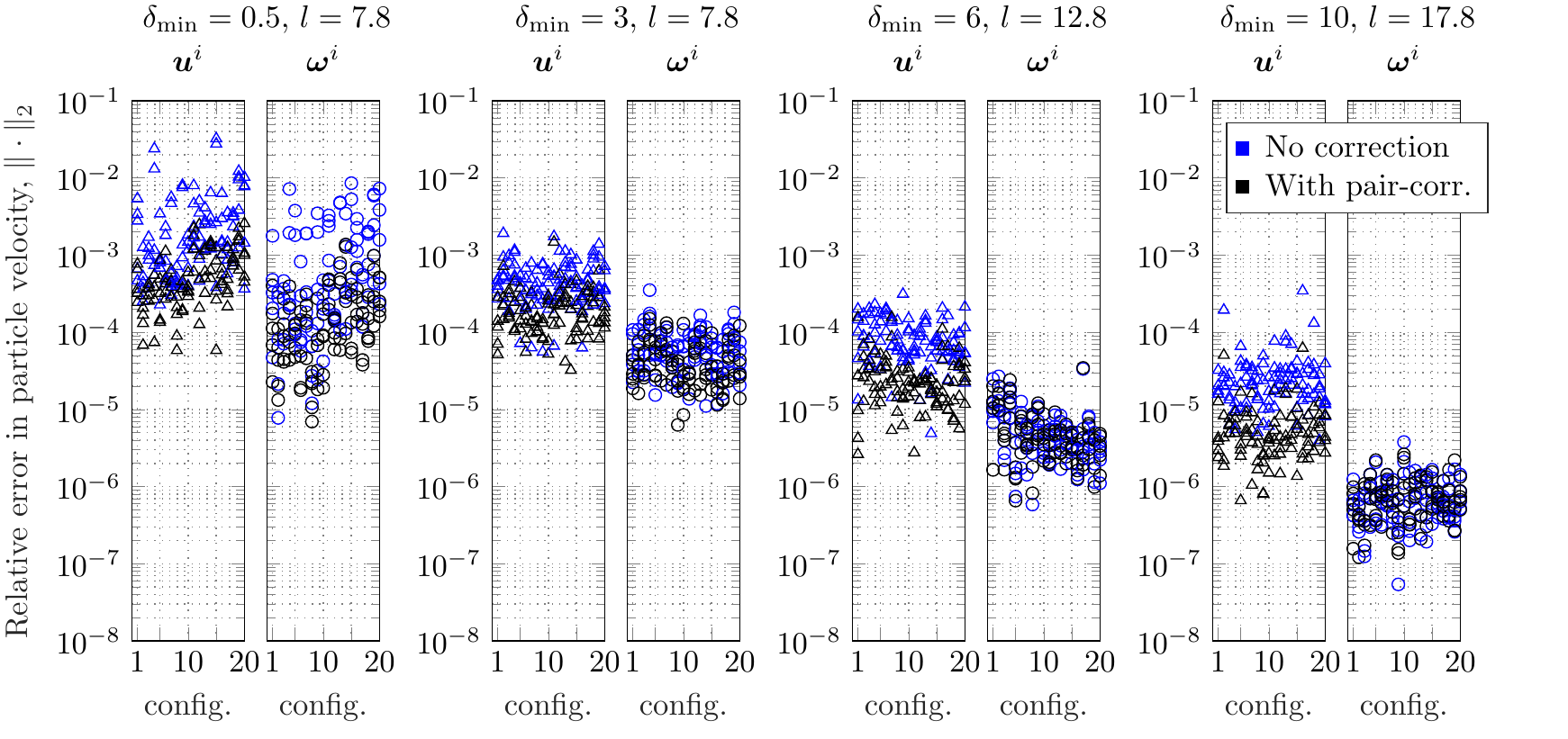}
	\caption{As in Figure \ref{random_spheres}, but comparing the results with and without pair-corrections for the discretisation with 12 blobs. }
	\label{random_corr_12}
	%\caption{Results for a chain of rods of aspect ratio 20.}
	%	\label{rods20}
\end{figure}

In Figure \ref{random_corr_12}, we apply pair-corrections to the resistance matrix obtained with the optimised grid for the coarsest blob resolution of the sphere. We can conclude that pair-corrections  improve on the accuracy for all studied ranges of particle separations and the improvement is approximately one order in magnitude. By comparing Figures \ref{random_spheres} and \ref{random_corr_12}, we can conclude that a similar improvement of the accuracy can be obtained with pair-corrections as by increasing the blob-resolution of the spheres.
%\clearpage

%%%%%%%%%%%%%%%%%%%%%%%%%%%%%%%%%%%%%%%%%%%%%%%%%%%%%%%%%%%%%%%%%%%%%%%%%%%%%%%%%%%%%%%
\clearpage
\subsection{Rods}\label{sec:num}
\noindent For illustration purposes, rods of two different aspect ratios are studied, a fat rod with $L/R = 4$ and a slender rod with $L/R = 20$. We remind of the relative error in the single particle mobility coefficients for the rt-grid, referred to as the self-interaction error, found in Table \ref{ar20_tab}, and of the cross-errors in the r- and t-grids (which together gives the combined solve) in Table \ref{cross_ar}, for four chosen discretisations of the rod for each rod size. These errors predict the errors for sufficiently separated particles in a multi-particle setting. For some separation distance for each discretisation and grid, the error in the self-interaction will be dominating the pair-interaction error. The cross-errors affect the accuracy of the combined solve, but only enter in off-diagonal blocks of the mobility matrix (not in the self-interaction $3\times 3$ blocks on the main diagonal).

In tests where the pair-correction is applied, the correction is constructed from BIE-results involving two isolated rods. To the best of our knowledge, pair-wise corrections have not previously been implemented for the rod geometry as no exact analytical lubrication expressions exist for other shapes than spheroidal particles. For each new relative configuration of two rods (among a larger set), a new accurate mobility matrix has to be computed for the pair. In practice, the column $j$ in the mobility matrix can be computed by solving the mobility problem with a unit force/torque vector with elements $\vec F^j_i=\delta_{ij}$. For two interacting particles, it then takes 12 Stokes solves to determine the form of the mobility matrix. To avoid a large number of such (costly) computations, we first stick to geometries where the pair can be obtained from translation and/or rotation of some basic configurations. We will however consider a random configuration of rods in the final numerical example. 

\begin{table}[h!]
	\centering
	\begin{tabular}{c|l|c}
		\textbf{Geometry} & \textbf{Studied properties} & \textbf{Section} \\ \hline \hline
	%	\multicolumn{3}{c}{\textbf{Rods}} \\ \hline
		Two parallel rods & \textbullet\quad The need of optimising for the blob grid geometry: A comparison of the\\&\medspace\medspace\quad rt-grid and a \emph{comparative} grid where we only optimise for $(R_g,L_g)$, but & \\ & \medspace\medspace\quad  set $a_h/s= 0.5$.& \ref{parallel} \\ & \textbullet\quad Effect of the self-interaction error in the rt-grid at large separations. &  \\ & \textbullet\quad Error levels with the rt-grid vs. a combined solve for varying $n_b$. &  \\ \hline
		
		Sweep test & \textbullet\quad As in \ref{parallel}, but for general relative particle orientations, visualising the & \ref{sweep_test} \\ for two rods &\medspace\medspace\quad gain from the smaller self-interaction error with a combined solve & \\&\medspace\medspace\quad compared to the larger self-interaction error using the rt-grid.& \\&\textbullet\quad The interplay between the self-interaction error and the pair-interaction & \\ &\medspace\medspace\quad error.   &\\ \hline
		Twisted rod chain & \textbullet\quad As in \ref{sweep_test}, with  pair-corrections applied to the rt-grid or to the  & \ref{sec:twist_three} \\ 
		-- three rods & \medspace\medspace\quad combined solve for neighbouring particles, with coarse and fine meshes.& \\&\textbullet\quad  The dependence of the pair-correction quality on the self-interaction & \\ &\medspace\medspace\quad  error of the underlying grid. & \\ \hline
		Twisted rod chain & \textbullet\quad As in \ref{sec:twist_three}, but applied to a larger set of particles, displaying the error & \ref{sec:twist_eight} \\ -- eight rods &\medspace\medspace\quad in $\vec v^i$, $\vec u^i$ and $\vec\omega^i$. & \\ \hline
		Random rods & \textbullet\quad As in \ref{sec:twist_eight}, but for general and asymmetric particle configurations,  & \ref{random_rods}  \\ &\medspace\medspace\quad  displaying the accuracy in $\vec v^i$ with and without corrections to the rt-grid\\ & \medspace\medspace\quad  or to the combined  solve for slender rods of a coarse and a fine resolution. & \\ & \textbullet\quad Comparing a pair-corrected combined solve for a coarsely resolved fat rod & \\ &\medspace\medspace\quad to a pair-corrected solution with a fine rt-grid. & \\ \hline\hline
	\end{tabular}
	\caption{Overview of numerical tests presented for rods with a progression of complexity between the consecutive tests.}
\label{sum_rods}
\end{table}

\clearpage
\subsubsection{Two parallel rods}\label{parallel}
\begin{figure}[h!]
	\centering
	\includegraphics[trim = {1.5cm 5.7cm 0cm 3.4cm},clip,width=1.05\textwidth]{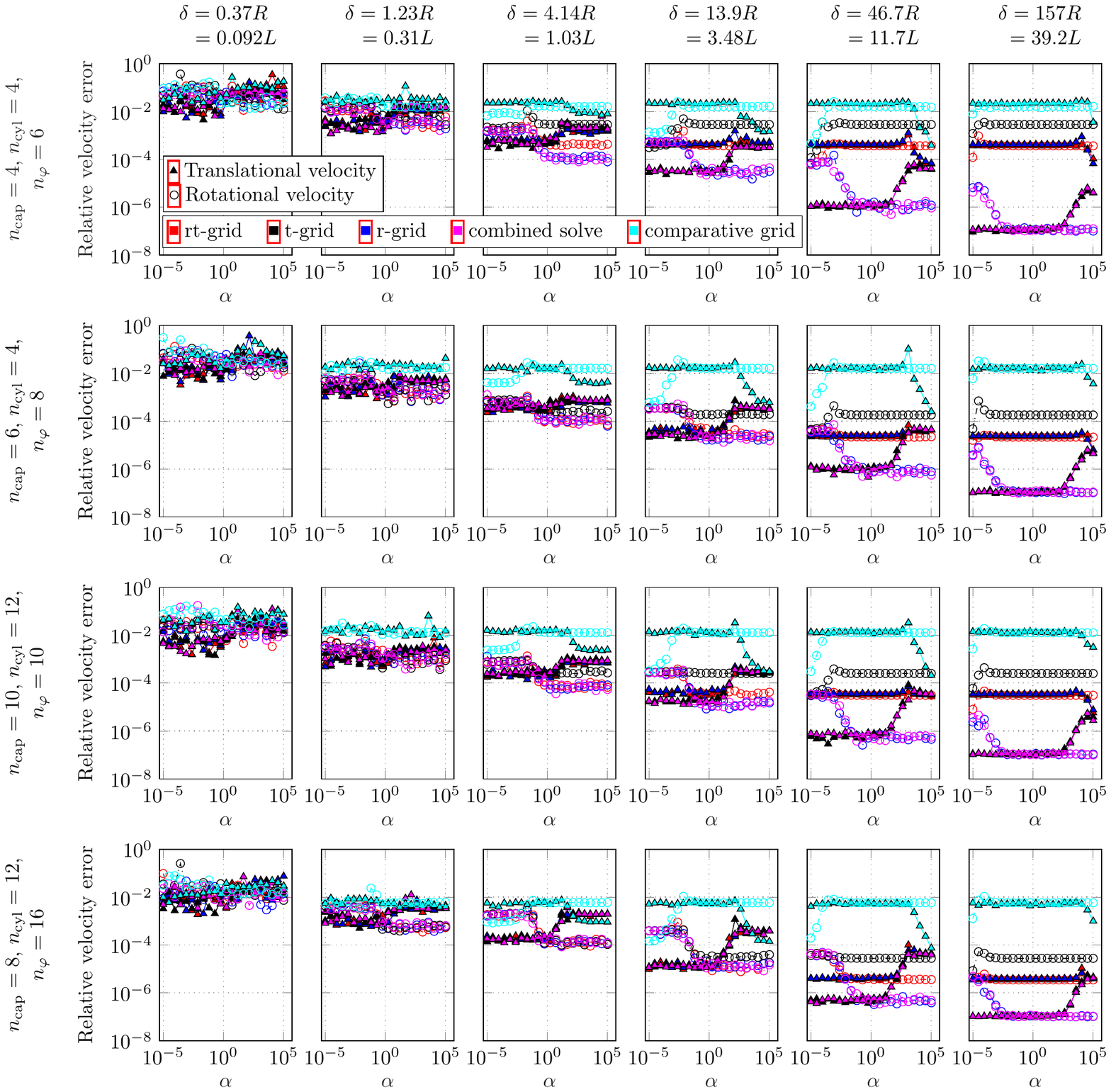}
	\caption{Two parallel rods of aspect ratio $L/R = 4$ with different blob grids $\lbrace n_{\text{cap}}, n_{\text{cyl}},n_{\varphi}\rbrace$ in the four rows and different gaps $\delta$ from left to right. For large gaps (see the rightmost panels), the error levels for the r-, t- and rt-grids are all predicted from Tables \ref{ar20_tab} and \ref{cross_ar}  and the error level for the comparative grid is predicted from Table \ref{compare_ar} (see bullet points in main text). For small gaps, the error in the pair-interaction is dominating, independently of the choice of grid.}
	\label{parallel_ar4}
	%\caption{Results for a chain of rods of aspect ratio 20.}
	%	\label{rods20}
\end{figure}

In this test, we explore the relation between the error in the mobility coefficients with a certain grid and the multi-particle errors at large particle-particle distances. The setting is two parallel rods of aspect ratio $L/R = 4$ with increasing separation distance. We consider five different strategies for solving the mobility problem: (i) using the optimised rt-grid where the error in all four mobility coefficients are minimized simultaneously, (ii) using the t-grid separately, where only translational mobility coefficients are matched, (iii) using the r-grid separately (matching rotational mobility coefficients), (iv) using the combined solve, where the t-grid is used to compute the translational velocities and the r-grid to compute the rotational velocities and, finally, (v) using a comparative grid, where we optimise for $(R_g,L_g)$, but set $a_h/s=0.5$. This last solution strategy is included for comparison to show the importance of optimisation of all parameters. For each inter-particle distance, the quotient between the force and torque magnitude is varied, with $\|\vec t^i\| = \alpha \| \vec f^i\|$. For each $\alpha$, 100 sets of randomly oriented forces and torques are assigned to the particles, drawn from a uniform distribution on the unit sphere.  In Figure \ref{parallel_ar4}, the maximum relative error in the translational and rotational velocity, as compared to the corresponding BIE-solution, is visualised for each $\alpha$ and each choice of $\lbrace n_{\text{cap}}, n_{\text{cyl}}, n_{\varphi}\rbrace$. For the rightmost panels in Figure \ref{parallel_ar4}, compare with Tables \ref{ar20_tab} and \ref{cross_ar} and note that

\begin{itemize}	
	\item For the t-grid, the rotational velocity plateaus at the single particle rotational error level as represented by  the t-grid cross-error, $|1-\xi_r/\hat{\xi}_r(R_g^t,L_g^t,a_h^t)|$. For the r-grid, the error for the translational velocity  plateaus at the translational error level represented by the r-grid cross-error,  $|1-\xi_t/\hat{\xi}_t(R_g^r,L_g^r,a_h^r)|$.  
	\item The rotational velocity is resolved to an error level of $10^{-7}$ for the r-grid and the translational velocity is resolved to the same level for the t-grid. This is due to the tolerance $\epsilon$ chosen for the r- and t-grid optimisation problems in \eqref{eq8}-\eqref{eq9}. For large $\alpha$, such that the force is dominant, and for small $\alpha$, where the torque is dominant, the cross-errors will have an impact on the solution. 
	\item For the original rt-grid, both the relative error for the translational and for the rotational velocity plateau at the error level presented by the relative error in the mobility coefficients (the self-interaction error for the rt-grid).
	\item For the comparative grid, considerably larger error levels are obtained than with the rt-grid for all choices of discretisation parameters. The error level corresponds to the self-interaction error for the comparative grid, as presented in Table \ref{compare_ar}.
	\item For the combined solve, the results are better than those obtained with the rt-grid and the error plateaus at a lower level for discretisation sets $\lbrace n_{\text{cap}}, n_{\text{cyl}}, n_{\varphi}\rbrace$ where the error level with the rt-grid is large. This however comes at the cost of solving two mobility problems instead of one.  
\end{itemize}

As the tolerance for solving the optimisation problems for the r- and t-grids in \eqref{eq8}-\eqref{eq9} is chosen to $\epsilon = 10^{-7}$, the error  in mobility problems with the combined solve will not decay below this level; This is the self-interaction error for the combined solve. Similarly as for the spheres, translational relative errors are the smallest when the force magnitude is dominating and rotational relative errors are the smallest when the torque is dominating. Effects from particle interactions are important for closely interacting particles, which is a reason to why the difference between different discretisations and solution strategies is small for small $\delta$ in Figure \ref{parallel_ar4}.

Similar results are obtained with the larger aspect ratio, $L/R = 20$, not included here. 
\clearpage
\subsubsection{Sweep test}\label{sweep_test}

\begin{figure}[h!]
	\centering
			\centering
	\includegraphics[trim = {16.4cm 1.5cm 15cm 5.0cm},clip,width = 0.4\textwidth]{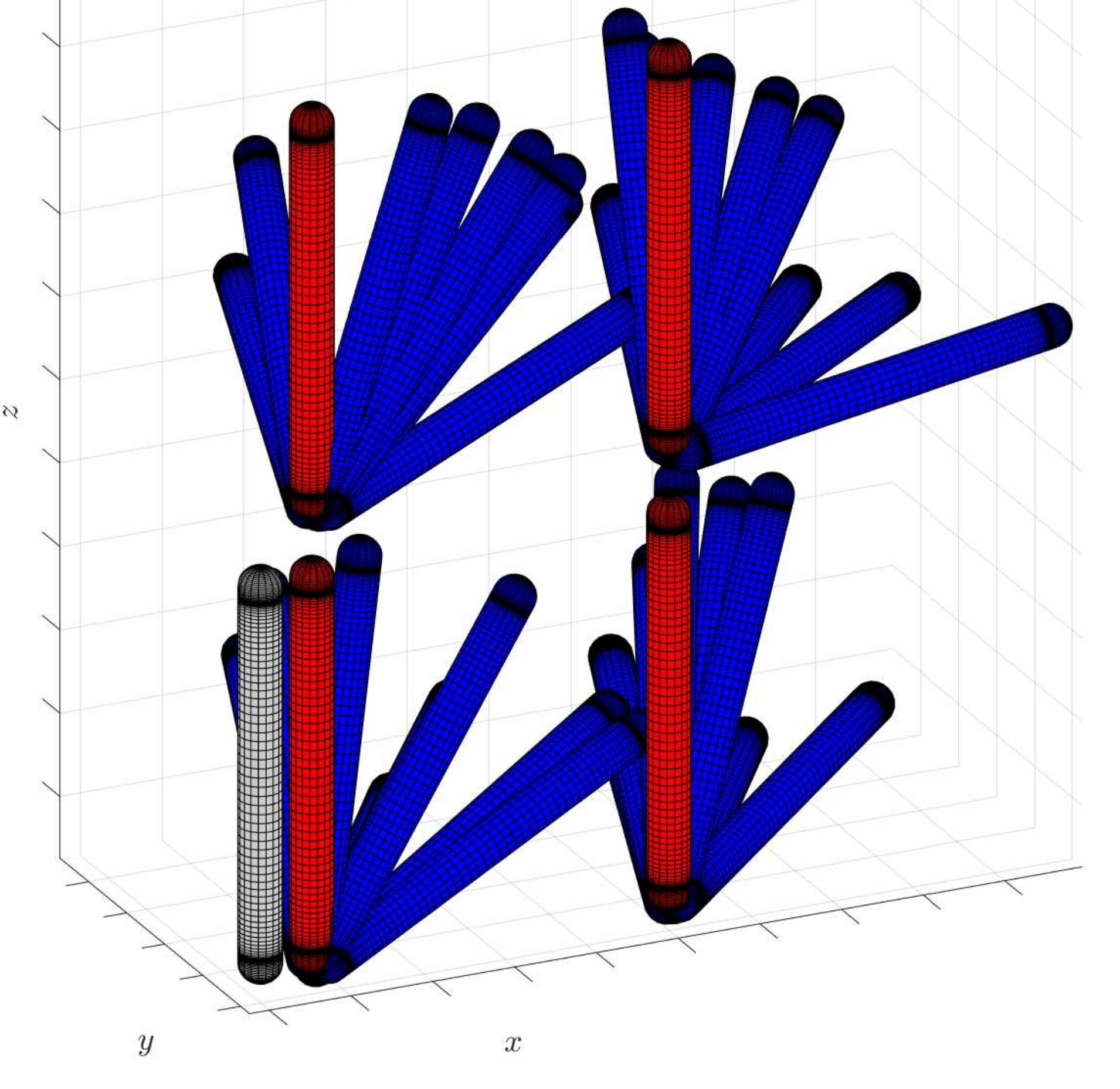}
	\caption{Test for two rods, where the second rod is placed in every node of a 2D-grid and randomly oriented to ten orientations for each new position. Grey indicates a fixed rod, whereas red indicates positions in a 2D grid for which the second rod is rotated in the first orthant (blue).}
	\label{grid_rod}
\end{figure}

In this numerical test, we investigate the worst velocity error for two neighbouring particles with varying relative orientations and compare results from using an rt-grid and a combined solve. This is a more general test than the one for two parallel rods in that we vary both distances and orientations, bearing in mind that parallel rods constitutes a particular geometry with a high degree of symmetry.
We perform a sweep test, with one rod fixed at the origin while the other rod is placed in a Cartesian 2D-grid (corresponding to the x-z-plane in Figure \ref{grid_rod}) relative to the first rod. For each new position, the second rod is oriented randomly in the first orthant with 10 different orientations, such that the minimum distance to the first rod is kept. See Figure \ref{grid_rod} for an illustration with the second rod in four different positions. 

\begin{figure}[h!]
	\centering
\begin{subfigure}[t]{0.45\textwidth}
			\centering
			\hspace*{-9ex}
			\includegraphics[trim = {0cm 13.5cm 9cm 3.4cm},clip,width=1.35\textwidth]{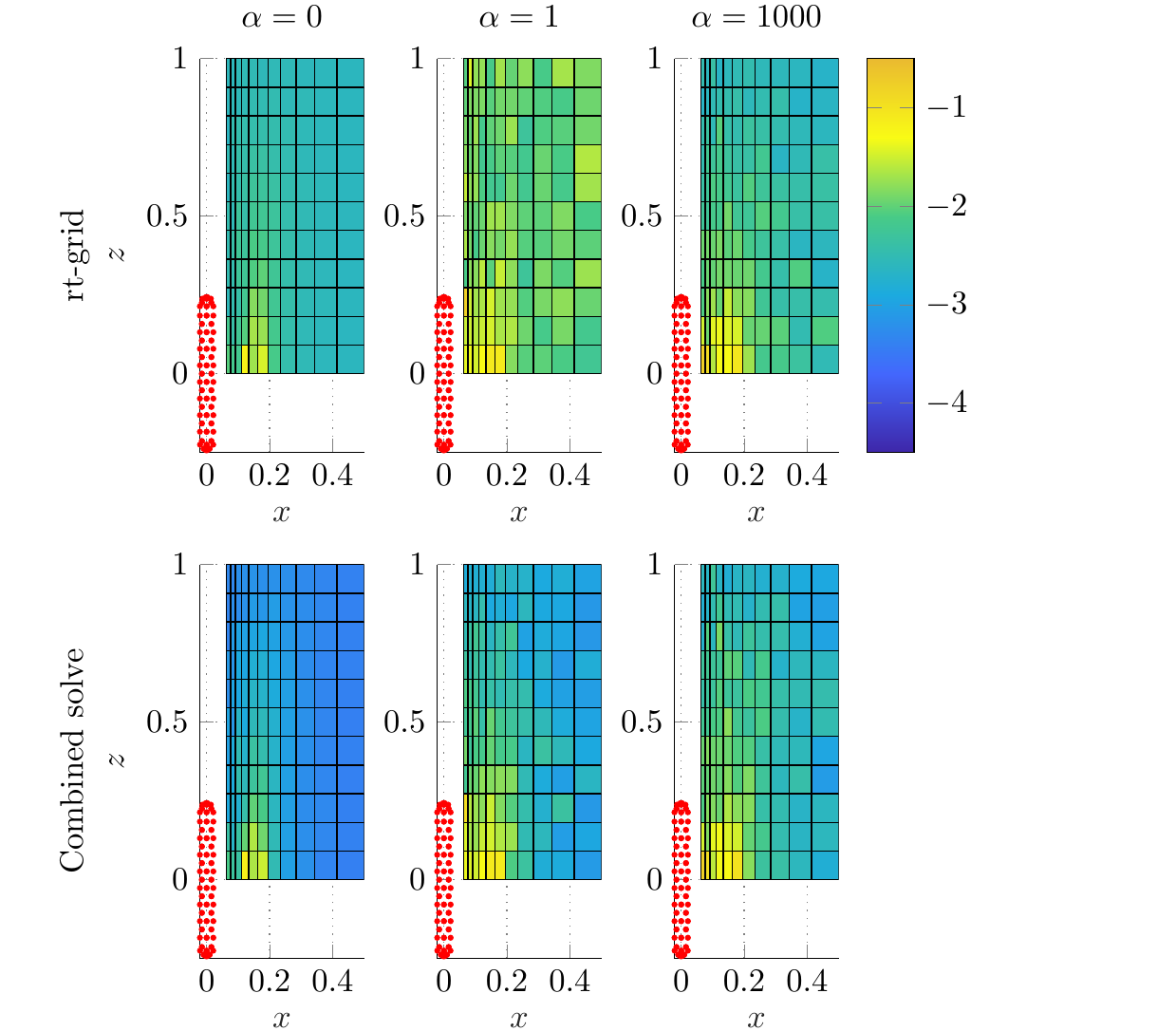}
			\caption{Rods discretised with $n_{\text{cap}}=4$, $n_{\text{cyl}}=15$, $n_{\varphi}=4$ and twisted layers of blobs in the axial direction (displayed in red).}
			%\caption{Results for a chain of rods of aspect ratio 20.}
		\label{sweep_ar20_1}
		\end{subfigure}~
		\begin{subfigure}[t]{0.45\textwidth}
			\centering
			%\hspace*{-5ex}
			\includegraphics[trim = {0cm 13.5cm 9cm 3.4cm},clip,width=1.35\textwidth]{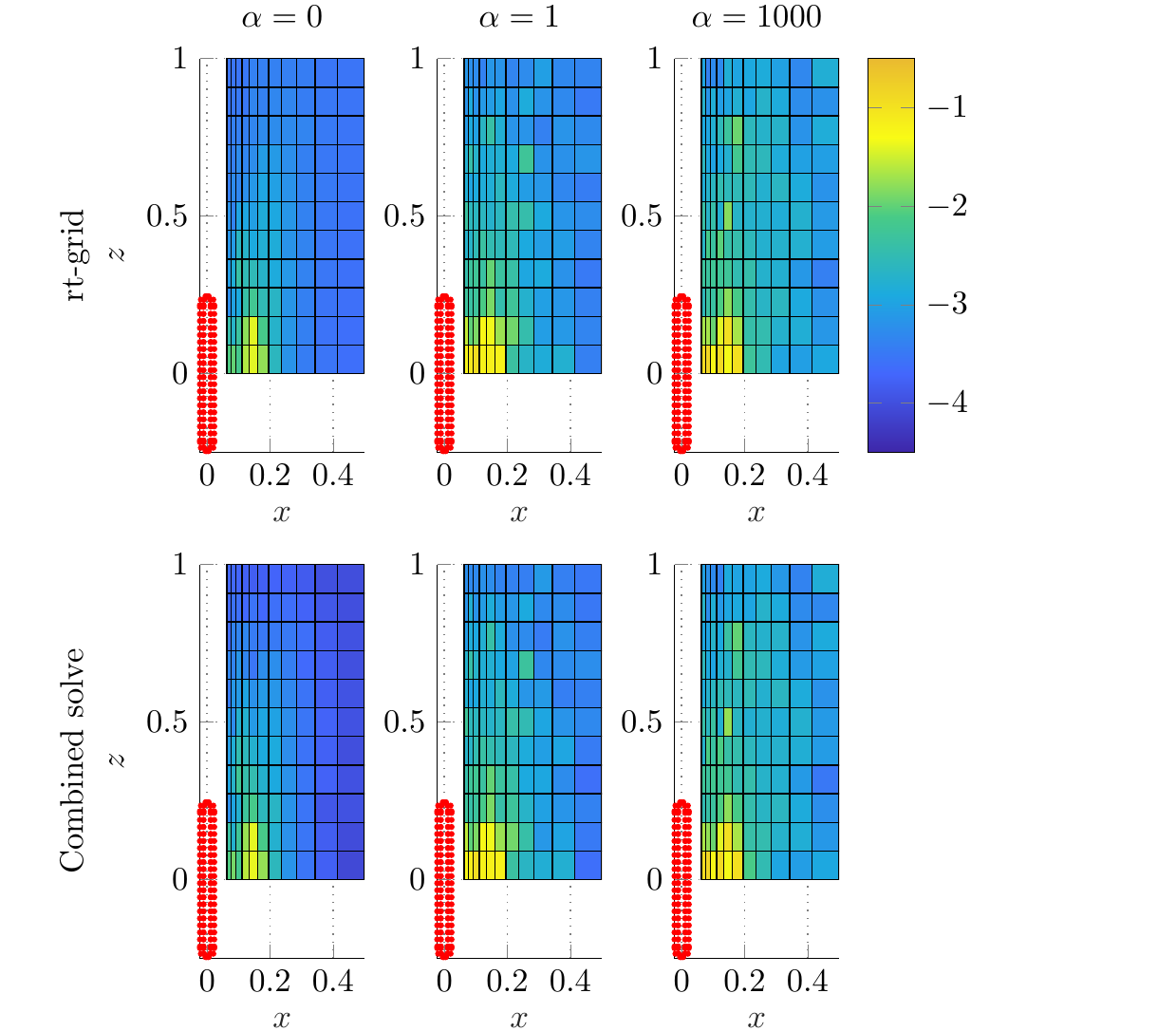}
			\caption{Rods discretised with $n_{\text{cap}}=3$, $n_{\text{cyl}}=20$, $n_{\varphi}=6$ and aligned layers of blobs in the axial direction (displayed in red).}
			\label{sweep_ar20_2}
			%\caption{Results for a chain of rods of aspect ratio 20.}
			%	\label{rods20}
		\end{subfigure}
		\caption{Sweep test for two rods of aspect ratio $L/R = 20$. The relative error in the velocity at the tip of the two particles is measured and the maximum over all sweeps and 200 different force/torque sets is displayed in each Cartesian coordinate ($\log_{10}$). Three different settings are considered, with $\|\vec t^i\| = \alpha\|\vec f^i\|$, where $\alpha \in\lbrace 0,1,1000\rbrace$. In the second and third column of plots, the angular velocity is more dominant, due to the choice of $\alpha$. For the coarsest choice of $\lbrace n_{\text{cap}},n_{\text{cyl}},n_{\varphi}\rbrace$ in (a), a clear improvement is noted if the combined solve is used instead of the rt-grid, as the self-interaction error is reduced to a level smaller than the pair-interaction error. With the finer grid in (b), the self-interaction error is already smaller than the pair-interaction error. Hence, we cannot further reduce this error by doing a combined solve.}	
		\label{sweep_ar20_all}
\end{figure}

The mobility matrix for the pair of rods is determined for each position and orientation. For each such configuration, 200 sets of randomly sampled forces and torques with $\|\vec t^i\| = \alpha\|\vec f^i\|$ are assigned to the pair and the resulting velocity is computed at the tip of each rod.  The maximum error over the orientations in the tip velocity, as compared to the corresponding BIE-solution for the same particle configuration, is taken as a measure of the accuracy in that node of the Cartesian grid. Results are depicted in Figure \ref{sweep_ar20_all} for rods of aspect ratio $L/R = 20$ discretised with two different grids, $\lbrace n_{\text{cap}}=4, n_{\text{cyl}}=15,n_{\varphi}=4\rbrace$ and $\lbrace n_{\text{cap}}=3, n_{\text{cyl}}=20,n_{\varphi}=6\rbrace$. The test is conducted both with the rt-grid and the combined solve.  %and in Figures \ref{sweep_20_tip_1}-\ref{sweep_20_tip_3} for rods of aspect ratio $L/R = 20$. 

There is an interplay between the self-interaction error and the error due to the interaction between rods. Similarly as for the two parallel aligned rods, the error flattens out to the error level of the self-interaction for particles not too close to contact when using the coarse rt-grid, $\lbrace n_{\text{cap}}=4, n_{\text{cyl}}=15,n_{\varphi}=4\rbrace$, see Figure \ref{sweep_ar20_1}. With the combined solve, which has a smaller self-interaction error, the error level decreases with separation and it is instead the pair-interaction error that is dominant. For the finer grid in Figure \ref{sweep_ar20_2}, the pair-interaction error is already dominant and nothing is gained by doing a combined solve for the studied range of relative positions. A combined solve is then only inferring an increased computational work as two mobility problems are solved instead of one to determine each velocity vector, given a vector of forces and torques.
\clearpage

\subsubsection{Twisted rod chain - three rods}\label{sec:twist_three}
This numerical example illustrates the relation between the mobility coefficient errors (the self-interaction error) with the rt-grid or the combined solve and the multi-particle error at large distances, similarly to the test for parallel rods and the sweep test. Here, we investigate the effect of also adding pair-corrections, which can be done with two different approaches: directly to the rt-grid or to the combined solve. In this test, these techniques are compared and we show numerically that for a pair-corrction  to improve on the errors due to particle interactions, the error in the self-interaction error has to be sufficiently small. If small self-interaction errors cannot be obtained with the rt-grid, a combined solve might be needed. 

A chain of rods is considered, where for each chain, a unit direction vector $\vec d\in \mathbb R^3$ and rotation $(\theta,\phi)$ are drawn at random from the first orthant. The chain is constructed with the first rod placed in the origin with orientation coinciding with the $z$-axis. The consecutive rods are obtained from the previous by translating the center coordinate by $\beta\vec d$ in the coordinate frame of the previous particle, with the constant $\beta(\delta_{\min})$ determining the magnitude of the translation such that the smallest distance between a pair of particles is $\delta_{\min}$, and then rotating by $(\theta,\phi)$. See Figure  \ref{chain2} in Section \ref{sec:twist_eight} in the appendix for chains of rods of aspect ratio $L/R=20$. 

\begin{figure}[h!]
	\centering
	\hspace*{8ex}
	\includegraphics[trim = {0cm 5.8cm 0cm 3.4cm},clip,width=1.1\textwidth]{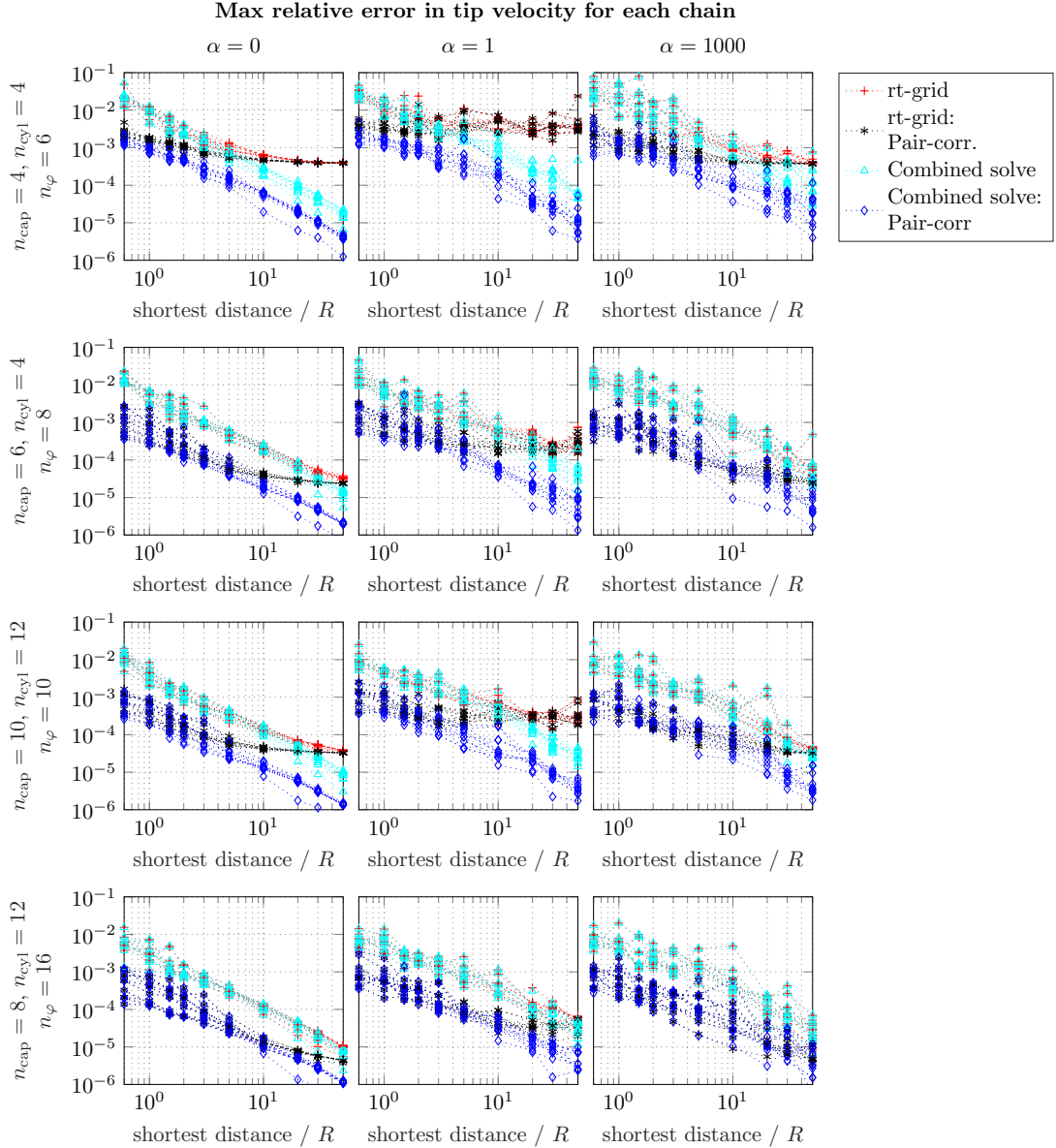}
	\caption{Twisted chains of three rods of aspect ratio $L/R=4$. The maximum relative tip velocity error is displayed for each of the ten chains, given 50 force/torque sets for each $\alpha$. Note that pair-corrections are only applied to neighbouring particles in the chain. The error level is correlated with the error in the self-interaction for the rt-grid for all three choices of $\alpha$. For overlapping red and cyan curves, the pair-interaction error is dominant - no differences can be seen between the solution with the rt-grid and the combined solve as both approaches imply a self-interaction error below the error level in the interaction between particles.}
	\label{twist_ar4_all}
\end{figure}

We first do the test for a small system of three rods and generalise the test to eight rods in Section \ref{sec:twist_eight}. For the three rods, ten different directions $\vec d$ and rotations $(\theta,\phi)$ are used to generate different configurations. For each chain and inter-particle distance, 50 different sets of randomly sampled forces and torques are applied to the particles (to make the test as general as possible), with the directions drawn from the unit sphere.  The error is quantified as the maximum relative error in the tip velocity of any of the three particles for each chain. If the self-interaction error in the rt-grid is large, a lot can be gained by applying the combined solve instead, to decrease the self-interaction error to a level below the error due to the interaction with other particles. Even more is gained by applying a pair-correction to the combined solve so that this interaction error is reduced. If the self-interaction error in the rt-grid is small however, not much is gained by using the combined solve and it suffices to do a pair-correction using the rt-grid.  We see that pair-corrections to the rt-grid is beneficial for small to moderate distances, for which the error due to particle interactions is larger than the self-interaction error. The exact distance for when this happens depends on the choice of discretisation. For larger inter-particle distances, the self-interaction error starts to become dominant, which means that the error in the pair-corrected solution plateaus at the same level as without corrections. 

\begin{figure}[h!]
	\centering
	\hspace*{8ex}
	\includegraphics[trim = {0cm 4.7cm 0cm 3.4cm},clip,width=1.1\textwidth]{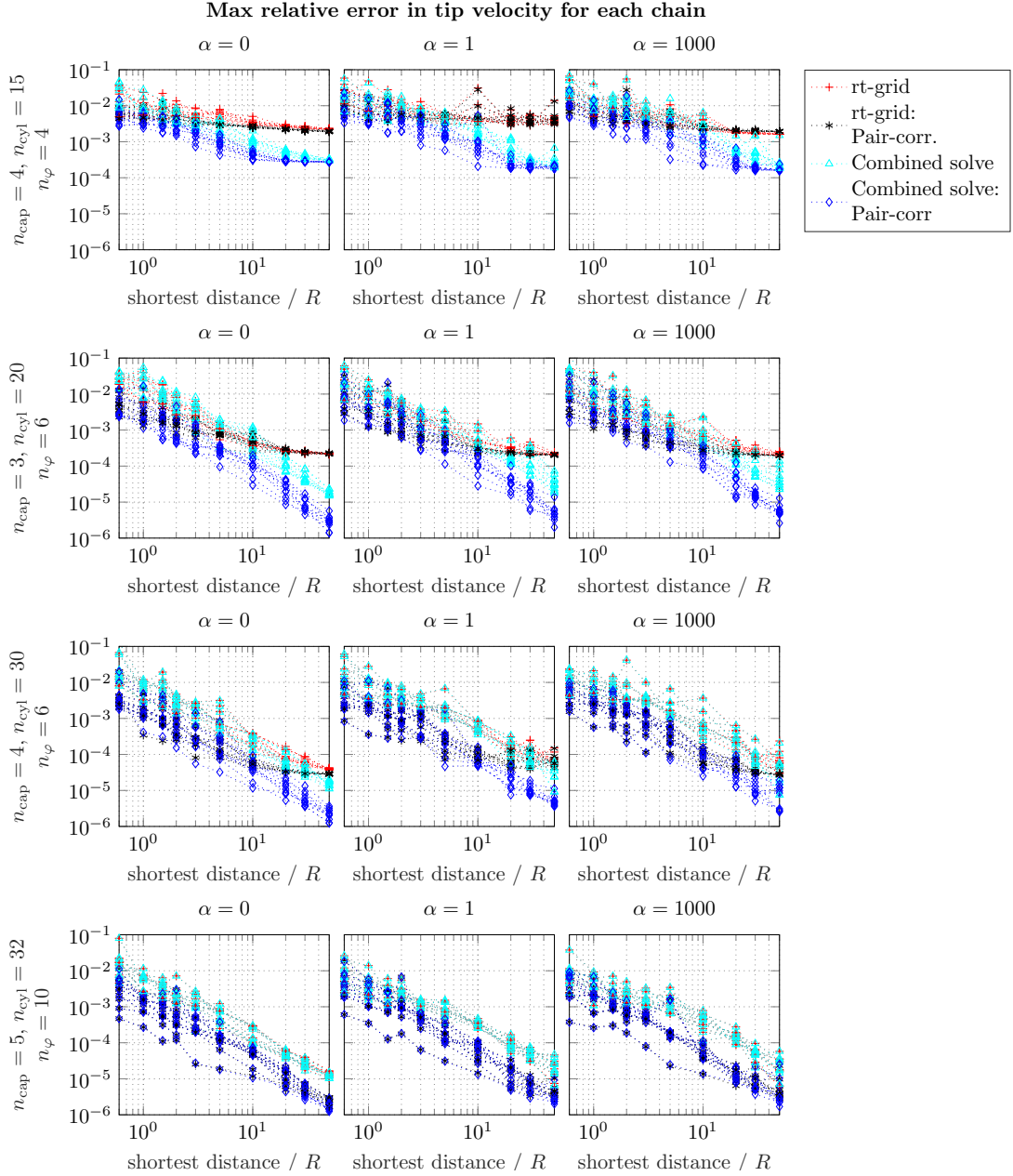}
	\caption{The same test as in Figure \ref{twist_ar4_all} (twisted chains of three rods), but for rods of aspect ratio $L/R=20$. Errors plateau at approximately the same constant level independently of $\alpha$, corresponding to the self-interaction error. For the coarsest grid in the top row of panels, the large cross-errors will affect the accuracy of the solution for the combined solve. Note that these might be due to the very small $n_{\varphi}$ used for this grid.}
	\label{twist_ar20_all}
\end{figure}

The same conclusion can in general be drawn for all relations $\alpha$ between the magnitudes of the forces and torques, with $\|\vec t^i\| = \alpha\|\vec f^i\|$, but some differences can be noted. In Figure \ref{twist_ar4_all}, solutions for $L/R=4$ with the larger error coarse grid $\lbrace n_{\text{cap}}=4,n_{\text{cyl}}=4,n_{\varphi}=6\rbrace $ and the coarse grid $\lbrace n_{\text{cap}}=6,n_{\text{cyl}}=4,n_{\varphi}=8\rbrace $ that has one order of magnitude smaller self-interaction error are compared with the finer grids $\lbrace n_{\text{cap}}=10,n_{\text{cyl}}=12,n_{\varphi}=10\rbrace $ and $\lbrace n_{\text{cap}}=8,n_{\text{cyl}}=12,n_{\varphi}=16\rbrace $, which has the best accuracy (two orders of magnitude larger self-interaction error than the coarsest grid). Note that the direction of the torque relative to the orientation of the particles matters a lot here. For certain chains and torque directions, the error from the translation and rotation add upp for $\alpha = 1$.  For the rods of aspect ratio $L/R = 20$, with results displayed in Figure \ref{twist_ar20_all}, errors plateau at approximately the same constant level independently of $\alpha$, corresponding to the self-interaction error. For the coarsest grid, the large cross-errors will affect the accuracy of the solution for the combined solve as a result of the very small $n_{\varphi}$ used for this grid. The slight difference in the error curves for $\alpha = 1$ between rod types could be explained by the relative magnitudes of the resistance coefficients, $\xi_t^{\parallel},\xi_t^{\perp},\xi_r^{\parallel},\xi_r^{\perp}$, where for instance $\xi_r^{\parallel}$ is very large for the slender rod, see Table \ref{coeffs} in Appendix \ref{sec:BIE}.

Similar conclusions can be drawn for longer chains of rods, see the discussion in Section \ref{sec:twist_eight} in the appendix and compare the results of Figures \ref{twist_ar20_all} and \ref{chain_8p}.

\begin{figure}[h!]
	\centering
	%\hspace*{8ex}
	\includegraphics[trim = {0.5cm 15.5cm 0.2cm 3.4cm},clip,width=0.95\textwidth]{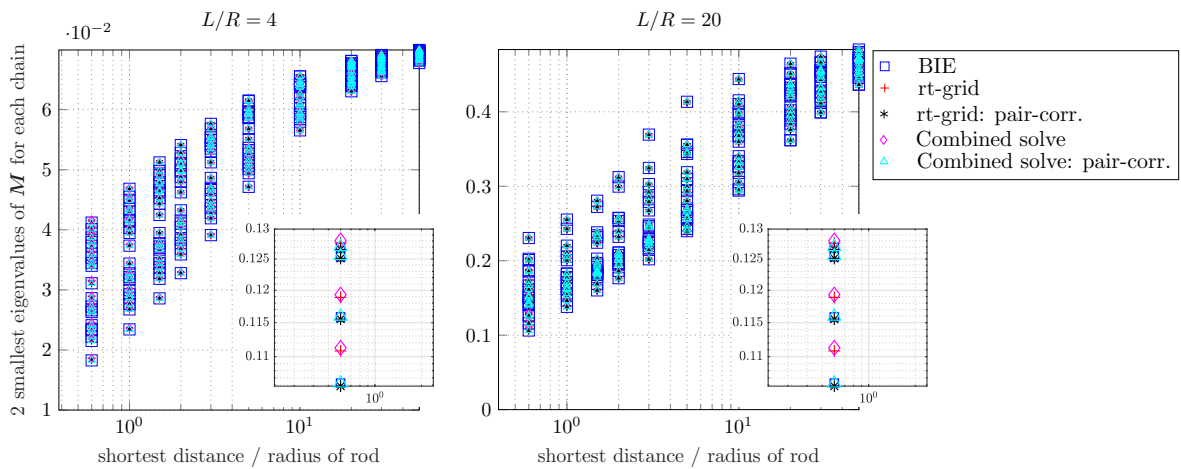}
	\caption{For each distance, the two smallest eigenvalues are displayed for each of the ten chains, so that a total of 20 eigenvalues are shown for each distance and solution strategy. The very smallest eigenvalues are displayed in the insets. The coarsest blob discretisation is used for each aspect ratio. The smallest eigenvalues from a pair-corrected mobility matrix coincide with the smallest eigenvalues of the reference BIE-matrix, in eye-ball norm. No eigenvalues from any of the multiblob strategies of solving the mobility problem is smaller than the corresponding eigenvalue with the BIE-solution and all of the mobility matrices are positive definite, as required.}
	\label{twist_eigs}
\end{figure}

The smallest eigenvalues of the mobility matrix for the different solution strategies are displayed for the coarsest blob grids in Figure \ref{twist_eigs} for the two rod types. It can be concluded that the smallest eigenvalues in $\vec M_{\text{BIE}}$ are correctly captured by using a pair-correction (both the one applied to the rt-grid and the correction applied to the combined solve), which indicates that lubrication effects are correctly represented. Note a small difference between the smallest eigenvalues for the pair-corrected and non-corrected solutions. The mobility matrix is positive definite for all separation distances and every tested solution strategy, which is an important property of the mobility matrix. 

\clearpage

\subsubsection{Random configurations of rods}\label{random_rods}
In the previous tests with chains of twisted rods in Section \ref{sec:twist_three} (and \ref{sec:twist_eight}), there is a lot of symmetry and pair-corrections are applied only to neighbouring particles. Now, we consider a more general setting, with random configurations of rods, with the smallest allowed distance between particle surfaces, denoted by $\delta_{\min}$, set to a given multiple of the rod radius. Each particle is repeatedly positioned and oriented in a cube of side length $l$ at random until its smallest distance to any other particle is larger than $\delta_{\min}$. The $\delta_{\min}$ and $l$ used for the two rod types are reported in Table \ref{boxsize}. Two example geometries are visualised in Figure \ref{example_geom} for the smallest $\delta_{\min}$ for each rod type.
\begin{table}[h!]
	\centering	
		\begin{tabular}{c |  c c c c  c c}
			\multicolumn{7}{c}{$L/R = 4$} \\ \hline \hline
			$\delta_{\min}$ & $0.5R$ & $3R$ & $6R$ & $10R$ \\  $l$ & 4.5 & 7.0 & 10.0 &  13.0 & & \\ \hline 
				\multicolumn{7}{c}{$L/R = 20$} \\ \hline \hline
				$\delta_{\min}$ & $R$ & $6R$ & $10R$ &  $20R$ & $50R$ & $100R$ \\
				$l$ & 0.375  & 0.75	& 1.0 & 1.6 & 3.0 & 5.5  \\
		\end{tabular}
	\caption{Choices of the parameters $l$ and $\delta_{\min}$ used in the example in section \ref{random_rods}: random configurations of seven rods.}
	\label{boxsize}
\end{table}
Note that, in this test, we choose to apply the pair-correction to all pairs of particles and not only to particles within some set cut-off. 

Each particle is assigned a net force and torque with each component independently drawn from a uniform distribution: for the slender rods, $\vec f^i\sim \mathcal U(-12\pi,12\pi)$ and $\vec t^i\sim\mathcal U(-1/8,1/8)$, while for the fat rods, $\vec f^i\sim \mathcal U(-1,1)$ and $\vec t^i\sim\mathcal U(-1,1)$. The difference is motivated by the fact that a large torque on slender rods has a large impact on the rotational velocities (see the mobility coefficients presented in Table \ref{coeffs} in Section \ref{sec:BIE} in the appendix). For each $\delta_{\min}$, ten random configurations are created and we study the relative errors in the tip velocities. 

\begin{figure}[h!]
	\centering
	\hspace{-0.5cm}
	\includegraphics[width=0.95\textwidth,trim = {3cm 20.9cm 2cm 2.5cm},clip]{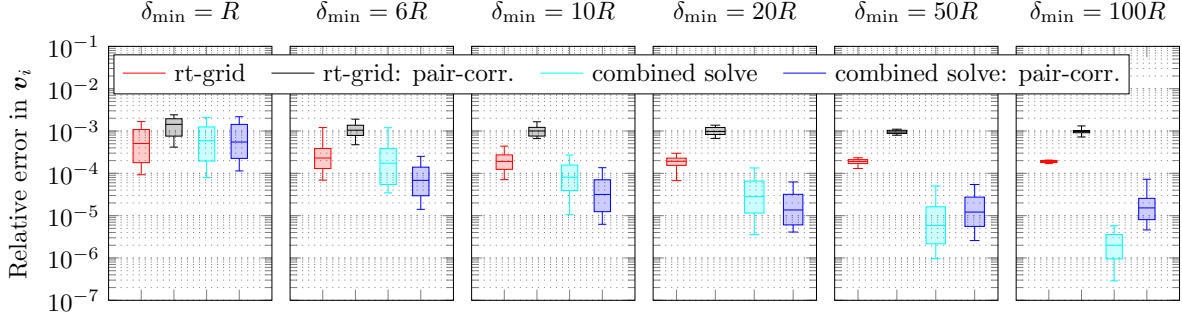}
	\caption{Random configurations of seven rods of aspect ratio $L/R=20$ positioned in a cube of length $l$ (see Table \ref{boxsize}). The rods are discretised with the coarser grid $n_{\text{cap}}=3$, $n_{\text{cyl}}=20$, $n_{\varphi}=6.$ Box plots show the relative error in the particle tip velocities $\vec v^i$ for all rods in ten different configurations, with whiskers displaying the minimum and maximum error, box edges displaying the 10th and 90th percentile of the error and the box center line displaying the median. The combined solve with or without pair-corrections is to be preferred over solutions with the rt-grid, as the self-interaction error with the rt-grid is large.}
	\label{random_ar20_2}
\end{figure}

\begin{figure}[h!]
	\centering
	\hspace{-0.5cm}
	\includegraphics[width=0.95\textwidth,trim = {3cm 20.9cm 2cm 2.5cm},clip]{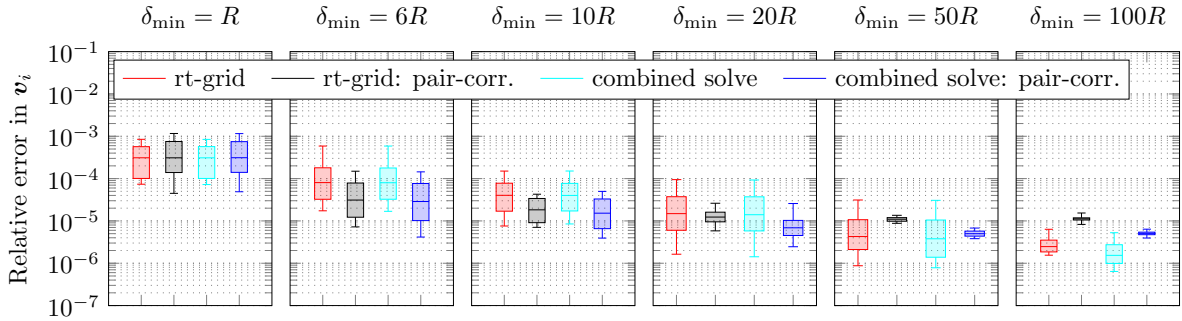}
	\caption{As in Figure \ref{random_ar20_2}, but for the finer discretisation $n_{\text{cap}}=5$, $n_{\text{cyl}}=32$, $n_{\varphi}=10.$ The self-interaction error with the rt-grid is almost as small as for the combined solve. For large $\delta_{\min}$, pair-corrections plateau at a level correlated with the self-interaction error in the underlying solution strategy, but at a level larger than the self-interaction error observed for the non-corrected rt-grid and combined solve, due to error pollution (see discussion in main text).}
	\label{random_ar20_4}
\end{figure}

Consider first configurations of slender rods, where the mobility problems are solved with and without pair-corrections for the rt-grid and the combined solve. Results are displayed in Figures \ref{random_ar20_2}-\ref{random_ar20_4} for a coarse and a fine discretisation. The errors decrease with $\delta_{\min}$ only up until the distance for which the self-interaction error dominates, which happens first for the coarse discretisations (this is the case already for $\delta_{\min} = R$ in Figure \ref{random_ar20_2}). Remember that $\delta_{\min}$ is only the shortest distance between particles and some pairs of particles will be considerably further apart. Hence, the self-interaction error is visible also for small $\delta_{\min}$. Note that the error level for the pair-corrected calculation plateaus at a constant level correlated with the self-interaction error in the underlying solution strategy (the rt-grid or the combined solve) and that self-interaction error is limiting the error level from below for all $\delta_{\min}$.  A small self-interaction error is essential for a pair-correction to improve on the error. Note that a pair-correction might result in an error larger than for the underlying solution strategy \emph{without} pair-corrections. This can be seen for the pair-correction applied to the rt-grid in Figure \ref{random_ar20_2} and for the pair-correction applied to the combined solve in the right-most panels of both Figure \ref{random_ar20_2} and \ref{random_ar20_4}, for which the self-interaction error is at the set level $\epsilon$ used in the optimisation problems to find r- and t-grids in \eqref{eq8}-\eqref{eq9}. An explanation is that the error in the self-interaction of the underlying solution strategy pollutes other blocks of the mobility matrix upon solving the linear system used to define the pair-correction. A pair-correction can hence never reduce a self-interaction error. Pair-corrections applied to the rt-grid and to the combined solve are comparable in accuracy for the minimum distances $\delta_{\min}$ for which the error due to interactions between particles is dominant. To conclude, we do not wish to apply pair-corrections for well-separated particles if the self-interaction error is large.


\begin{figure}[h!]
	\centering
%	\hspace*{-1cm}
	\hspace{0.5cm}
	\includegraphics[width=0.88\textwidth,trim = {3cm 20.5cm 3cm 2.5cm},clip]{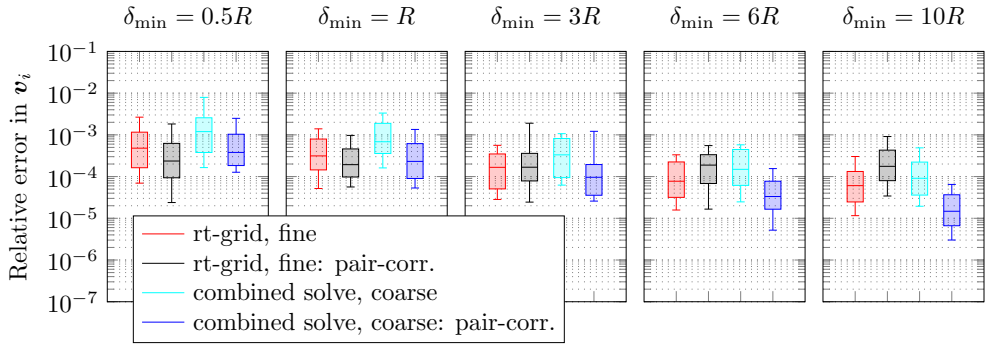}
	\caption{As in Figure \ref{random_ar20_2}, but for fat rods of aspect ratio $L/R=4$. We compare the combined solve for a coarse grid with $n_{\text{cap}}=4$, $n_{\text{cyl}}=4$, $n_{\varphi}=6$ and a solution with the rt-grid for the fine discretisation $n_{\text{cap}}=10$, $n_{\text{cyl}}=12$, $n_{\varphi}=10$. The accuracy is approximately of the same order for the non-corrected rt-grid and combined solve. With a pair-correction applied to the combined solve, smaller errors can be obtained than by pair-correcting the rt-grid with the fine resolution.}
	\label{random_ar4_2}
\end{figure}

Next, consider configurations of fat rods of aspect ratio $L/R=4$. We use a combined solve with and without pair-corrections for a coarse grid. Results are compared with a solution computed with the rt-grid with and without corrections using a fine grid. The fine grid has more than double the number of blobs as the coarse grid. From Figure \ref{random_ar4_2}, we can conclude that by solving a mobility problem twice with a coarse grid, we can obtain the same accuracy as by using a fine grid. The pair-corrected combined solve with a coarse blob resolution outperforms the pair-corrected rt-grid with a fine resolution. The key is the reduction of the self-interaction error obtained with the combined solve. An interpretation of the result is that even if the mobility problem is solved with linear complexity, we gain from using a coarse grid and a combined solve.

\begin{figure}[h!]
	\centering
	%\begin{subfigure}[b]{0.85\textwidth}
	\includegraphics[width=1\textwidth,trim = {3cm 16.7cm 2.5cm 2.5cm},clip]{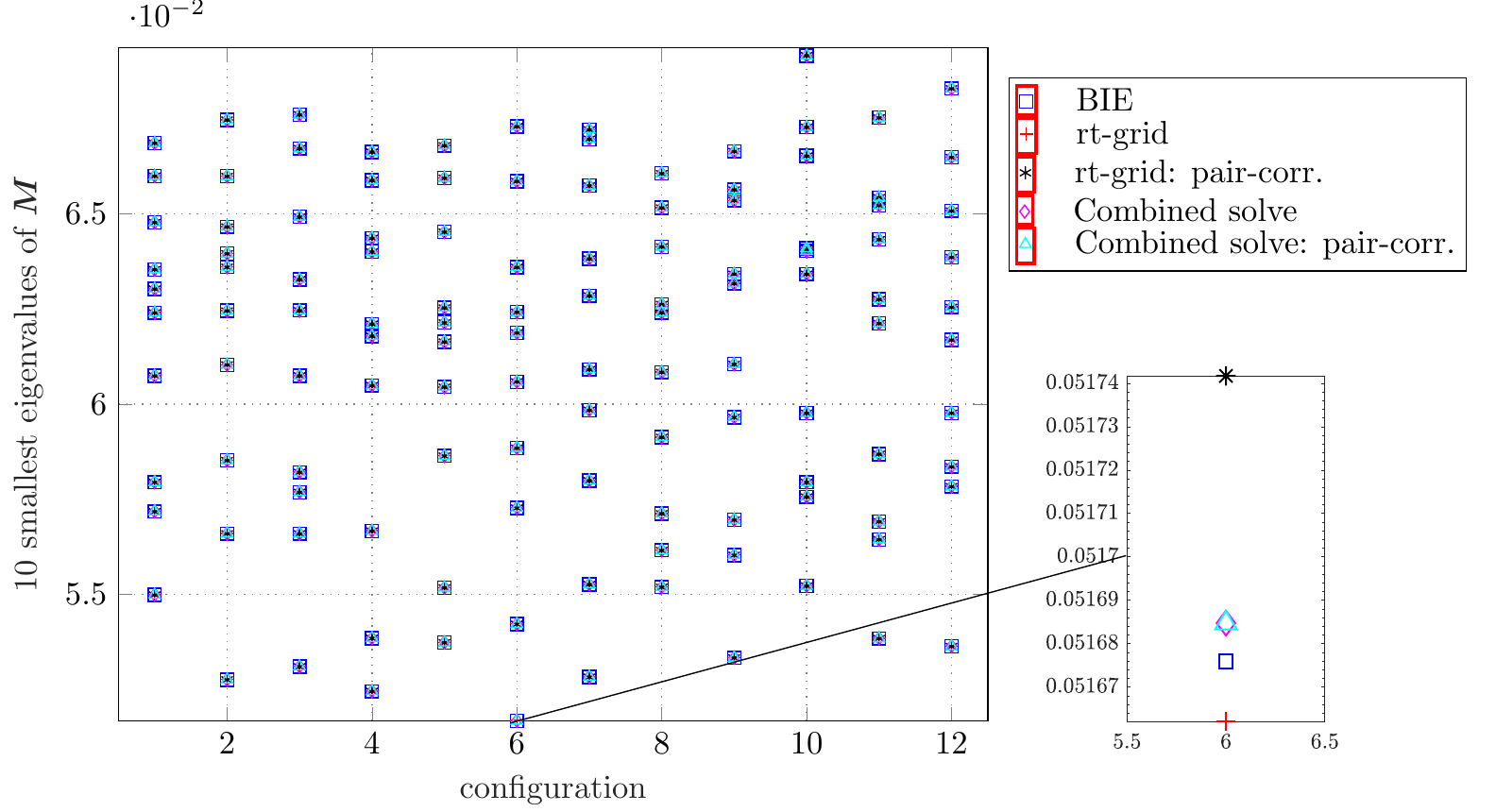}
	%\end{subfigure}
%\begin{subfigure}[b]{0.14\textwidth}
%	
%		\includegraphics[width=1.1\textwidth,trim = {0cm 0cm 0cm 0cm},clip]{figures/random_rods_eigs_zoom.pdf}
%		\end{subfigure}	
	\caption{The ten smallest eigenvalues of the mobility matrix for each of the configurations of rods of aspect ratio $L/R=4$ in Figure \ref{random_ar4_2} corresponding to $\delta_{\min}=6R$ and the coarse grid $n_{\text{cap}}=4$, $n_{\text{cyl}}=4$, $n_{\varphi}=6$. For all solution strategies with the multiblob method, the eigenvalues are very close to the eigenvalues of the reference BIE-matrix (see inset to the right for the very smallest eigenvalue in the larger left plot). This means that the corrected matrix is positive definite in all cases.}
	\label{eigs_ar4}
\end{figure}

Finally, we consider the eigenvalues of the corrected mobility matrix in one specific case: rods of aspect ratio $L/R=4$ and all ten particle configurations with $\delta_{\min}=6$, to visualise that the corrected matrix is positive definite. The smallest eigenvalues of the mobility matrices corresponding to all four multiblob solution strategies are displayed in Figure \ref{eigs_ar4}. Upon comparison with the eigenvalues of the reference BIE-matrix, it can be concluded that the ten smallest eigenvalues agree if a pair-correction is applied and that the corrected matrices are positive definite.

%\begin{figure}[h!]
%	\centering
%	\hspace*{-8ex}
%	\includegraphics[width=1.03\textwidth,trim = {2cm 17.2cm 1.5cm 2.5cm},clip]{figures/random_rods_ar4.pdf}
%	\caption{The example in Section \ref{Ex:random_rods}: Random configurations of seven rods of aspect ratio $L/R=4$ placed in a box such that the minimum inter-particle distance is $\delta_{\min}$. Relative error in the particle translational velocities, $\vec u^i$, and rotational velocities, $\vec{\omega}^i$ are displayed for all rods in 12 different configurations for each $\delta_{\min}$. The worst error in the original multiblob method is improved for all configurations, with a larger improvement for large $\delta_{\min}$. }
%	\label{random_ar4}
%\end{figure}
%\clearpage
%%%%%%%%%%%%%%%%%%%%%%%%%%%%%%%%%%%%%%%%%%%%%%%%%%%%%%%%%%%%%%%%%%%%%%
\section{Recommended solution strategies}
In this work, techniques for solving the mobility problem have been demonstrated with numerical examples presenting systems of rods or spherical particles. We would like to complete these numerical examples with a suggested solution procedure for a given general axisymmetric particle type where we assume that accurate mobility coefficients are known. We also assume that the multiblob geometry is determined by a set of geometric parameters that we can optimise for together with the blob radius ratio $a_h/s$. Proceed as follows: 
\begin{enumerate}
	\item Solve the optimisation problem in \eqref{opt} to minimise the maximum relative error in the four mobility coefficients $\lbrace 1/\xi_t^{\parallel},1/\xi_t^{\perp},1/\xi_r^{\parallel},1/\xi_r^{\perp}\rbrace$ to find the rt-grid.  Choose the number of blobs along different directions of the particle such that this error is small. This error level will bound the error from below in a multi-particle suspension of the given particle type.  The total number of blobs, $n_b$, should be chosen so that the resulting optimised blob radius does not lead to overlapping blobs for inter-particle distances where the model is typically used. For very closely interacting particles, accuracy can be gained by picking a slightly larger number of blobs, as this would reduce the pair-interaction error. This however comes at an increased cost in solving mobility problems.
	\item For some combinations of blob geometry, total number of blobs and discretisation of the particle, it might be impossible to choose a grid with small errors in both translational and rotational mobility coefficients. As it is crucial to push down the self-interaction error, we then propose a different approach. If errors are not sufficiently small using the rt-grid, we can choose to optimise for two different grids, the r- and t-grids, matching for the rotational and translational mobility separately as in \eqref{eq8}-\eqref{eq9}, and combine solutions from two separate mobility problems. As two solves then are required to get a solution to a single mobility problem, we need to reduce $n_b$ for this approach. 
	\end{enumerate}
There is hence an important choice to make in how to reduce the self-interaction error -- one could try to obtain this by increasing $n_b$ and compute an optimal rt-grid or by doing a combined solve for a smaller $n_b$. 
\begin{enumerate}
\setcounter{enumi}{2}
\item Given a small self-interaction error using either an rt-grid or a combined solve, pair-corrections can be applied to reduce errors for interacting particles, assuming that we have an accurate method at hand that can be applied quickly for a small system of two particles only. Apply pair-corrections only for pairs of particles sufficiently close to each other by introducing a cut-off, below which the error due to interactions between particles is larger than the self-interaction error.
\end{enumerate}

%%%%%%%%%%%%%%%%%%%%%%%%%%%%%%%%%%%%%%%%%%%%%%%%%%%%%%%%%%5
\section{Conclusions}
The accuracy of the multiblob method for the Stokes mobility problem has been studied in the deterministic setting for spherical and rod-like particles. Comparisons have been made to a reference solution accurately computed with a numerical method based on a boundary integral equation of the second kind equipped with QBX-quadrature. A particle discretised with a number of spherical blobs determining its surface is seen as a model for some ideal particle of a certain shape. Blobs are placed to match the single particle mobility matrix, referred to as the self-interaction, of the ideal particle, motivated by the fact that in a very dilute suspension, the hydrodynamic self-interaction for a particle is dominating the effects due to the presence of other particles. This dominance is more pronounced the larger the particle separations. We employ an optimisation procedure to match this self-interaction.  The resulting optimised grid geometry (parameters needed to determine the geometric surface where blobs are to be placed, together with the blob radius) is referred to as the rt-grid. The error level obtained upon optimising for the rt-grid will set the velocity error level for any multi-particle suspension: For sufficiently large particle-particle distances, the error plateaus at the self-interaction error level from the rt-grid. The self-interaction error sets a baseline also for denser suspensions, even though we cannot isolate the error from the interaction error with other particles in this case. We would like to stress that by solving an optimisation problem, we can obtain the smallest possible self-interaction error, given a number of blobs and a means of discretising the particle. If such an optimisation strategy is not employed and a reasonable blob-grid geometry is simply hand-picked, the self-interaction error might become several orders larger in magnitude. We have demonstrated this with numerical examples both for the sphere and the rod geometry. The  level of the self-interaction error is very sensitive to the precise choice of the grid geometry and even a slight change to the optimal parameters can lead to a drastic error increase. 

For spheres, the self-interaction error level for the rt-grid is in the order of $10^{-8}$ and negligible independently of the blob grid resolution compared to errors stemming from interactions between particles. We showed that in contrast to choosing the blob grid as in \cite{USABIAGA2016}, where only the translational hydrodynamic radius is matched, small errors can be obtained in both translational and rotational velocities with our optimisation technique. Due to the smaller optimised hydrodynamic radii $a_h$ of the blobs compared to the specific choice made in \cite{USABIAGA2016}, lubrication effects are not as accurately captured as the fluid ``leaks'' in between blobs, distorting the repelling forces between particles at very small distances. This effect is however small and notable only for distances where blobs start to overlap. Hence, we cannot expect a sphere model to be accurate in this regime of particle-particle distances anyway.

For rods, the self-interaction error in the rt-grid is the dominant source of error for coarse discretisations. A different choice is to match for rotation and translation mobility components separately, such that we obtain one optimal blob-grid for translation and one for rotation. We can then solve the mobility problem twice and combine the solutions, extracting translational velocities from the translationally matched grid and rotational velocities from the rotationally matched. This strategy is referred to as a combined solve. The negative aspect of applying the combined solve is that we have to solve each mobility problem twice, hence doubling the cost. On the other hand, the self-interaction error can be reduced significantly, and is for large particle gaps mainly characterised by a parameter $\epsilon$ used in the optimisation procedure to adjust the interplay between different contributions to the error in the single particle mobility.

We also showed numerically that a pair-correction inspired by Stokesian dynamics can be applied for  clusters of particles to improve the accuracy further. Pair-corrections applied to the rt-grid improve on the error for all gaps small enough that the self-interaction error is not the dominating contribution to the total error. For larger separations, the self-interaction error of the rt-grid limits the velocity errors from below on a certain level. Pair-corrections can hence only be successfully applied if pair-interaction errors are larger than the self-interaction error level. For coarse discretisations, a pair-correction applied to the combined solve is therefore beneficial, as the self-interaction error of the combined solve is lower. Said differently, a well-matched self-interaction is a necessity for pair-corrections to improve on the accuracy.

The mobility problem has been studied for a small number of particles with varying relative distances and orientations. In conclusion, coarsely resolved particles of the studied type can successfully be used to approximate the solution to mobility problems, with controllable accuracy compared to a set of reference particles, where the mobility problem is solved with a much more accurate but also computationally expensive method.  In future work, the multiblob method allows for the study of heterogeneous suspensions containing particles of different shapes and/or aspect ratios, each with an optimised grid geometry. Another future research direction is to extend the blob grid optimisation strategy to systems with boundaries and/or periodicities.

In contrast to the case for Stokesian dynamics, the  additivity assumption inherent in the pair-corrected method does not violate the accuracy. From the construction of the pair-corrections, we however have no technique to guarantee the positive-definiteness of the correction matrix or the resulting mobility matrix. In all numerical tests performed with the pair-correction, the resulting (corrected) mobility matrix is positive definite, which has been checked by explicitly computing its eigenvalues. For the pair-correction to be competitive in large scale simulations,  a multivariate interpolant over relative orientations and particle distances is required for the fast evaluation of the pair mobility matrix for non-spherical particles. Such a pre-computed function is to be developed in future work. If Brownian motion is accounted for in the system, it is desired that the correction built from this interpolant is positive definite -- a property that also has to be considered in the sphere case.

% An alternative to incorporating lubrication forces via a pair-wise correction as in Stokesian dynamics is the lubrication correction proposed by Lefebvre-Lepot et al.~\cite{Lefebvre-Lepot2015}. This correction however requires an evaluation of the lubrication field for all particles not in the pair, requiring a multi-variate interpolant.  

In this work, a smaller number of particles have been studied, allowing for comparisons to an accurate method where the surface of the particles is well resolved. In future work, we plan to use the same framework but move from the deterministic setting to study the diffusive behaviour in Brownian systems containing a large particle number. Control of the deterministic error  in the multiblob approach, as provided by the techniques presented in this work, is a necessary piece in understanding also other contributions to the total error in a dynamic simulation, such as a time-discretisation error and a statistical error in a Brownian setting. %Specifically of interest is to study effective rotational and translational diffusion in heterogeneous suspensions and in suspensions where particles are affected by realistic interaction potentials stemming from physical experiments or MD-simulations. 

\section*{Acknowledgements}
We acknowledge the support by the research environment grant ``Interface'' from the Swedish Research Council. 

%\section*{Declaration of competing interest}
%There are no competing interests to declare.

\addcontentsline{toc}{section}{References}
\def\url#1{}
\bibliographystyle{myIEEEtran} %including doi
\bibliography{multiblob}

\appendix
\clearpage
\section{BIE-parameters and reference mobility coefficients}\label{sec:BIE}
Parameters used to compute the reference solution with the BIE-method equipped with QBX are presented in Table \ref{qbx_params}, with the discretisation described in detail by Bagge \& Tornberg in the appendix of \cite{Bagge2021}.
\begin{table}[h!]
	\centering	
	\begin{tabular}{c|c|c|c}
		Aspect ratio & $n^{QBX}_{\text{cap}}$ & $n^{QBX}_{\text{cyl}}$ &  $n^{QBX}_{\varphi}$  \\ \hline%& Quadrature tolerance
		$L/R =4$ & 40 & 10 & 25\\ \hline% & $10^{-6}$ 
		$L/R =20$ & 35 & 60 & 18 %& 	$10^{-6}$ \\ 
	\end{tabular}
\caption{Parameters for computing the reference solution for rods using the BIE-solver with QBX}
\label{qbx_params}
\end{table}

For spheres, 60 Gauss-Legendre points are used in the axial direction (from pole to pole) and 60 equidistant points are chosen in the azimuth direction for each spherical particle.

The computed accurate mobility coefficients for rods of aspect ratio $L/R = 4$ and $L/R=20$ are reported in Table \ref{coeffs}. 

\begin{table}[h!]
	\centering	
	\begin{tabular}{l|c|c|c|c}
		& $1/\xi_t^{\parallel}$ & $1/\xi_t^{\perp}$ & $1/\xi_r^{\parallel}$ & $1/\xi_r^{\perp}$ \\ \hline
		$L/R = 4$ &  $0.0713091832 % 0.071309183183800;
		$& $0.0804688924%2402250; 0.080468892402250;
		$& $0.0836722897% 0.083672289717590;
		$& $0.153993117$% 0.153993117291970;  
		\\ \hline
		$L/R = 20$ & $0.504333544%4180318 0.504333544180318 \\
		$ & $0.701528202$ %1526653 % 0.701528201526653 \hline	
		& $14.4347584 %14126913 % 14.434758414126913	
		$ & $251.837496%5157767e+02 % 2.518374955157767e+02;
		$ \\
	\end{tabular}
\caption{Mobility coefficients computed with the BIE-method for a single rod. These will be used as reference when optimising for the blob grid.}
\label{coeffs}
\end{table}
%%%%%%%%%%%%%%%%%%%%%%%%%%%%%%%%%%%%%%%%%%%%%%%%%%%%%%%%%%
\section{Grid parameters}\label{sec:opt_param}
Optimal blob geometries corresponding to the r- and t-grids are presented in Table \ref{rt_geometry} for selected discretisation triplets $\lbrace n_{\text{cap}}, n_{\text{cyl}}, n_{\varphi}\rbrace$ for the slender rod with $L/R = 20$ and the fat rod with $L/R = 4$. 
\begin{table}[h!]
	\centering	
	\begin{tabular}{c| c| c | c| c|c | c|c|c|c}
		
		$n_{\text{cap}}$	& $n_{\text{cyl}}$ & $n_{\varphi}$ & aligned & $L_g^t$  & $R_g^t$  & $L_g^r$  & $R_g^r$ & $a_h^t/s$ & $a_h^r/s$  \\	 \hline \hline
		\multicolumn{10}{c}{$L/R = 20$} \\ \hline 	
		4 & 15 & 4 & No & 0.484& 0.0213 & 0.486 & 0.0214 & 0.359 & 0.359 \\ 
		3 & 20 & 6 & Yes & 0.493   & 0.0229 & 0.492 & 0.0228 & 0.329 & 0.336	\\ 	
		4 & 30 & 6 & No & 0.495 & 0.0232 & 0.495 & 0.0232 & 0.267 & 0.267\\ 		
		5 & 32 & 10 & Yes & 0.496 & 0.0235 & 0.496 & 0.0235 & 0.342 & 0.342 \\	 \hline \hline
		\multicolumn{10}{c}{$L/R = 4$} \\ \hline
		4  & 4 & 6 & No & 1.929 &  0.477 & 1.933 & 0.476 &  0.222 &  0.223 \\ 			
		6 & 4 & 8 & No & 1.953 &  0.479 & 1.952 & 0.479 & 0.221 & 0.221 \\ 		
		10 & 12 & 10 & No& 1.967 & 0.483 & 1.966 & 0.484 & 0.202 & 0.201\\ 	
		8 & 12 & 16 & Yes & 1.966 & 0.483  & 1.966  & 0.483 & 0.305 & 0.305\\		
	\end{tabular}
\caption{Geometry obtained for the r- and t-grids upon solving the optimisation problems in \eqref{eq8} and \eqref{eq9}, corresponding to the error levels reported in Table \ref{cross_ar} in Section \ref{rod-rt}. For finer discretisations, the difference in the parameters $(L_g,R_g,a_h/s)$ is small for the r- and t-grids; they are however not identical even if it appears so in the table.}
\label{rt_geometry}
\end{table}
\clearpage
\noindent For the introductory numerical example in Figure \ref{important_opt}, parameters are presented in Table \ref{circle_params}.
\begin{table}[h!]
	\centering	
	\begin{tabular}{c|c|c|c}
		Grid & $L_g$ & $R_g$ & $a_h/s$ \\ \hline
		Optimised &  1.953 &  0.479  & 0.221\\     
		Non-optimised & 1.972 & 0.484 & 0.223
	\end{tabular}
\caption{Grid parameters for each of the five fat rods of aspect ratio $L/R=4$ forming a circle in the example presented in Figure \ref{important_opt}. Each rod is discretised with $n_{\text{cap}}=6$, $n_{\text{cyl}}=4$ and $n_{\varphi}=8$ and the non-optimised grid is the optimised grid perturbed by $1\%$.}
\label{circle_params}
\end{table}

%%%%%%%%%%%%%%%%%%%%%%%%%%%%%%%%%%%%%%%%%%%%%%%%%%%%%%

\section{Pair-corrections for particle chains}
\subsection{A twisted chain of spheres}\label{Ex:twist_spheres}
\begin{figure}[h!]
	\centering
	\begin{subfigure}[t]{0.2\textwidth}
		\centering
		\hspace*{-9ex}
		\includegraphics[trim = {2.1cm 0.35cm 3.5cm 0.3cm},clip,width=\textwidth]{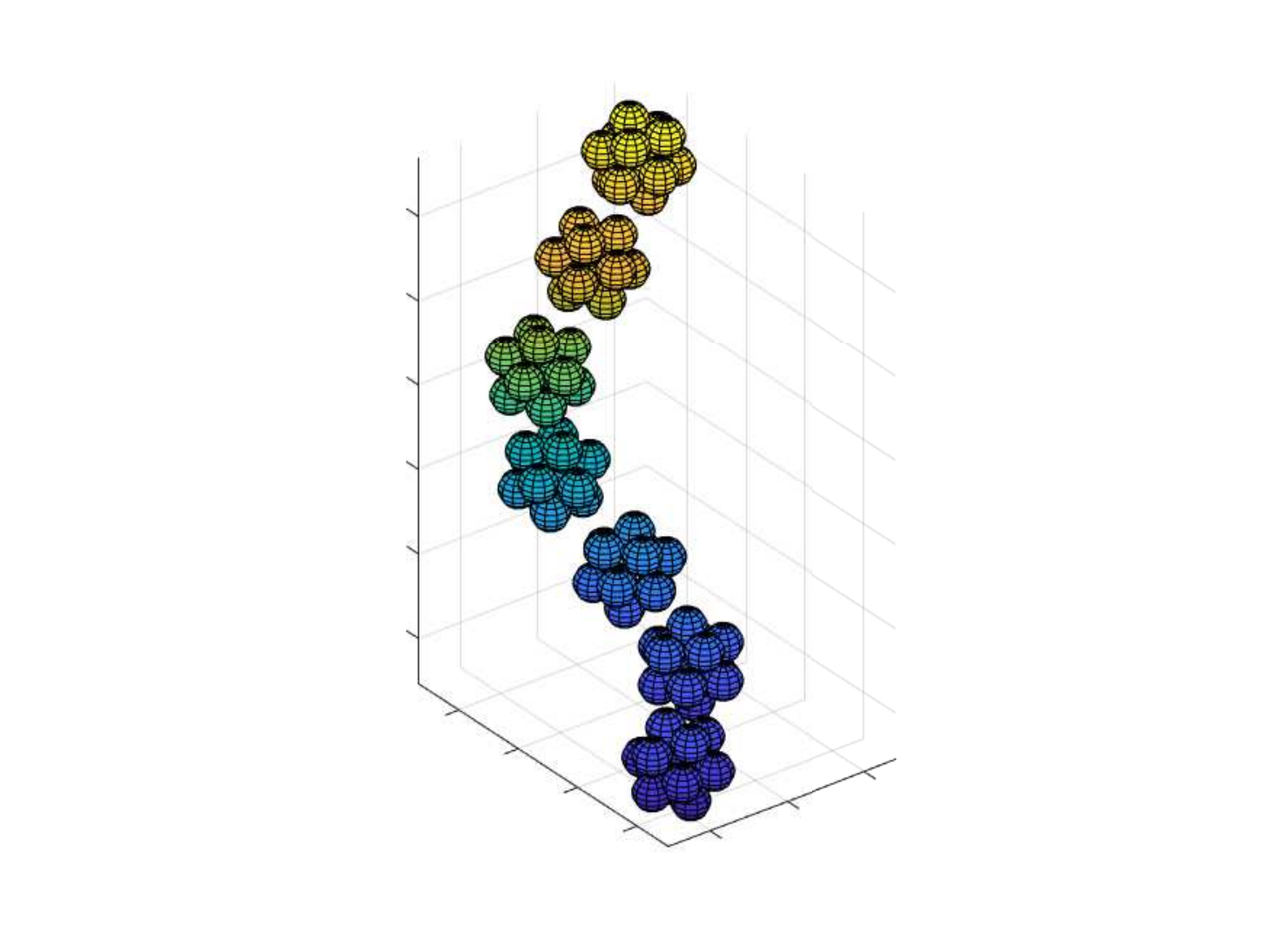}
		\caption{Example geometry: 12 blobs per sphere}
		%\caption{Results for a chain of rods of aspect ratio 4.}
		%	\label{rods20}
	\end{subfigure}~
	\begin{subfigure}[t]{0.2\textwidth}
		\centering
		\hspace*{-3ex}
		\includegraphics[trim = {1.2cm 0.3cm 2.0cm 0.3cm},clip,width=\textwidth]{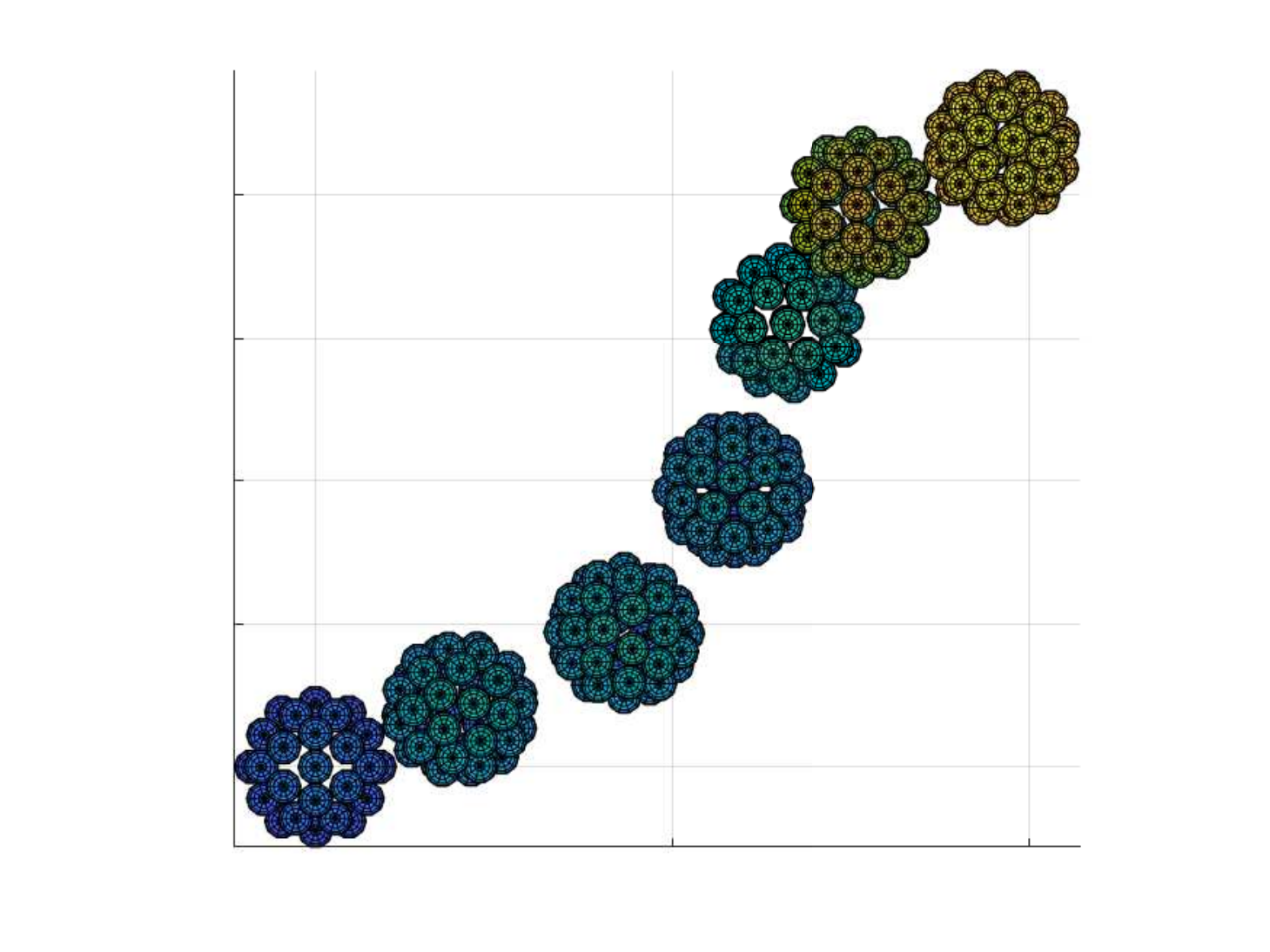}
		\caption{Example geometry: 42 blobs per sphere}
		%\caption{Results for a chain of rods of aspect ratio 20.}
		%	\label{rods20}
	\end{subfigure}~
	\begin{subfigure}[t]{0.23\textwidth}
		\centering
		%\hspace*{-9ex}
		\includegraphics[trim = {1.2cm 0.3cm 2.0cm 0.3cm},clip,width=\textwidth]{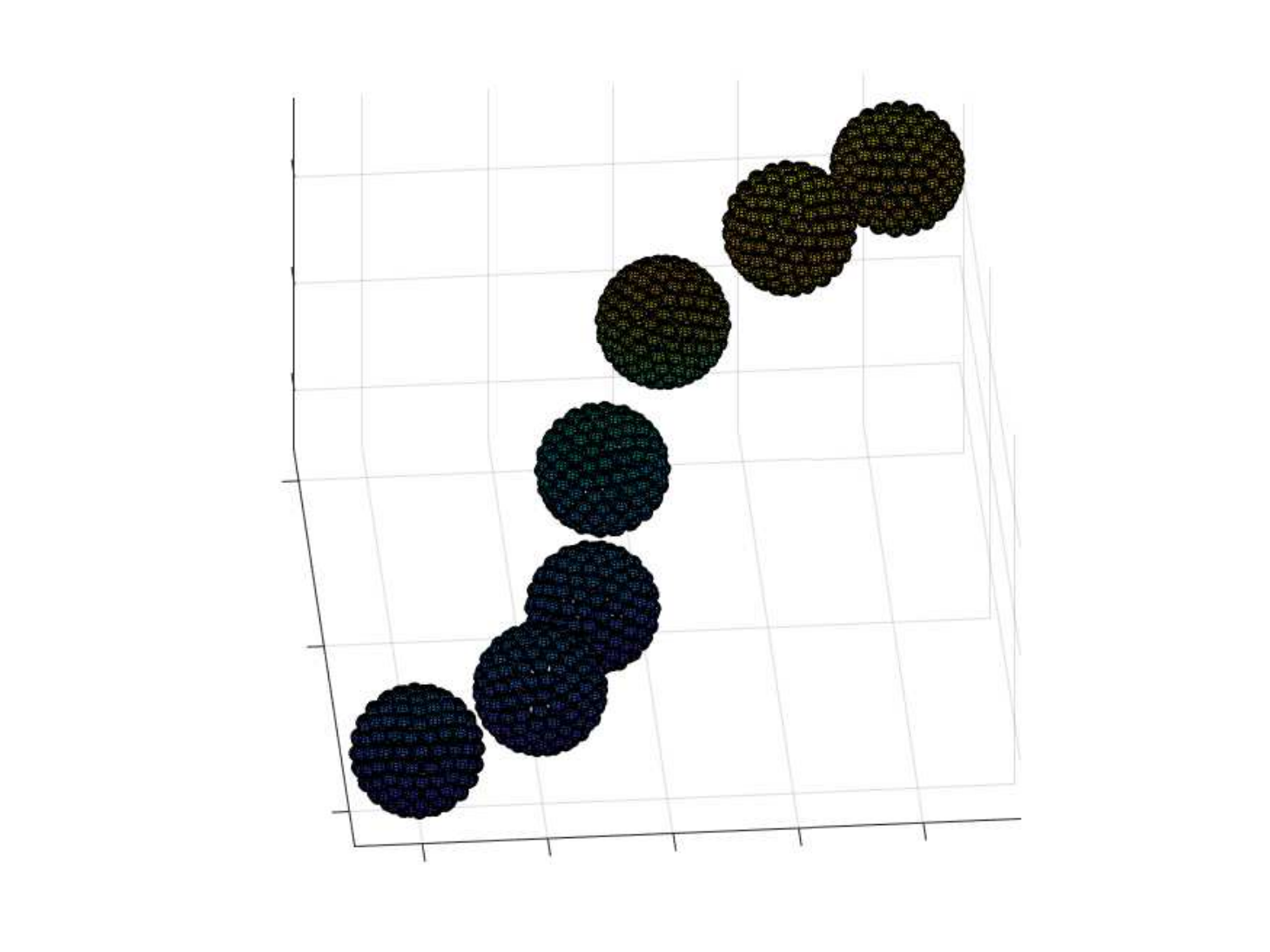}
		\caption{Example geometry: 162 blobs per sphere}
		%\caption{Results for a chain of rods of aspect ratio 20.}
		%	\label{rods20}
	\end{subfigure}
	\caption{The example in Section \ref{Ex:twist_spheres}: Chains of seven spheres with the same relative transformation used between two consecutive particles.}
	\label{chain_spheres}
\end{figure}

\begin{figure}[h!]
	\centering
	\hspace*{2ex}
	\makebox[\textwidth][c]{%
		\begin{subfigure}[t]{0.3\textwidth}
			\centering
			\includegraphics[trim = {3cm 19.5cm 12.5cm 2.4cm},clip,width=\textwidth]{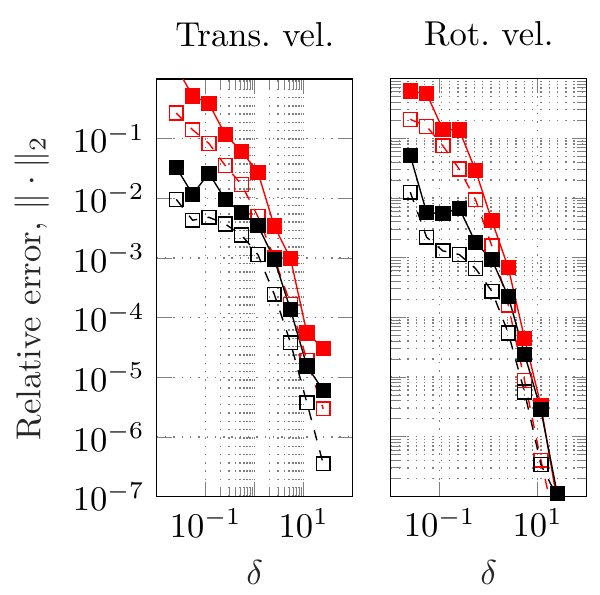}
			\caption{12 blobs per sphere}
			%\caption{Results for a chain of rods of aspect ratio 4.}
			%	\label{rods20}
		\end{subfigure}
		\begin{subfigure}[t]{0.3\textwidth}
			\centering
			%\hspace*{-9ex}
			\includegraphics[trim = {3cm 19.5cm 12.5cm 2.4cm},clip,width=\textwidth]{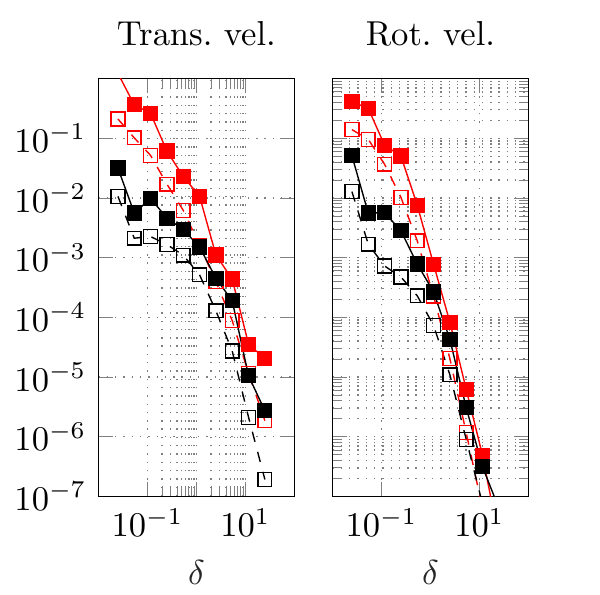}
			\caption{42 blobs per sphere}
			%\caption{Results for a chain of rods of aspect ratio 20.}
			%	\label{rods20}
		\end{subfigure}
		
		\begin{subfigure}[t]{0.3\textwidth}
			\centering
			\hspace*{-7ex}
			\includegraphics[trim = {3cm 19.5cm 12.5cm 2.4cm},clip,width=\textwidth]{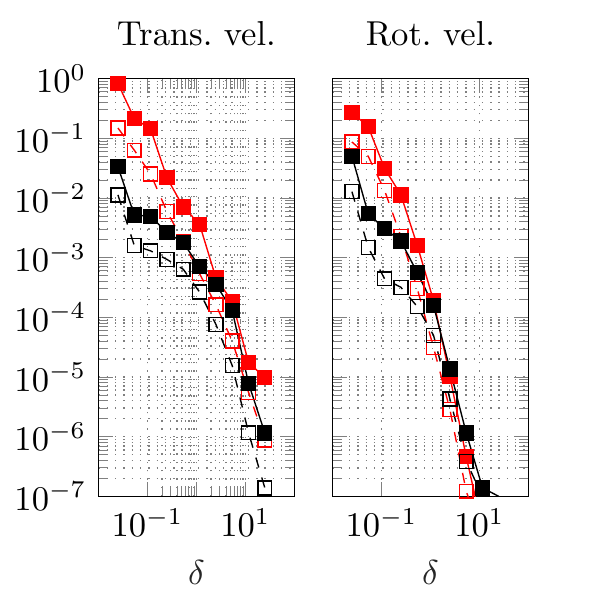}
			\caption{162 blobs per sphere}
			%\caption{Results for a chain of rods of aspect ratio 20.}
			%	\label{rods20}
		\end{subfigure}

		\begin{subfigure}[t]{0.18\textwidth}
			\centering
			\vspace{-25ex}
			\hspace*{-7ex}
			%\vspace*{-20ex}
			\includegraphics[trim = {9.5cm 19.5cm 3cm 4cm},clip,width=2.3\textwidth]{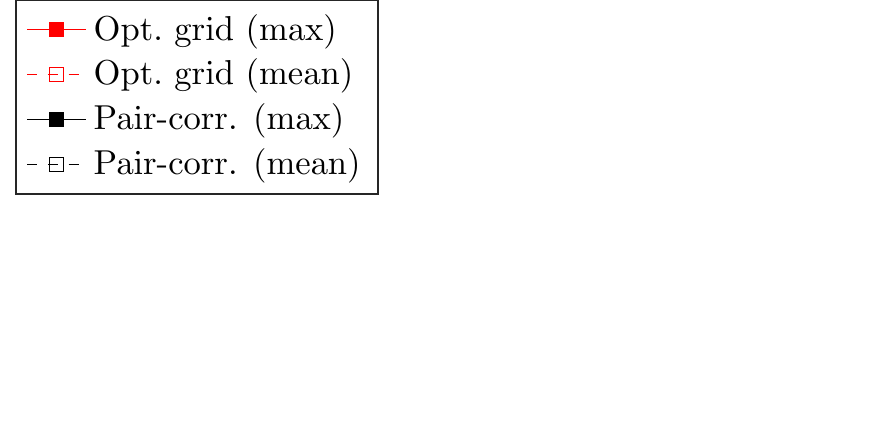}
			%\caption{162 blobs per sphere}
			%\caption{Results for a chain of rods of aspect ratio 20.}
			%	\label{rods20}
		\end{subfigure}
	}
	\caption{Velocity errors for data accumulated from all seven spheres in ten different randomised particle chains, with example chains displayed in Figure \ref{chain_spheres}, formed with the same relative transformation between two consecutive particles. The optimised multiblob grids defined by $(R_g,a_h)$ are used for each resolution. The mean and max velocity error is displayed for the translational velocity and rotational velocity, versus the inter-particle distance $\delta$ for different number of blobs in the discretisation.   Pair-corrections improve on the error if $\delta$ is sufficiently small.}
	\label{corr_sphere_chain}
\end{figure}

We consider a setting where seven spheres form a chain: The first sphere is placed at the origin, with the next placed at a prescribed distance $\delta$ in the randomized direction $\vec v$ from the first particle, rotated by $(\theta,\varphi)$.  Here, $\vec v$ is a vector in the first orthant and $(\theta,\varphi)$ are sampled from the first orthant. The next spheres in the chain are placed accordingly, by translating the center coordinate by $\delta \vec v$ in the coordinate frame of the previous particle and then rotating. Example geometries are visualised in Figure \ref{chain_spheres}. For each chain, a particular force and torque is set for each sphere, chosen with each component independently drawn from $\mathcal U(-1,1)$ resulting in a somewhat arbitrary force/torque vector with $\|\vec f^i\|\approx\|\vec t^i\|$. This choice means that we mainly look at the error in the main diagonal blocks of the mobility matrix as these blocks will be dominant. The mean and maximum relative error in the particle velocity for three different resolutions of the multiblob spheres are displayed in Figure \ref{corr_sphere_chain} with and without pair-corrections. An improvement in accuracy can be noted for the finest blob grid compared to the coarsest and pair-corrections improve on the accuracy for the small inter-particle distances.

% Applying a pair-correction between all pairs of spheres, the result in Figure \ref{corr_sphere_chain} is obtained. 
%\begin{figure}[h!]
%	\centering
%	\includegraphics[trim = {3cm 17.2cm 6.2cm 2.5cm},clip,width=0.65\textwidth]{figures/twist_nocorr.pdf}
%	\caption{Velocity errors for data accumulated from all seven spheres in ten different randomised particle chains where the optimised multilob grids are used for each resolution. The chains are formed with the same relative transformation used between two consecutive particles. The mean and max velocity error is displayed for the translational velocity (left) and angular velocity (right), versus the inter-particle distance $\delta$ for different number of blobs in the discretisation.}
%	\label{sphere_chain}
%\end{figure} 

%%%%%%%%%%%%%%%%%%%%%%%%%%%%%%%%%%%%%%%%%%%%%%%%%%%%%%%%%%%%%%%%%%%%%%%%%%%%%%%%%%%%%%%%%%
\subsection{Twisted rod chains - eight rods}\label{sec:twist_eight}
\begin{figure}[h!]
	\centering
	%\hspace*{-20ex}
	\includegraphics[trim = {1.6cm 18.9cm 1.8cm 3.4cm},clip,width=\textwidth]{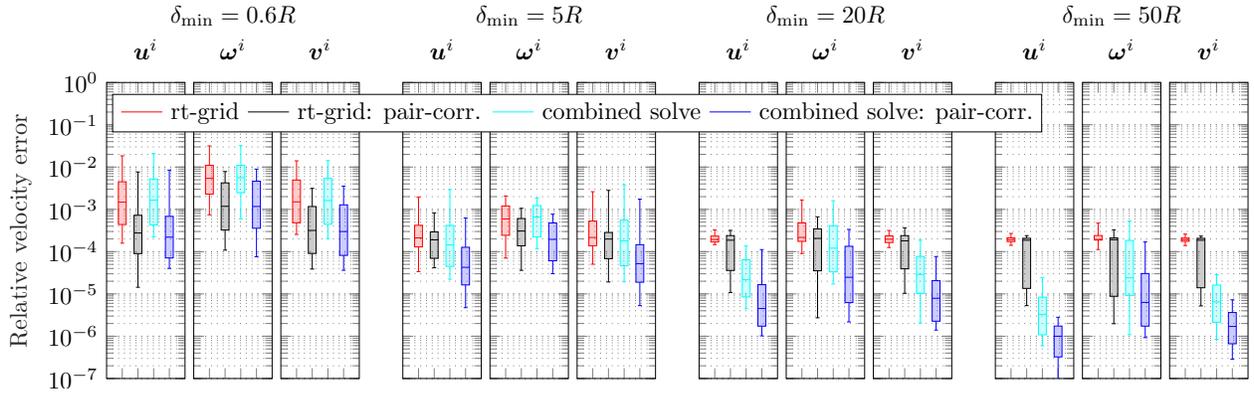}
	\caption{Relative error in the translational, rotational and tip velocity for each rod in each of the ten chain configurations displayed in Figure \ref{chain2} for slender rods with $L/R = 20$ discretised with the grid $n_{\text{cap}}=3$, $n_{\text{cyl}}=20$, $n_{\varphi}=6$. Whiskers in box plots display the minimum and maximum error, box edges display the 10th and 90th percentile of the error and the box center line display the error median. For larger particle distances (in the rightmost panels), the error using the rt-grid with or without pair-corrections is dominated by the self-interaction error, and plateaus at this level. The self-interaction error can be improved upon with a combined solve and a pair-correction applied to the combined solve can reduce interaction errors. For small separation distances (in the leftmost panels), a pair-correction can be applied directly to the rt-grid to enhance accuracy, as the interaction error is the dominating contribution to the error here. Note that the pair-corrections are only applied to neighbouring particles in this test. }
	\label{chain_8p}	
\end{figure}

We repeat the test with rod chains, as described in Section \ref{sec:twist_three}, but now for eight particles and a specific force/torque vector $\vec F$, with each component independently drawn from $\mathcal U(-1,1)$. This means that here, $\|\vec f^i\| \approx \|\vec t^i\|$. Statistics for the relative error in the translational, rotational and tip velocities for all particles in ten different chains is visualised in Figure \ref{chain_8p}, for rods with $L/R=20$ and a semi-coarse discretisation. Pair-corrections are only applied to neighbouring particles. If applied to the rt-grid, error levels are decreased for small $\delta_{\min}$, whereas for distances $\delta_{\min}$ where the self-interaction start to become dominant in the mobility matrix relative to the errors due to particle interactions, the errors corresponding to pair-corrections plateau at the same level as when using the non-corrected rt-grid. If the combined solve is used instead, the accuracy can be further enhanced with a pair-correction, as self-interaction errors then are small.  

%\begin{figure}[h!]
%	\centering
%	\includegraphics[trim = {0cm 0cm 0cm 0cm},clip,width=\textwidth]{figures/twisted_nocorr_np8_ar4_tip_N648.pdf}
%	\caption{Error in the velocity at the tip of each rod.  Aspect ratio $L/R = 4$ and the grid $n_{\text{cap}}=6$, $n_{\text{cyl}}=4$ and $n_{\varphi}=8$. Colors: \colorbox{cyan}{rt-grid with pair-correction}, \colorbox{magenta}{rt-grid}, black: pair-correction with old grid, \colorbox{blue}{\quad}self-correction with old grid,  \colorbox{red}{old grid}}	
%\end{figure}

%\begin{figure}[h!]
%	\centering
%	\includegraphics[trim = {0cm 0cm 0cm 0cm},clip,width=\textwidth]{figures/twisted_nocorr_np8_ar20_tip_N3206.pdf}
%	\caption{Error in the velocity at the tip of each rod.  Aspect ratio $L/R = 20$ and the grid $n_{\text{cap}}=3$, $n_{\text{cyl}}=20$ and $n_{\varphi}=6$. Colors: \colorbox{cyan}{rt-grid with pair-correction}, \colorbox{magenta}{rt-grid}, black: pair-correction with old grid, \colorbox{blue}{\quad}self-correction with old grid,  \colorbox{red}{old grid}}	
%\end{figure}

\begin{figure}[h!]
	\centering
	\begin{subfigure}[t]{0.2\textwidth}
		\centering
		\hspace*{-5ex}
		\includegraphics[trim = {6cm 9cm 6cm 9cm},clip,width=\textwidth]{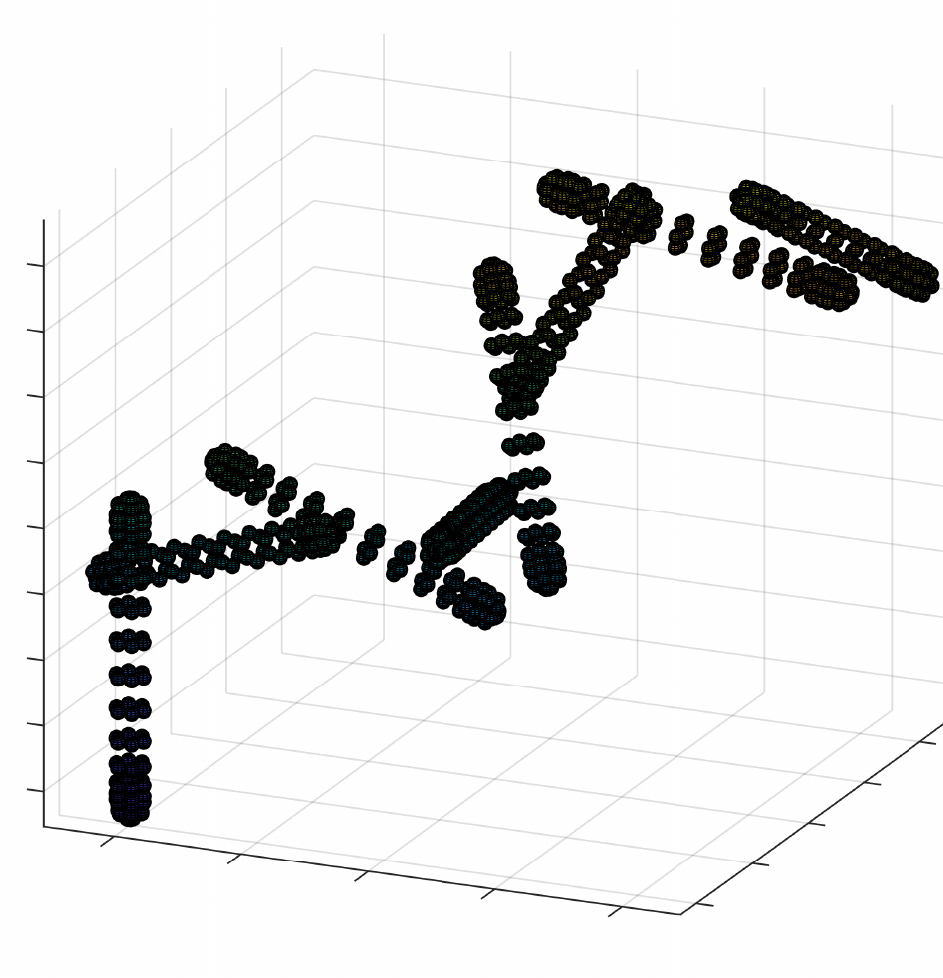}
		\caption{}
		%\caption{Results for a chain of rods of aspect ratio 4.}
		%	\label{rods20}
	\end{subfigure}~
	\begin{subfigure}[t]{0.19\textwidth}
		\centering
		\hspace*{-2ex}
		\includegraphics[trim = {6cm 9cm 7cm 9cm},clip,width=\textwidth]{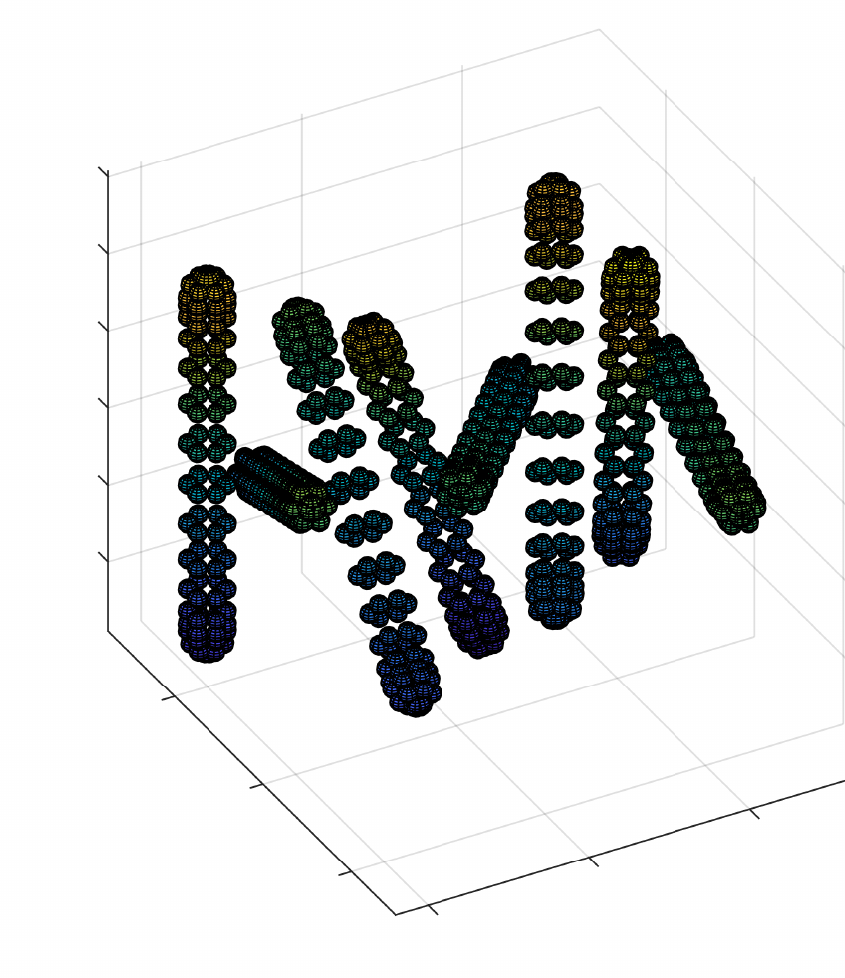}
		\caption{}
		%\caption{Results for a chain of rods of aspect ratio 20.}
		%	\label{rods20}
	\end{subfigure}
	\begin{subfigure}[t]{0.16\textwidth}
		\centering
		%\hspace*{-9ex}
		\includegraphics[trim = {7cm 9cm 8cm 9cm},clip,width=\textwidth]{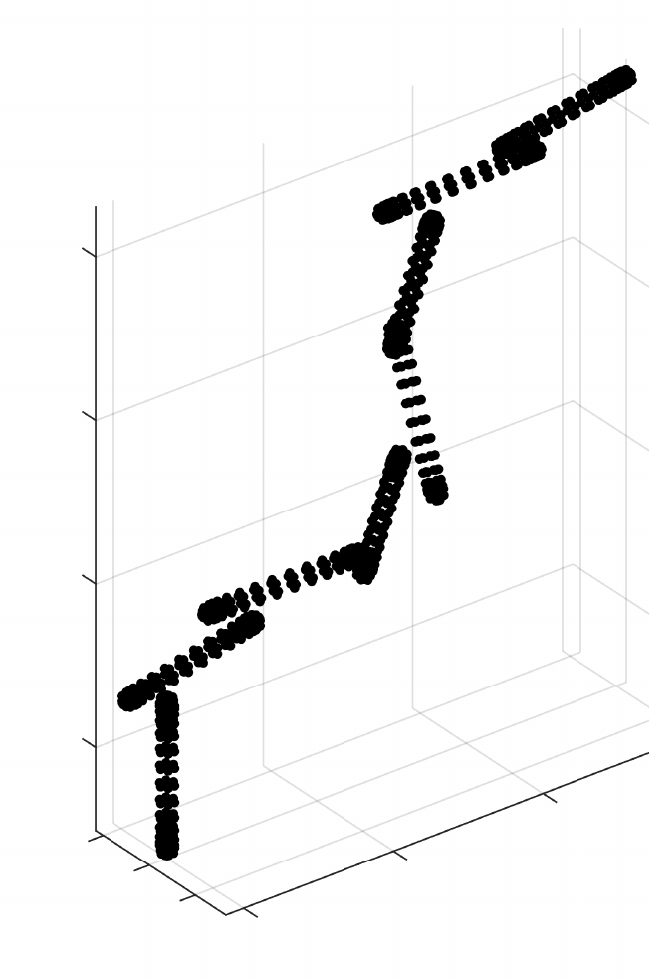}
		\caption{}
		%\caption{Results for a chain of rods of aspect ratio 20.}
		%	\label{rods20}
	\end{subfigure}
	\begin{subfigure}[t]{0.22\textwidth}
		\centering
		%\hspace*{-9ex}
		\includegraphics[trim = {6cm 9cm 6cm 10cm},clip,width=\textwidth]{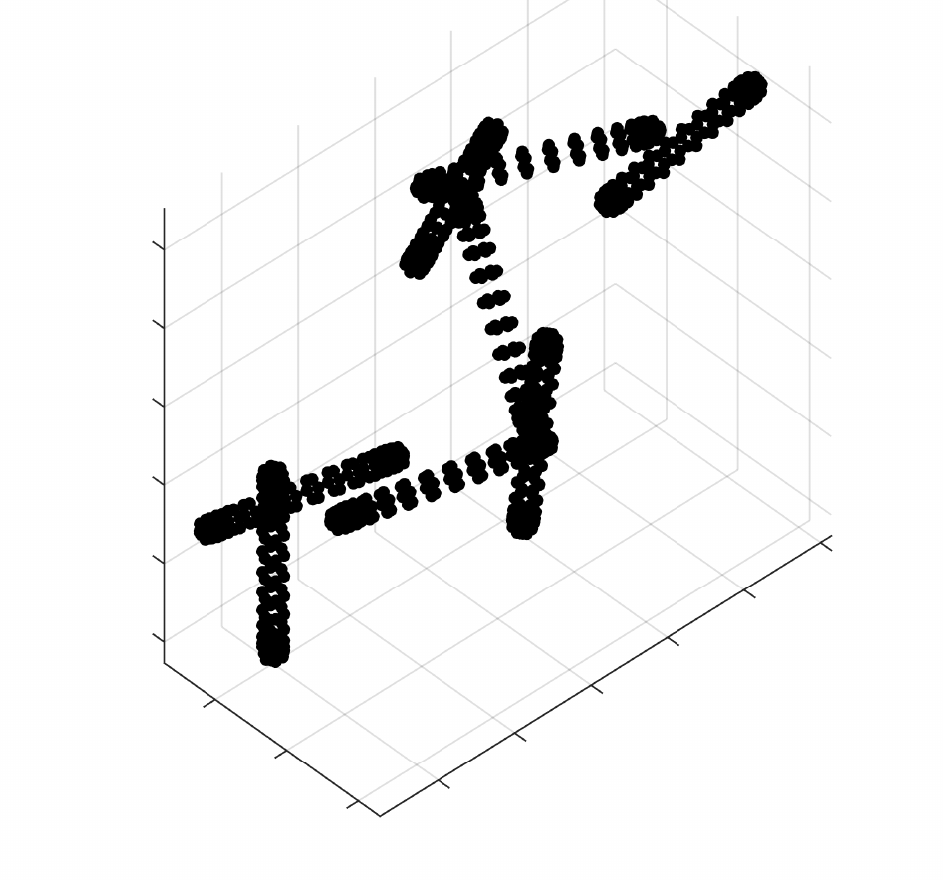}
		%\caption{Results for a chain of rods of aspect ratio 20.}
		\caption{}
		%	\label{rods20}
	\end{subfigure}
	\begin{subfigure}[t]{0.16\textwidth}
		\centering
		%\hspace*{-9ex}
		\includegraphics[trim = {6cm 9cm 6.5cm 9cm},clip,width=\textwidth]{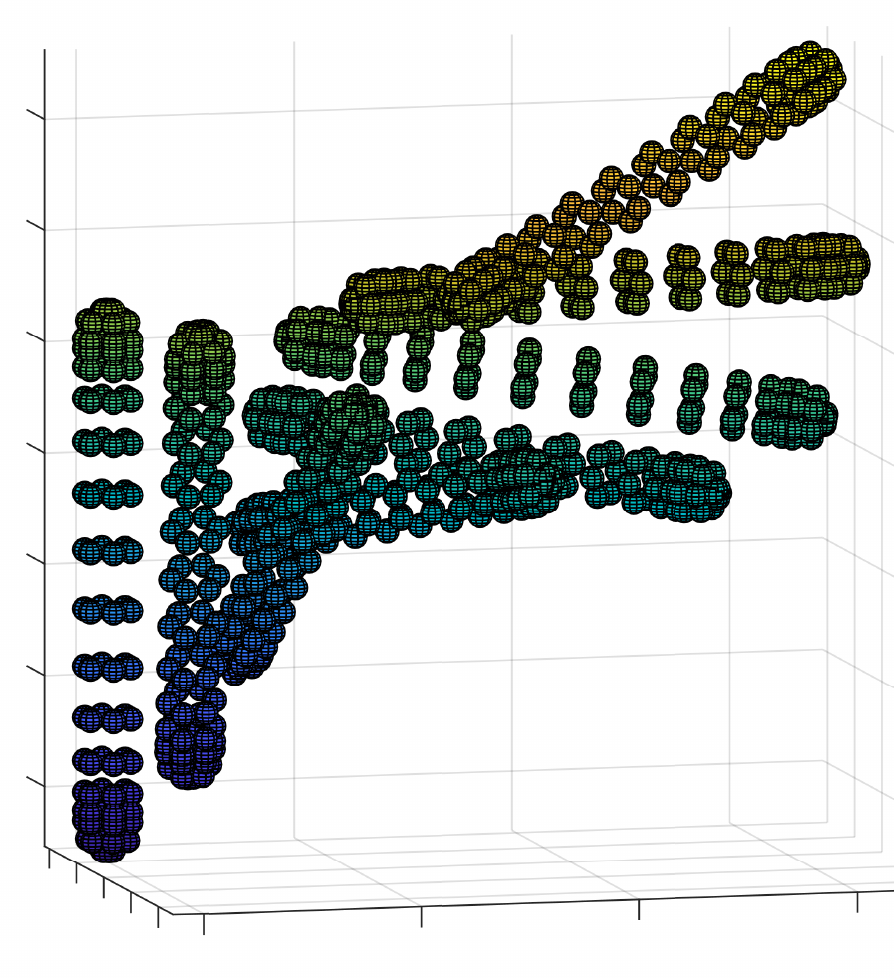}
		\caption{}
		%\caption{Results for a chain of rods of aspect ratio 20.}
		%	\label{rods20}
	\end{subfigure}	
	
	\begin{subfigure}[t]{0.18\textwidth}
		\centering
		%\hspace*{-9ex}
		\includegraphics[trim = {6cm 9cm 6cm 9cm},clip,width=\textwidth]{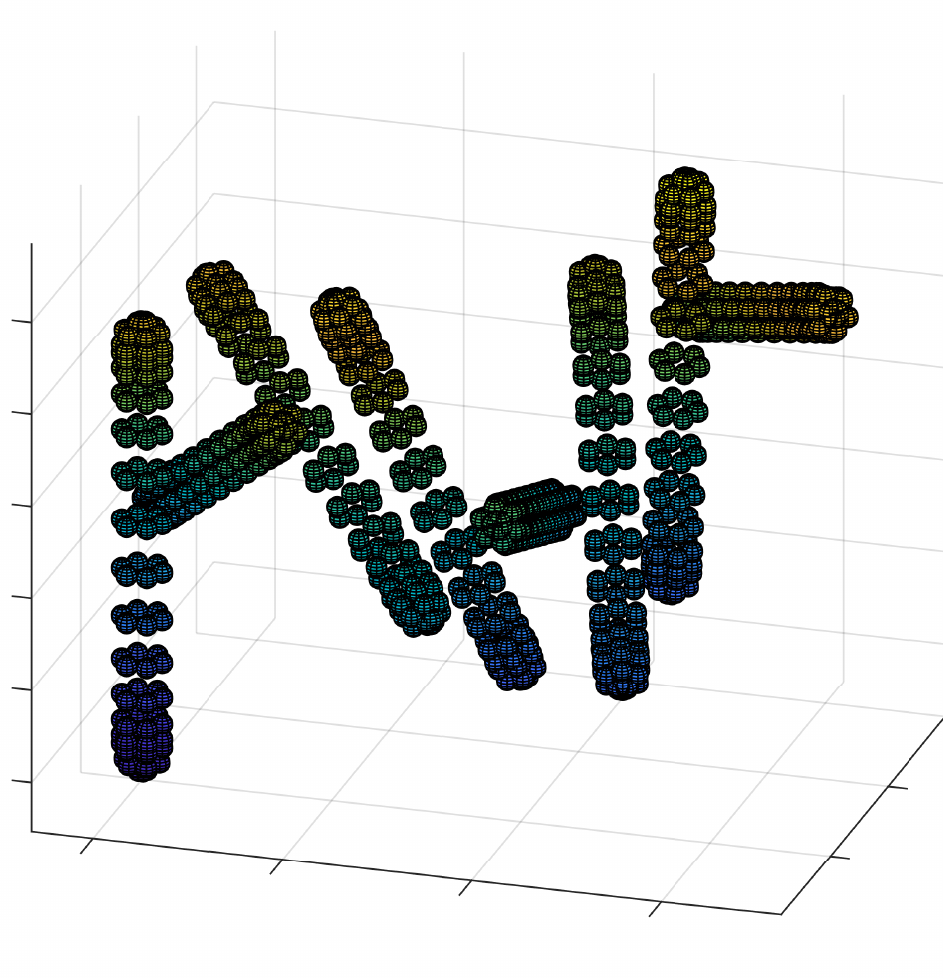}
		\caption{}
		%\caption{Results for a chain of rods of aspect ratio 20.}
		%	\label{rods20}
	\end{subfigure}
	\begin{subfigure}[t]{0.21\textwidth}
		\centering
		%\hspace*{-9ex}
		\includegraphics[trim = {6cm 9cm 6cm 10cm},clip,width=\textwidth]{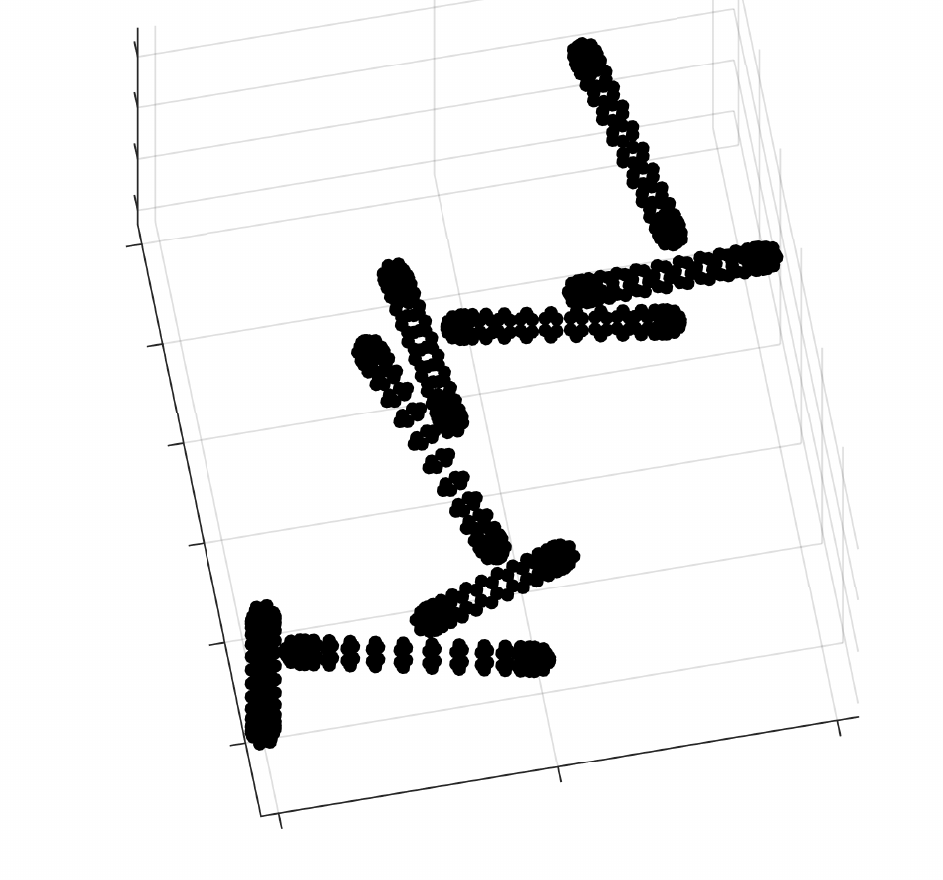}
		\caption{}
		%\caption{Results for a chain of rods of aspect ratio 20.}
		%	\label{rods20}
	\end{subfigure}	
	\begin{subfigure}[t]{0.2\textwidth}
		\centering
		%\hspace*{-9ex}
		\includegraphics[trim = {6cm 9cm 6cm 9cm},clip,width=\textwidth]{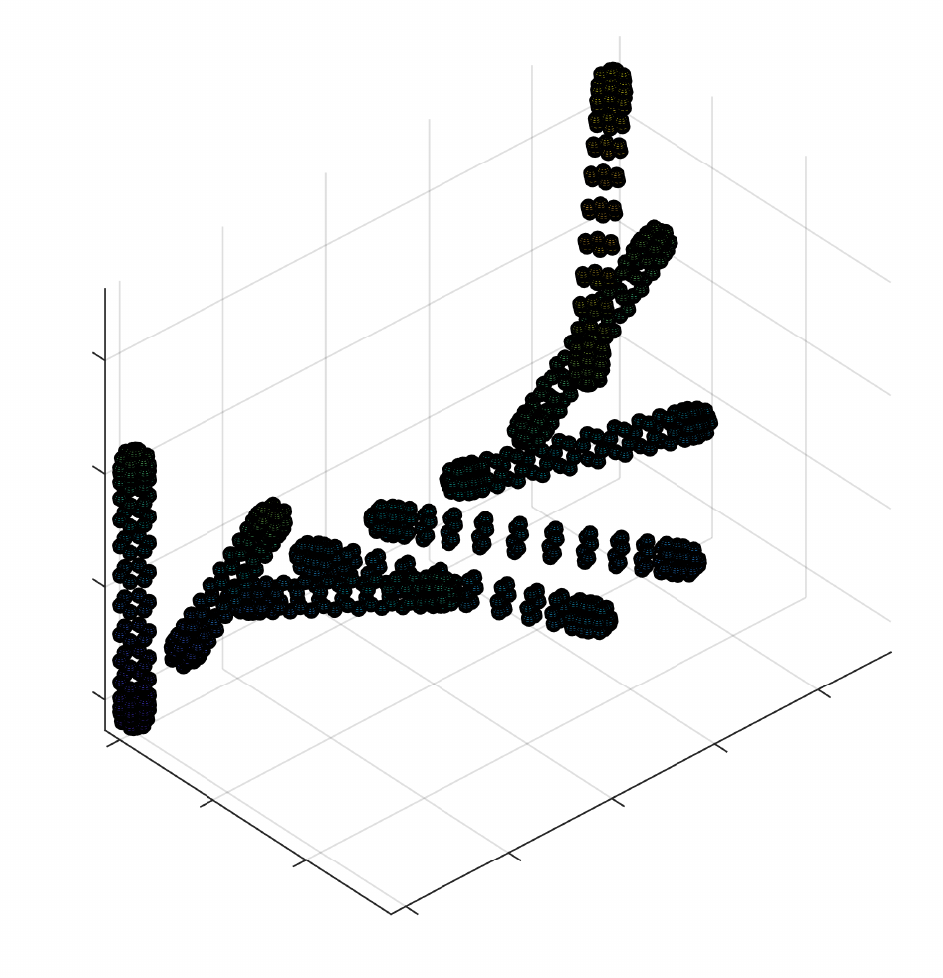}
		\caption{8}
		%\caption{Results for a chain of rods of aspect ratio 20.}
		%	\label{rods20}
	\end{subfigure}	
	\begin{subfigure}[t]{0.16\textwidth}
		\centering
		%\hspace*{-9ex}
		\includegraphics[trim = {7cm 9cm 7.5cm 8cm},clip,width=\textwidth]{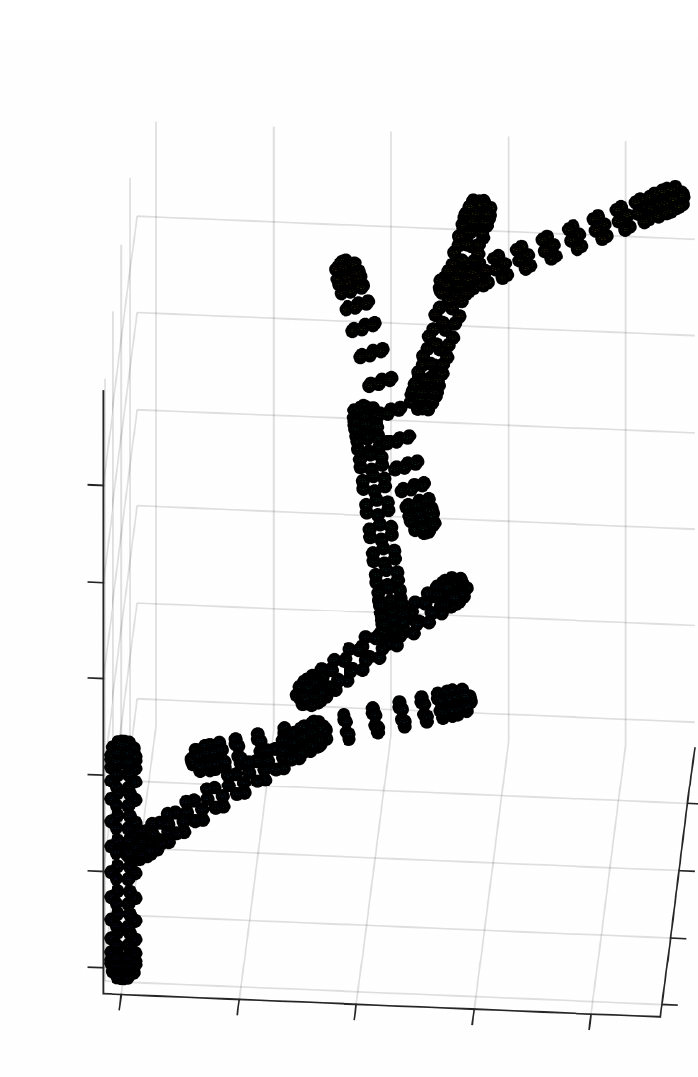}
		\caption{9}
		%\caption{Results for a chain of rods of aspect ratio 20.}
		%	\label{rods20}
	\end{subfigure}	
	\begin{subfigure}[t]{0.14\textwidth}
		\centering
		%\hspace*{-9ex}
		\includegraphics[trim = {7cm 9cm 8cm 8cm},clip,width=\textwidth]{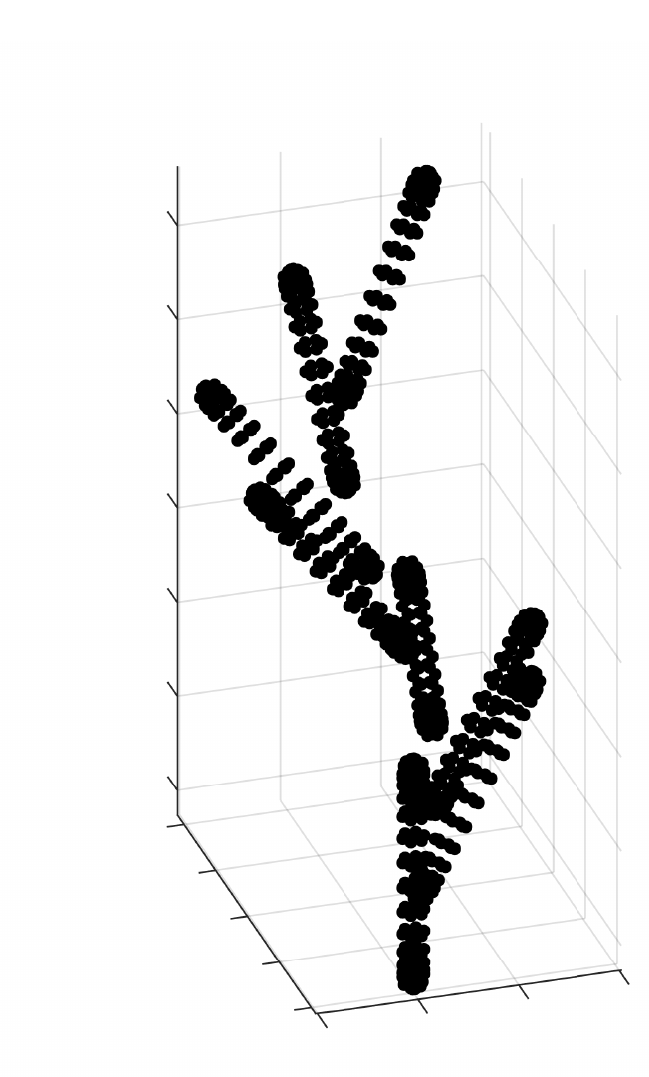}
		\caption{}
		%\caption{Results for a chain of rods of aspect ratio 20.}
		%	\label{rods20}
	\end{subfigure}	
	\caption{Chains of eight rods of aspect ratio $L/R = 20$ with the same relative transformation used between two consecutive rods considered in the example in Section \ref{sec:twist_eight}.  Similar chains of three rods are considered in Section \ref{sec:twist_three}.}
	\label{chain2}
\end{figure}

\end{document}